\documentclass[12pt,letterpaper,oneside,american,english]{amsart}
\usepackage{lmodern,textcomp,amsthm,graphicx,setspace,fullpage,pdflscape,amsmath,babel,courier, ragged2e,booktabs,dcolumn,multirow,enumitem,afterpage,bbm,bigdelim}

\usepackage[T1]{fontenc}
\usepackage[latin9]{inputenc}
\usepackage[authoryear]{natbib}
\usepackage[justification=centering,font=scriptsize]{caption}
\usepackage[flushleft]{threeparttable}
\usepackage[percent]{overpic}

\usepackage{pdfcolmk,rotfloat,textcomp,mathrsfs,amsthm,amstext}
\usepackage[usenames, dvipsnames]{xcolor}
\PassOptionsToPackage{normalem}{ulem}
\usepackage{ulem}
\providecolor{lyxadded}{rgb}{0,0,1}
\providecolor{lyxdeleted}{rgb}{1,0,0}

\theoremstyle{plain}

\newtheorem*{prop*}{\protect\propositionname}
\addto\captionsamerican{\renewcommand{\propositionname}{Proposition}}
\addto\captionsenglish{\renewcommand{\propositionname}{Proposition}}
\providecommand{\propositionname}{Proposition}
\theoremstyle{definition}

\makeatletter

\def\@biblabel#1{\hspace*{-\labelsep}}
\makeatother
\def\sym#1{\ifmmode^{#1}\else\(^{#1}\)\fi}

\newcolumntype{L}[1]{>{\raggedright\let\newline\\\arraybackslash\hspace{0pt}}m{#1}}
\newcolumntype{C}[1]{>{\centering\let\newline\\\arraybackslash\hspace{0pt}}m{#1}}
\newcolumntype{R}[1]{>{\raggedleft\let\newline\\\arraybackslash\hspace{0pt}}m{#1}}
\newcommand{\MONTH}{\ifcase\the\month \or January \or February \or March \or April \or May \or June \or July \or August \or September \or October \or November \or December \fi}
\usepackage{etoolbox}
\makeatletter
\patchcmd{\@settitle}{\uppercasenonmath\@title}{}{}{}
\patchcmd{\@setauthors}{\MakeUppercase}{}{}{}
\makeatother

\pdfpageheight\paperheight
\pdfpagewidth\paperwidth
\usepackage[hidelinks]{hyperref} 

\begin{document}

\title{{\large{The direct and spillover effects of large-scale affirmative action \\ at an elite Brazilian university}}}

\author{\vspace{0.3in} \normalsize{Cecilia Machado \\  Germ\'an Reyes \\ Evan Riehl \\ \\ 
 \MONTH \the\year\\  \\}
 \vspace{0.5in} }
 
\thanks{{\footnotesize{Machado: FGV EPGE, Praia de Botafogo, 190 11\textsuperscript{o} andar, Rio de Janeiro, RJ 22250-900, Brazil (email: cecilia.machado@fgv.br). Reyes: Department of Economics, Cornell University, 404 Uris Hall, Ithaca, NY 14853, United States (email: gjr66@cornell.edu). Riehl (corresponding author): Department of Economics and ILR School, Cornell University, 266 Ives Hall, Ithaca, NY 14853, United States (email: eriehl@cornell.edu). For helpful comments we thank Jim Berry, Sandra Black, Fernanda Estevan, Chao Fu, Laura Gee, Adam Kapor, Josh Kinsler, Michael Lovenheim, Tatiana Mocanu, Ana Ribeiro, Eduardo Pontual Ribeiro, Seth Sanders, Fabio Schanaider, Miguel Urquiola, Russell Weinstein, and participants at various seminars and conferences. We are grateful to Universidade do Estado do Rio de Janeiro (UERJ) for assistance with the data. Cecilia Machado acknowledges funding from the Coordena\c{c}\~ao de Aperfei\c{c}oamento de Pessoal de N\'ivel Superior - Brasil (CAPES) - Finance Code 001. The opinions expressed in this paper are those of the authors alone.}}}

\begin{abstract}
\begin{singlespace}
\noindent We examine the effects of an affirmative action policy at an elite Brazilian university that reserved 45 percent of admission slots for Black and low-income students. We find that marginally-admitted students who enrolled through the affirmative action tracks experienced a 14 percent increase in early-career earnings. But the adoption of affirmative action also caused a large decrease in earnings for the university's most highly-ranked students. We present evidence that the negative spillover effects on highly-ranked students' earnings were driven by both a reduction in human capital accumulation and a decline in the value of networking.

\end{singlespace}
\end{abstract}
\maketitle
\newpage{}

\pagenumbering{arabic}

\clearpage


Top universities face growing pressure to increase their students' racial and socioeconomic diversity. \citet{chetty2020income} argue that large-scale income-based affirmative action at selective colleges could significantly increase intergenerational mobility in the United States. Consistent with this, \citet{bleemer2022affirmative} finds that race-based affirmative action at the University of California increased underrepresented minority students' earnings. \citeauthor{bleemer2022affirmative} argues that the policy improved allocative efficiency because displaced students were not worse off.

These arguments about affirmative action's benefits assume selective universities can increase diversity without reducing their value added.\footnote{\enskip \citet{chetty2020income} write: ``[W]e also assume that [our] estimated causal effects do not change under our counterfactual student reallocations, in particular ignoring potential changes in value-added that may arise from having a different group of students (peer effects)'' (p.\ 1626). Similarly, \citet{bleemer2022affirmative}'s claim that affirmative action improved allocative efficiency relies on the untested assumption that the policy did not reduce the returns of inframarginal white and Asian students.} Yet the value of attending a top college may depend on its student body composition. At schools with high-achieving students, professors can teach courses at an advanced level \citep{duflo2011}, and individuals may learn from peers with similar academic preparation \citep{carrell2013,arcidiacono2016}. Schools with wealthy student bodies provide access to peers and alumni in high-paying sectors of the economy \citep{zimmerman2019,michelman2022old}. Employers' recruiting and hiring decisions may depend on the expected ability of a school's students \citep{macleod2017,weinstein2018employer}. If these mechanisms are important, significantly increasing the scale of affirmative action can negatively affect all students' outcomes.  There is little compelling evidence on the existence and magnitude of such spillovers because isolating variation in the composition of a college's student body is challenging.

We examine the direct and spillover effects of large-scale affirmative action at Rio de Janeiro State University (UERJ), one of Brazil's most prestigious universities. UERJ consistently ranks among the top 15 universities nationally. In some years, over 100,000 students take UERJ's entrance exam, competing for roughly 5,000 admissions. Thus, UERJ's national prestige and selectivity are comparable to elite U.S.\ private colleges.

UERJ was among the first Brazilian universities to adopt affirmative action. It did so on a large scale. Historically, white students from private high schools were disproportionately likely to gain admission through UERJ's entrance exam. Starting in 2004, UERJ reserved 45 percent of slots in each major for Black and public high school students from low-income families. This policy suddenly and dramatically increased the racial and socioeconomic diversity of UERJ's students

We collected data on the schooling and labor market outcomes of students who applied to UERJ before and after the adoption of affirmative action (AA). Our base dataset includes entrance exam scores and admission outcomes for all UERJ applicants in 1995--2001 (pre-AA) and 2004--2011 (post-AA). We link these data to UERJ enrollment/graduation records, and to Brazil's national employer-employee dataset for the years 2003--2019.

Our analysis exploits variation in exposure to UERJ's affirmative action policy across majors. Admission to UERJ is major-specific, and while the fraction of slots reserved for affirmative action was the same in each major, the \emph{take-up} of these slots varied. In UERJ's most prestigious programs, affirmative action students made up 45 percent of the incoming class because the number of applicants typically exceeded the reserved quotas. The quotas often went unfilled in less-selective programs, and UERJ would fill open seats from the general applicant pool. Thus the share of enrollees who were from an affirmative action track was 10--20 percent in some programs.

We use two empirical strategies to identify the effects of affirmative action on its intended beneficiaries and on other UERJ students. In majors with high take-up of affirmative action, we use a regression discontinuity (RD) design that compares applicants above and below admission score cutoffs \citep{hoekstra2009,kirkeboen2016}. Our RD design identifies the returns to attending UERJ for marginally-admitted applicants in each track.

Our second strategy exploits variation in affirmative action take-up to identify the policy's spillover effects on other UERJ students. We use a difference-in-differences (DD) design that estimates changes in outcomes between pre- and post-AA cohorts, and across majors with higher and lower take-up. This analysis focuses on a sample of \emph{top enrollees} whose entrance exam scores were high enough to gain admission regardless of whether affirmative action existed in their cohort. Our DD design identifies the effects of a 19 percentage point increase in the share of top enrollees' classmates who were from an affirmative action track.

We have two main findings. First, for marginally-admitted affirmative action students, enrolling in UERJ led to a 14 percent increase in early-career hourly wages. We find no effects of UERJ enrollment on college degree attainment, but affirmative action enrollees were significantly more likely to obtain jobs at high-paying firms affiliated with UERJ alumni. This suggests that their early-career earnings gains were primarily driven by networking mechanisms. We find that the earnings and networking benefits decreased as affirmative action students' careers progressed, but our later-career results are less powered.

Second, the adoption of affirmative action lowered the earnings of UERJ's highly-ranked students. In our DD analysis, top enrollees' hourly wages decreased by 14 percent in majors with high affirmative action take-up relative to those with lower take-up. This effect persisted up through the end of our data range. We also find declines in earnings for highly-ranked underrepresented minority students who could have gained admission to UERJ in absence of affirmative action. We do not find significant changes in the characteristics and admission scores of top enrollees in more- vs. less-affected majors, although point estimates suggest that compositional changes could play a small role in our results. Instead, we find evidence that the negative spillover effects on earnings were driven by both networking and learning mechanisms. The adoption of affirmative action reduced the likelihood that top enrollees obtained jobs at high-paying firms affiliated with UERJ peers and alumni, and it reduced the performance of UERJ's top students on a college exit exam.

Our findings show that elite universities face a tradeoff between promoting upward mobility for disadvantaged students and maintaining sources of their value added that stem from admitting high-achieving and wealthy students. Further, our results suggest that disadvantaged students with the highest admission scores may have been better off with a smaller-scale affirmative action policy.

Our early-career results for affirmative action students are consistent with other evidence that disadvantaged students benefit from attending selective universities. There is a large literature on university affirmative action, but there is limited evidence its earnings impacts \citep{arcidiacono2015}.\footnote{\enskip Other research on affirmative action looks primarily at impacts on diversity or graduation rates \citep{cortes2010,backes2012,hinrichs2012,kapor2015,arcidiacono2016,bagde2016}. This is true of most work on affirmative action in Brazil \citep{francis2012using,ribeiro2016affirmative,estevan2019redistribution,vieira2019,otero2021,ribeiro2021longterm,mello2022centralized}.} \citet{bertrand2010} and \citet{bleemer2022affirmative} find earnings gains for disadvantaged students who were given admission preference at selective colleges in India and the United States. Similarly, \citet{francis2018black} find earnings benefits for male students admitted through reserved quotas at the University of Bras\'ilia. Related work finds earnings gains for low-income or minority students who were marginally-admitted to U.S.\ public university systems \citep{zimmerman2014,smith2020,bleemer2021}. Our estimate of the early-career earnings return for affirmative action students---a 14 percent---is much smaller than analogous estimates from many of these papers.\footnote{\enskip \citet{zimmerman2014} finds that \emph{admission} to the Florida State University system increased the likelihood of enrolling by roughly 50 percent, and it increased earnings by 22 percent. \citet{bleemer2022affirmative} finds that an affirmative action ban decreased minority students' enrollment in selective University of California colleges by eight percentage points, and earnings fell by 0.05 log points. These estimates imply returns to selective college enrollment of roughly 44--87 percent.\label{fn:magnitudes}} This may be because affirmative action did not affected the educational attainment of UERJ applicants, whereas these papers often find effects on bachelor's degree attainment.

Our findings are new in showing that affirmative action can benefit disadvantaged students through networking. \citet{zimmerman2019} and \citet{michelman2022old} find that networking is an important mechanism for the long-run earnings benefits of attending elite universities, but that only students from advantaged backgrounds benefit from networking. Our data is unique in measuring early-career employment in a broad set of firms. Our results suggest that affirmative action students can also benefit from access to high-wage firms affiliated with alumni, at least early in their careers. We find similar effects for marginal enrollees from the general track, suggesting that networking is an important mechanism in research on the returns to college selectivity \citep[e.g.,][]{dale2002}.\footnote{\enskip There is a large literature on the earnings returns to attending selective colleges and/or majors \citep{hoekstra2009,saavedra2009,hastings2013,kirkeboen2016,canaan2018,hoxby2018,anelli2020returns,sekhri2020,ng2022returns}. These papers typically cannot examine both job networks and earnings. We contribute to a small literature on network formation in college \citep{marmaros2002peer,mayer2008old,zhu2023beyond}.}

Lastly, our paper is novel in identifying spillover effects of large-scale affirmative action. Several papers examine the efficiency effects of admission policies that benefit disadvantaged students by comparing earnings returns for students who were ``pulled in'' and ``pushed out.'' The evidence is mixed; \citet{bleemer2022affirmative} finds efficiency gains, \citet{bertrand2010} and \citet{riehl2023exams} find efficiency losses, and \citet{black2023winners} find limited earnings effects in either group.\footnote{\enskip A related literature examines student/college match effects in graduation and earnings outcomes \citep{andrews2016,arcidiacono2016,dillon2020consequences,mountjoy2020returns}.} A full evaluation of the efficacy of affirmative action must also consider spillover effects on untargeted students \citep{durlauf2008}. Several papers examine how a university's racial or socioeconomic diversity affects other students' earnings \citep{daniel2001racial,arcidiacono2010does,hinrichs2011effects}, but this work relies on strong selection-on-observables assumptions. We find negative earnings spillovers under weaker assumptions, and we present evidence on both learning and networking mechanisms. The existence of spillovers means that the true effects of large-scale admission reforms may differ from those estimated using existing student/college allocations, as in, for example, \citet{chetty2020income}'s ``need-affirmative'' counterfactual enrollment scenario.


\section{Context and data}\label{sec:context}

\subsection{UERJ and higher education in Brazil}

Our setting is an elite public university in Brazil called Rio de Janeiro State University, or UERJ (\emph{Universidade do Estado do Rio de Janeiro}). It is one of the oldest and most prestigious universities in Brazil; UERJ ranked 11\textsuperscript{th} nationally in a 2012 ranking by the newspaper \emph{Folha}. UERJ is part of Brazil's system of \emph{state universities}, which are funded by the governments of each state. Brazil also has a system of \emph{federal universities}. State and federal universities are highly-regarded and tuition-free, and admissions are highly competitive. The number of UERJ applicants is often 10--20 times greater than the number of slots. Most Brazilian students attend one of the nation's 2,000+ private colleges, which tend to be moderately selective or open enrollment.

UERJ offers 40--50 undergraduate majors each year in a variety of fields. Students apply to specific programs. Admission is determined by a two-round entrance exam that the university administers near the end of each year. The first round consists of a qualifying exam that is common to all applicants. Students who pass the qualifying exam take field exams in several subjects that depend on their desired major. Admissions are based on a weighted average of field exam scores. The top-scoring applicants are admitted up to a cutoff determined by the program's capacity. 

\subsection{Data}

Our analysis matches two UERJ datasets to national employer-employee records. Our base dataset includes all individuals who applied to UERJ in 1995--2001 and 2004--2011 \citep{uerj2020applicants}.\footnote{\enskip UERJ does not have application records for the 2002--2003 cohorts.} We focus on applicants who passed the first-round exam, which is the relevant sample of potential admits for our analyses. We observe the program individuals applied to, their overall admission score, and their admission outcome. In some cohorts, we observe demographic characteristics and field exam subject scores.\footnote{\enskip Appendix B.1 provides details on our variable definitions and data availability.} Our second dataset contains students who enrolled in UERJ from 1995--2011 \citep{uerj2020enrollees}. This dataset includes the student's program, enrollment date, status as of 2020 (graduated, dropped out, or still enrolled), and final year.

Lastly, we use the 2003--2019 years of Brazil's employer-employee dataset, the \emph{Rela\c{c}\~{a}o Anual de Informa\c{c}\~{o}es Sociais}, or RAIS \citep{rais2021}. This dataset covers the universe of formal-sector jobs in Brazil. Worker variables include demographics, educational attainment, occupation, hours worked, and monthly earnings. Firm variables include the firm's industry, location, and number of employees.

We merge the UERJ and RAIS datasets using national ID numbers. For individuals with missing ID numbers, we merge using names and birthdates. See Appendix B.2 for details.

\subsection{Affirmative action at UERJ}\label{sub:uerj_aa}

Historically, Black, low-income, and public high school students were underrepresented at state and federal universities, partly because they typically earned lower scores on the schools' entrance exams.\footnote{\enskip Other factors likely contributed to limited diversity at selective colleges, such as access to information about the admission process \citep{hoxby2013,machado2021centralized}.} The lack of diversity was contentious because these universities are publicly-funded and tuition-free.

UERJ was one of the first Brazilian universities to address this disparity through affirmative action. In 2003, the state government of Rio de Janeiro passed a law that required UERJ to reserve seats for students from underrepresented groups. Only two other public universities had affirmative action at the time, and both were located in other states \citep{junior2014}. Other universities adopted race- and/or income-based quotas in subsequent years \citep{ferman2005,vieira2019}, and a 2012 national law mandated quotas at all federal universities. But UERJ was the only university in Rio de Janeiro with affirmative action for much of the 2000s.

UERJ's policy reserved 45 percent of seats in each program for low-income applicants from disadvantaged groups. Historically there was one admission track for each major. In 2004, UERJ added three affirmative action tracks per major.\footnote{\enskip UERJ introduced affirmative action in the 2003 cohort following the state law. There were two admission tracks in 2003---low-income and general---and each track reserved some seats for Black applicants. The quota system described in the text was in place for all of 2004--2011.} 20 percent of slots in each major were reserved for public high school applicants. Another 20 percent of slots were reserved for Black applicants. Five percent of slots were reserved for other disadvantaged groups (e.g., disabled and indigenous applicants). To apply through an affirmative action track, applicants also had to be from a low-income family as verified by tax records.\footnote{\enskip In 2004, for example, applicants' per capita family income had to be below R\$300 per month \citep{zoninsein2008}, which was 40 percent of national GDP per capita.} Applicants who did not meet these criteria could apply through the general track, which governed the remaining slots.  Within each track, admissions were based solely on field exam scores.

Although the fraction of reserved slots was the same in each major, the \emph{take-up} varied significantly. Figure \ref{fig:variation_aa} plots the share of affirmative action enrollees in the 2004--2011 cohorts ($y$-axis) against a measure of each program's selectivity ($x$-axis). In highly-selective programs like Law and Medicine, the reserved quotas usually filled up, so affirmative action students made up 45 percent of the class. In less selective programs like Math and Teaching, the number of affirmative action applicants was frequently less than the quota, and UERJ filled open slots from the general track. Thus the share of affirmative action enrollees was as low as 10--20 percent in some programs. The low take-up is attributable to lower desirability of some programs and UERJ's strict criteria for affirmative action eligibility.

\begin{figure}[htpb]{\scriptsize
\begin{centering}
\includegraphics[width=.7\linewidth]{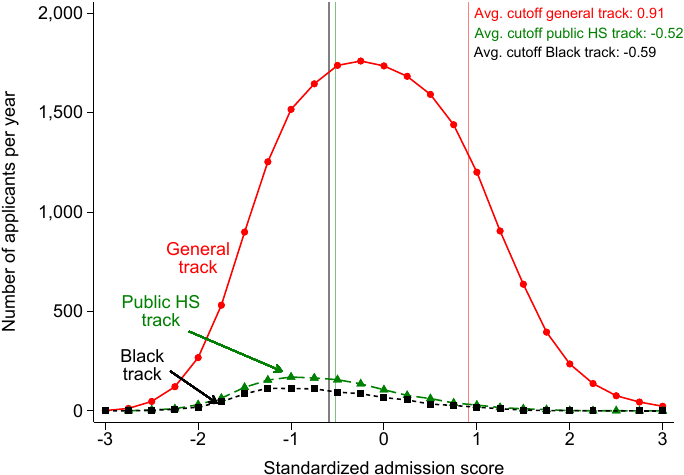}
\protect\caption{Admission score distribution and mean cutoff by application track (2004--2011)} \label{fig:dist}
\par\end{centering}
\singlespacing \justify	
\emph{Notes}: This figure shows the distribution of standardized admission scores for applicants in each application track. The sample includes the 24 programs in our RD sample (Panel A of Table \ref{tab:summ_stats}). We standardize scores to be mean zero and SD one in the population of all applicants in the same program/cohort, and plot distributions in  0.25 SD bins of the standardized score. Vertical lines represent the average admission cutoff in each track. 

}
\end{figure}

UERJ's policy gave a large implicit preference to affirmative action students. Figure \ref{fig:dist} plots the distribution of admission scores for 2004--2011 applicants in the Black, public school, and general tracks. Scores are standardized to be mean zero and standard deviation (SD) one among \emph{all} applicants to a given program/cohort. Vertical lines show the mean cutoff score in each track, which is the mean score of the last admitted students. The average cutoff is $-$0.5 in the public school track, $-$0.6 in the Black track, and $+$0.9 in the general track. Thus marginally-admitted affirmative action students typically scored 1.5 standard deviations below marginal admits in the general track.

\subsection{Samples}\label{sec:samp}

We use two samples to analyze the impacts of UERJ's affirmative action policy. In Sections \ref{sec:rd-spec}--\ref{sec:rd-returns}, we use a regression discontinuity (RD) design that compares admitted and rejected applicants. In Sections \ref{sec:dd-spec}--\ref{sec:dd-spillovers}, we use a difference-in-differences (DD) design that compares enrollees in programs with higher and lower take-up rates of affirmative action.

Our RD sample includes programs in which we can estimate returns for marginally-admitted affirmative action students. We cannot implement our RD design in cases where there were no rejected students, so we restrict our RD sample to programs where the Black and public school quotas typically filled up. Specifically, our RD sample includes 24 programs in which 30 percent or more of the 2004--2011 enrollees were from an affirmative action track (programs above the horizontal line in Figure \ref{fig:variation_aa}). In these programs, we also exclude any cohort/application-track pair with fewer than five applicants below the admission threshold (see Appendix Tables B2--B4).\footnote{\enskip We restrict to the same programs in our RD sample of general applicants so that it is comparable to the Black and public school samples. We exclude the disabled/indigenous track, as these quotas rarely filled. Appendix B.4 provides details on our sample construction.}

Our DD sample includes all programs that UERJ offered both before and after 2004.\footnote{\enskip UERJ re-organized a few programs during our sample period. Our DD analysis combines re-organized programs into one program. See Appendix Tables B2--B4.} This includes the 24 programs in our RD sample plus 19 other programs with lower rates of affirmative action take-up. We focus on a sample of top enrollees who could have attended UERJ regardless of whether affirmative action existed in their cohort (see Section \ref{sec:dd-spec}).
 
Table \ref{tab:summ_stats} shows summary statistics for our RD and DD samples. Panel A includes programs in both samples, and Panel B includes programs that are only in our DD sample. Our RD sample includes a wide variety of business, health, engineering, humanities, and social science majors. Our DD sample includes many teacher-training programs, but it also includes Economics, Math, and several engineering majors. Affirmative action applicants (columns C--E) were disadvantaged relative to general applicants (columns A--B) as measured by race, mother's education, and family income.


\section{RD specification}\label{sec:rd-spec}

\subsection{Regression model}
	
We use a two-stage least squares (2SLS) RD model to estimate the returns to enrolling in UERJ:
\begin{align}
E_{ip} &= \theta D_{ip} + \alpha x_{ip}+\psi D_{ip} x_{ip}+\gamma_p + \varepsilon_{ip} \quad \text{if } |x_{ip}|\le h^Y \label{rd_rf} \\
Y_{ip} &= \beta E_{ip} + \tilde{\alpha} x_{ip}+\tilde{\psi} D_{ip} x_{ip}+\tilde{\gamma}_p + \tilde{\varepsilon}_{ip} \quad \text{if } |x_{ip}|\le h^Y \label{rd_iv}.	
\end{align}
$Y_{ip}$ is an outcome for individual $i$ who applied to UERJ in application pool $p$. Application pools are defined by a program, cohort, and admission track. The endogenous treatment variable, $E_{ip}$, is an indicator that equals one if the applicant enrolled in the UERJ program and cohort that they applied to. We instrument for UERJ enrollment with an indicator for an admission score above the final cutoff for application pool $p$, $D_{ip}$.

We use a local linear specification to estimate returns for applicants on the admission margin. We include fixed effects for each application pool, $\gamma_p$, and an interaction between $D_{ip}$ and the running variable, $x_{ip}$, which is individual $i$'s admission score in application pool $p$. We normalize $x_{ip}$ so that it equals zero for the last admitted student and has SD one in the population of all applicants in a program/cohort. Our regression samples include only applicants whose admission scores are within $h^Y$ standard deviations of the admission threshold. Our benchmark results use the \citet{calonico2014} bandwidth computed separately for each outcome $Y$; Appendix Tables A4--A6 show that our main results are robust to different bandwidths. We cluster standard errors at the individual level, as some individuals apply to UERJ more than once.

We estimate equations \eqref{rd_rf}--\eqref{rd_iv} separately for three groups: pre-AA applicants (1995--2001), post-AA general track applicants (2004--2011), and affirmative action applicants. We pool across the Black and public school tracks to increase power. The estimates for affirmative action applicants show how UERJ's policy affected its targeted beneficiaries. The estimates for pre- and post-AA general applicants provide evidence on how the policy changed untargeted students' returns to attending UERJ. However, this evidence is not conclusive because the policy also implicitly raised admission thresholds in the general track.

\subsection{Identification assumptions and balance tests} \label{sub:ident}

The main RD identification assumption is that applicants' admission scores are effectively randomly assigned near the thresholds. Applicants have little scope to manipulate their scores, but non-random sorting could arise from waitlist admissions. UERJ fills declined seats through multiple rounds of waitlist offers to applicants with the next highest scores (see Appendix B.3 for details). Our instrument and running variable, $D_{ip}$ and $x_{ip}$, are defined by the final threshold in each application pool. Thus the last admitted student may be particularly likely to accept an admission offer, and this tendency may be correlated with potential outcomes.

Balance tests show no evidence that the RD assumption is violated for affirmative action applicants. Appendix Table A1 presents estimates from RD regressions that use demographic characteristics and qualifying exam scores as dependent variables. We cannot reject the hypothesis that these coefficients are jointly equal to zero ($p=0.88$). We find similar results combining these characteristics into an index of predicted wages (Appendix Figure A1). There is no evidence of a discontinuity in the density of admission scores using the \citet{mccrary2008} test (Appendix Figure A2). These results match our prior that waitlist admissions are unlikely to cause non-random sorting in the affirmative action tracks because most applicants accepted their admission offer.

We also find covariate balance for general applicants, but the \citeauthor{mccrary2008} test reveals a statistically significant decrease in the admission score densities at the pre- and post-AA general track thresholds. UERJ's yield was lower in the general track, so there was more scope for non-random sorting from waitlist admissions. Thus our RD results for general applicants should be interpreted with some caution. Reassuringly, our findings are similar in ``donut hole'' regressions that drop applicants near the cutoffs (Appendix Tables A4--A6).

We also make the standard instrumental variable and local average treatment effect (LATE) assumptions \citep{angrist1996}. Instrument relevance is satisfied because the UERJ enrollment rate increases sharply at the admission threshold (Table \ref{tab:rd_returns}, Panel A). The exclusion restriction requires that our instrument affects outcomes only through the channel of enrolling in UERJ. This could be violated if, for example, admission to UERJ caused individuals to apply to other schools. We cannot rule out this possibility, but we believe our results are primarily attributable to UERJ enrollment, particularly in the affirmative action tracks where the first-stage coefficient is large. The monotonicity assumption is plausible because it is unlikely that applicants would have attended UERJ if and only if they were \emph{below} the cutoff.

Under these assumptions, the $\beta$ coefficient from equation \eqref{rd_iv} can be interpreted as the average causal effect of attending UERJ for marginally-admitted compliers. Compliers are students who would have enrolled if and only if they scored above the cutoff. This estimand measures the returns to UERJ enrollment relative to the mix of educational choices that students would have made if they were rejected, which is relevant for evaluating the efficacy of affirmative action as a policy to reduce inequality. 


\section{Effects of affirmative action on marginal admits}\label{sec:rd-returns}

\subsection{Graduation and earnings}\label{sub:rd-labor}

We begin our RD analysis by examining the effects of UERJ enrollment on graduation rates and earnings. Table \ref{tab:rd_returns} presents results for pre-AA general applicants (columns A--B), post-AA general applicants (columns C--D), and affirmative action applicants (columns E--F). Columns (A), (C), and (E) show means of each dependent variable for applicants who scored just below the threshold (within 0.1 SD). Columns (B), (D), and (F) display RD coefficients. Panel A presents first-stage coefficients, $\theta$, from equation \eqref{rd_rf}. Panels B--C show 2SLS RD coefficients, $\beta$, from equations \eqref{rd_rf}--\eqref{rd_iv}. In Panel B, we measure outcomes 6--9 years after UERJ application to capture individuals' initial jobs after (potential) graduation. To examine longer-run effects, Panel C measures outcomes 10--13 years after application.\footnote{\enskip All of our RD regressions include one observation per applicant. We use the applicant's mean real earnings over the periods of 6--9 or 10--13 years after application. For binary outcomes, we use the maximum over each period, so our estimates reflect ever having a job with those characteristics. Most UERJ students who graduate do so in 4--6 years (see Appendix Figure A3).} Figure \ref{fig:rd} presents RD graphs for our main outcomes; these graphs show the reduced-form effects of UERJ admission by plotting means of each outcome in 0.1 SD bins of the standardized admission score. 

Panel A of Table \ref{tab:rd_returns} shows that crossing the admission threshold increased the likelihood that affirmative action applicants enrolled in UERJ by 69 percentage points (column F). The first stage for affirmative action applicants is more than double that for general applicants (columns B and D) because most other universities in Rio did not have affirmative action during 2004--2011. In the general track, marginal admits would typically have been competitive for admission to other top colleges in the area (see Section \ref{sub:enrol_oth}).

Our first finding is that marginal enrollees in the affirmative action tracks were only slightly less likely to graduate from UERJ than those in the general track. The first row of Panel B shows how enrolling in a UERJ program affected the likelihood of graduating from that program by nine years later. 64 percent of marginal affirmative action enrollees graduated by this time, as compared with 68--71 percent of marginal general track enrollees. The similarity of these graduation rates is striking since the admission scores of marginal affirmative action enrollees were 1.5 standard deviations lower on average.

UERJ enrollment did not significantly affect the likelihood that individuals worked in the formal sector. Our measures of formal employment are indicators for appearing in the RAIS at any time in 6--9 or 10--13 years after application. In the affirmative action tracks, the formal employment rates for marginally-rejected applicants are above 70 percent in both time periods (column E), and the 2SLS RD coefficients are close to zero (column F). For pre-AA applicants, we find a positive and significant effect on early-career formal employment (Panel B), but this effect does not persist into the later time period (Panel C).

Importantly, affirmative action students experienced an increase in early-career earnings from attending UERJ. UERJ enrollment caused a 14 percent increase in the mean hourly wages of affirmative action compliers measured 6--9 years after application. The gain in early-career monthly earnings was \$110 (in 2019 U.S.\ dollars). Panel C of Figure \ref{fig:rd} shows visual evidence of a discontinuity in the early-career hourly wages of marginally-admitted affirmative action students (black triangles). The RD coefficient for monthly earnings is roughly one-fifth of the earnings gap between marginally-rejected general and affirmative action applicants (\$1,391 vs.\ \$817). Thus UERJ's affirmative action policy meaningfully reduced early-career earnings inequality among applicants on the margin of admission.

We find some evidence that the initial earnings gain for affirmative action students declined as their careers progressed. Panel C of Table \ref{tab:rd_returns} shows that the effect of UERJ enrollment on affirmative action students' hourly wages declined to 0.024 log points measured 10--13 years later (see also Panel D of Figure \ref{fig:rd}). We reject equality of the early- and later-career wage coefficients at $p<0.05$ (Appendix Table A3). The gain in monthly earnings for affirmative action students also declined to \$56 in the later period, but this estimate is not statistically distinguishable from the early-career return. Appendix Figure A4 shows that the wage gains for affirmative action students decreased both over time (holding the sample of cohorts fixed) and across cohorts (holding potential experience fixed).

For general applicants, we find evidence of a \emph{negative} early-career return to attending UERJ in the cohorts with affirmative action. We find no significant earnings effects in the pre-AA cohorts (column B of Table \ref{tab:rd_returns}), but UERJ enrollment reduced the early-career hourly wages of 2004--2011 general applicants by eight percent (column D). Similarly, the 2SLS RD estimate for post-AA general applicants' monthly earnings is $-$153 USD, and this estimate is statistically significant at $p<0.05$. This suggests that the returns to attending UERJ for non-AA students may have been lower in the cohorts with affirmative action. But this evidence is not conclusive because both earnings coefficients decline in magnitude in the later time period (Panel C).

\subsection{College selectivity and major choice}\label{sub:enrol_oth}

To interpret our earnings results, it is important to understand which college programs UERJ enrollees would have attended if they were not admitted. UERJ is an elite school, but it exists in a highly-competitive market. The federal university in Rio de Janeiro, UFRJ, ranked 3\textsuperscript{rd} in a 2012 national ranking by the newspaper \emph{Folha}, while UERJ ranked 11\textsuperscript{th}. There are three other selective federal universities in the Rio suburbs, and more than five private universities in the city itself (see Appendix Table A8). UERJ applicants in the general and affirmative action tracks differed in the likelihood that they could gain admission to these other colleges during our sample period. Further, applicants to a particular UERJ program may have pursued a different major at another school. 

We examine effects on college and major choice using Brazil's higher education census \citep{inep2019superior}, which covers all colleges in the country. We do not have access to ID numbers in this dataset, so we match it to our sample of UERJ applicants using exact day of birth, gender, and year of enrollment.  These variables do not uniquely identify individuals, so we define our dependent variables as the \emph{total} number of students at a particular university or major that have the same birthdate, gender, and enrollment year as the UERJ applicant. We can only include 2009--2011 UERJ applicants in this analysis because individual-level census data does not exist prior to 2009. The fuzzy merge and smaller sample reduce the precision of our RD estimates for this analysis. (See Appendix B.5 for details.)

With these caveats, we find that UERJ's affirmative action policy allowed disadvantaged applicants to attend a more selective college. Panel A of Table \ref{tab:rd-census-enrol} displays $\theta$ coefficients from our reduced-form RD specification \eqref{rd_rf}, which estimates the effects of UERJ \emph{admission}. The number of UERJ enrollees in the census data increases by 0.88 at the affirmative action thresholds (column F), which is broadly similar to our first stage estimate of 0.69 in Table \ref{tab:rd_returns}. We do not find effects on enrollment in UFRJ, other federal universities in Rio, or private universities in the top 100 of the \emph{Folha} ranking. Instead, the number of enrollees in lower-ranked Rio universities falls by roughly 0.5 at the affirmative action thresholds. Although these estimates are imprecise, they match our prior that many affirmative action applicants would not have gained admission to other top universities, and thus often had less-selective private schools as their fallback option.

Admission to UERJ also altered the major choices of affirmative action applicants. In the last two rows of Panel A, our dependent variables measure the total number of enrollees in Rio de Janeiro universities with the same major as the one that the UERJ applicant applied to. The number of Rio enrollees with the applicant's major increases at the affirmative action thresholds by 0.35 using 2-digit major codes, and by 0.46 using 3-digit codes. In combination with the RD estimate for the total number of UERJ enrollees (0.88), these coefficients suggests that roughly half of affirmative action compliers would have chosen a different major if they were not admitted to UERJ. These changes in field of study are an important mechanism for our RD earnings results, but such changes are relevant for evaluating affirmative action policies in any context where individuals may pursue different majors at different schools.

For general applicants, admission to UERJ reduced the likelihood of enrolling in other top federal and private universities in Rio (Table \ref{tab:rd-census-enrol}, Panel A, column D). Thus most general track compliers would likely have attended other selective universities if they had been rejected.

\subsection{Educational attainment}

We next examine whether UERJ enrollment affected the likelihood that individuals earned \emph{any} college or postgraduate degree. We use the RAIS to define three binary measures of educational attainment: 1) a college degree during the period of 6--9 years after UERJ application; 2) a college degree by 2019; and 3) a postgraduate degree by 2019. Panel B of Table \ref{tab:rd-census-enrol} shows 2SLS RD estimates for these outcomes using regression samples that include all applicants who appear in the RAIS.\footnote{\enskip We find no evidence that the observable characteristics of UERJ applicants who appear in the RAIS change discontinuously at the admission thresholds (Appendix Table A1).}

We find no effects on educational attainment for both affirmative action and general applicants. Most notably, UERJ enrollment did not affect the likelihood that affirmative action applicants earned a college or postgraduate degree (column F). 71 percent of marginally-rejected Black and public school applicants earned a college degree by 2019 (column E), which is a very high rate by Brazilian standards. This reflects the fact that UERJ's affirmative action applicants were high-achieving, even though they were disadvantaged relative to general UERJ applicants. In the general tracks, college degree attainment rates were even higher (columns A and C), and we also find no effects on educational attainment (columns B and D).

\subsection{Employment with UERJ alumni}\label{sub:rd_alumni}

As a final potential mechanism, we consider the effects of networking with UERJ peers and alumni. Elite university networks can improve students' access to high-paying jobs through many channels \citep{rivera2016pedigree}, including on-campus recruiting \citep{weinstein2022firm}, referrals \citep{calvo-armengol2004}, and school reputation \citep{macleod2015}. 

To test for network mechanisms, we use the RAIS to define outcome variables that indicate when UERJ applicants obtained jobs at firms affiliated with other UERJ alumni. Specifically, consider a UERJ applicant $i$ who applied to major $m$. We define applicant $i$ as obtaining a job at an \emph{alumni firm} if their firm employed another individual $j$ who graduated from major $m$ (the ``alum''). Our simplest network outcome is an indicator equal to one if the applicant's firm ever hired another alum. We define different versions of this variable based on the alum's characteristics, the timing of their employment, and the concentration of alumni at the firm. We use major-specific networks because students in the same program often take classes together and work in similar labor markets.\footnote{\enskip All of our network outcomes are leave-individual-out; even if an applicant completed a UERJ degree, these variables equal one only if there is \emph{another} alum affiliated with that firm. Our variable definitions allow applicants to be beneficiaries or benefactors of UERJ's alumni network. For example, an applicant could work at an alumni firm if they got a job from an alum's referral or if they referred an alum.}

Attending UERJ significantly increased the likelihood that affirmative action students obtained jobs at firms affiliated with other UERJ alumni. Panel A of Table \ref{tab:rd_job} shows that marginal affirmative action enrollees were 13.7 percentage points more likely to work at a firm affiliated with any UERJ alum in the period of 6--9 years after application (see also Panel E of Figure \ref{fig:rd}). This is a 29 percent increase from the mean rate of alumni firm employment for marginally-rejected applicants (47.7 percent). Affirmative action enrollees were more likely to work with both general and AA alumni (second and third rows of Panel A). Similarly, attending UERJ increased the \emph{proportion} of UERJ alumni at affirmative action applicants' firms by 8.5 alumni per 1000 workers (fourth row of Panel A). We also find large effects on early-career employment at alumni firms for general applicants (columns B and D).

\begin{figure}[t]{\scriptsize
\begin{centering}
\includegraphics[width=.65\linewidth]{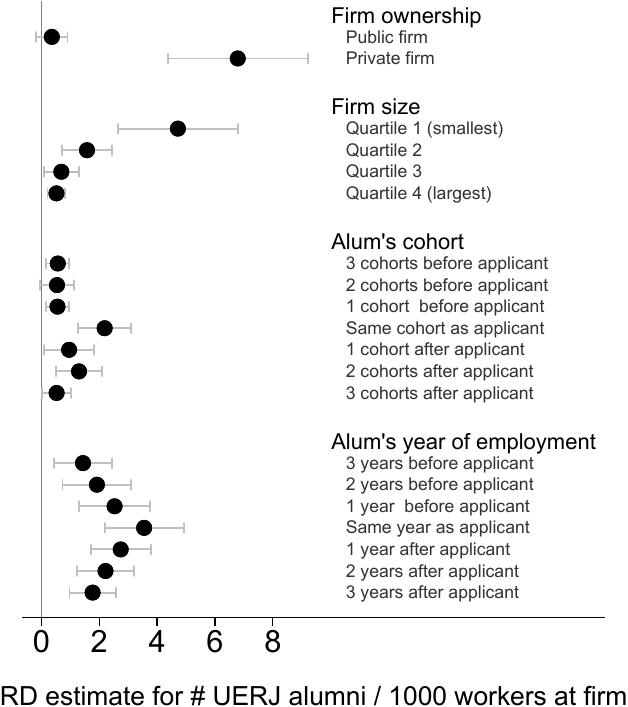}

\par\end{centering}
\protect\caption{Heterogeneity in RD estimates for alumni per 1000 workers at the firm} \label{fig:network_het}
\singlespacing\justify
\emph{Notes}: This figure displays RD estimates for the effects of UERJ enrollment on the mean number of alumni per 1000 workers at the applicants' firms measured 6--9 years after application. These estimates are analogous to those in the fourth row of Table \ref{tab:rd_job} (Panel A), but we pool across all applicant groups. We use four types of dependent variables:
\begin{itemize}
\item \underline{Firm ownership}: The firm's mean number of alumni per 1000 workers interacted with dummies for public and private firms;
\item \underline{Firm size}: The firm's mean number of alumni per 1000 workers interacted with dummies for quartiles of firm size (number of workers);
\item \underline{Alum's cohort}: The firm's mean number of alumni per 1000 workers computed separately using alumni who enrolled in UERJ in each cohort from $3$ years before to $3$ years after the applicant's cohort;
\item \underline{Alum's year of employment}: The firm's mean number of alumni per 1000 workers computed separately using alumni who worked at the firm in each year from $3$ years before to $3$ years after the applicant.
\end{itemize}
Markers depict RD coefficients, $\beta$, from equation \eqref{rd_iv}. Horizontal bars are 95 percent confidence intervals using standard errors clustered at the individual level.

}
\end{figure}

Figure \ref{fig:network_het} presents evidence that the results in Table \ref{tab:rd_job} are partly driven by networking mechanisms. It is possible that the above employment effects reflect major-specific human capital accumulation rather than networking since admission to UERJ affected major choices (Table \ref{tab:rd-census-enrol}). To distinguish between these mechanisms, Figure \ref{fig:network_het} displays heterogeneity in RD estimates for the number of UERJ alumni per 1000 workers at the applicant's firms (pooling across all applicant groups). We find larger estimates in cases where networking is likely more important. The RD coefficients are larger for employment in small private firms than in large public firms.\footnote{\enskip Networking is likely more important at small private firms because most public firms in Brazil use exams to hire workers \citep{mocanu2022designing}.} Similarly, the employment effects are largest for alumni from the applicant's cohort, and for alumni who work at the firm at the same time as the applicant. This variation is consistent with referral and recruiting mechanisms, and it is hard to reconcile with major-specific human capital. Appendix Table A10 shows that enrolling in UERJ increased access to firms with UERJ alumni even \emph{within} groups of firms in the same location and industry. In other words, the presence of UERJ alumni is a strong predictor of an applicant's employment outcome even among firms in the same narrowly-defined labor market. 

Mean wages at firms affiliated with UERJ alumni were 0.44 log points higher than those at other firms in our sample (Appendix Table A12), suggesting that affirmative action enrollees benefited from increased access to these firms.\footnote{\enskip Appendix Table A9 provides examples of alumni firms. Firms with the highest alumni concentration include financial organizations like Accenture and the Brazilian Development Bank, as well as branches of the multinational petroleum company Petrobras.} Consistent with this, attending UERJ increased the mean wage at affirmative action compliers' early-career firms by 0.11 log points (last row of Panel A, column F), which is similar in magnitude to the individual-level wage coefficient (0.13 log points). Notably, UERJ enrollment \emph{reduced} the average wage at post-AA general applicants firms' by 0.10 log points (column D), consistent with their negative earnings effects in Table \ref{tab:rd_returns}.

Yet the benefits of accessing UERJ's alumni network decreased as individuals' careers progressed. Panel B of Table \ref{tab:rd_job} shows RD estimates for the same alumni firm outcomes as in Panel A, but instead measured 10--13 years after UERJ application. For all outcomes and all applicant groups, the RD estimates are smaller in the later period, and many are not statistically different from zero. This suggests that alumni networks are most important for initial job placement, and that their influence declines as individuals progress in the labor market. 

\subsection{Discussion}\label{sub:rd_discuss}

Our graduation results show that most affirmative action students succeeded academically at UERJ. Related work argues that affirmative action may cause disadvantaged students to drop out or switch majors---particularly in STEM fields---because it places them in schools where they are less-prepared than their classmates \citep{arcidiacono2016}. UERJ graduation rates are high by Brazilian standards, and most programs in our RD sample are in non-STEM fields (Table \ref{tab:summ_stats}). Thus relative academic preparation may be less important for degree completion in our setting.\footnote{\enskip Affirmative action students' early-career earnings gains were driven by UERJ's health and business programs, which also have high graduation rates. In STEM programs, affirmative action students graduated at much lower rates, and we find no evidence of positive returns. See Appendix Table A7.} On the other hand, we do not find that affirmative action increased the likelihood that disadvantaged students earned a college degree, as other work has found \citep{bleemer2022affirmative}. One possibility is that negative effects of mismatch in academic preparation were offset by positive effects of UERJ's greater resources, yielding a zero net effect on degree attainment.

Our findings suggest that the early-career earnings gains for affirmative action students were driven partly by networking mechanisms. UERJ's affirmative action policy increased disadvantaged students' access to higher-paying firms affiliated with its alumni (Table \ref{tab:rd_job}). Taken together with the OLS wage premium for alumni firms (0.44 log points), the RD estimate for alumni firm employment (14pp) can explain nearly half of affirmative action students' early-career wage gains (0.13 log points). Our paper differs from \citet{zimmerman2019}'s and \citet{michelman2022old}'s findings that the benefits of networking at elite universities accrue \emph{only} to students from advantaged backgrounds. This difference may arise because our data includes early-career outcomes in a broader set of firms, and our results are unclear about whether initial networking benefits are persistent. Further, our estimate of the early-career return for affirmative action students is substantially smaller than that in \citet{bleemer2022affirmative} (see footnote \ref{fn:magnitudes}), which may be because we find no effects on college degree attainment.

For general track enrollees, we find some evidence that the early-career return to attending UERJ declined from the pre-AA to post-AA cohorts (Table \ref{tab:rd_returns}, Panel B). This suggests that there may have been negative spillover effects of affirmative action on other UERJ students. In the post-AA cohorts, marginally-admitted general applicants were less likely to work at high-paying firms and more likely to work with affirmative action alumni (Table \ref{tab:rd_job}). Thus affirmative action may have reduced the value of networking at UERJ because the new disadvantaged students tended to obtain lower-paying jobs. But our RD analysis cannot conclusively identify spillover effects because affirmative action also affected the characteristics of marginally-admitted general applicants.\footnote{\enskip Appendix Table A2 shows that post-AA general track compliers were more likely to be non-white and younger than pre-AA compliers, although the magnitude of these differences is relatively modest.} To present more compelling evidence on spillover effects of UERJ's policy, we turn to our second empirical strategy.


\section{DD specification}\label{sec:dd-spec}

\subsection{Top enrollee sample}

To estimate the effects of affirmative action on other UERJ students, we construct a sample of \emph{top enrollees} who could have attended UERJ regardless of whether affirmative action existed in their cohort. For each major $m$, we define $N_m$ to be the minimum number of students who enrolled through the \emph{general} track in any cohort in 1995--2011.\footnote{\enskip In other words, we define $N_m = \min_{c \in \{1995,\dots,2011\}} N_{mc}$, where $N_{mc}$ is the number of general track enrollees in major $m$ and cohort $c$.} Our top enrollee sample is a balanced panel at the major level that includes the $N_m$ enrollees with the highest admission scores in each cohort. Since 55 percent of slots were reserved for general applicants, this sample contains roughly the top half of the class.

\subsection{Regression model}

\begin{figure}[H]{\scriptsize
\begin{centering}
\includegraphics[width=.7\linewidth]{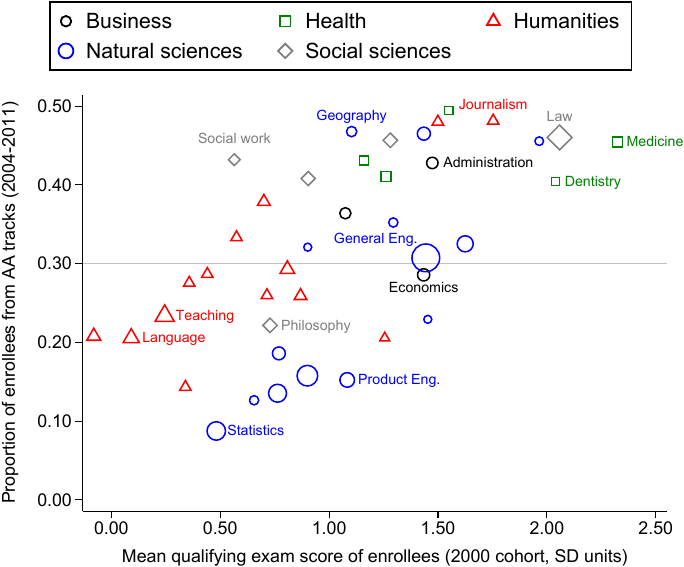} 
\protect\caption{Take-up of affirmative action and program selectivity} \label{fig:variation_aa}
\par\end{centering}
\singlespacing \justify
\emph{Notes}: This figure plots exposure to affirmative action ($y$-axis) and selectivity ($x$-axis) for each UERJ program in our sample. The $y$-axis displays the fraction of enrollees in the 2004--2011 cohorts who entered through an affirmative action track. The $x$-axis displays the mean score on the 2000 qualifying exam for enrollees in each program. We compute each applicant's average score across all exam subjects and standardize to mean zero and SD one in the population of qualifying exam takers. The figure omits two programs for which we do not have scores in the 2000 qualifying exam (mechanical engineering and production engineering). Marker sizes are proportional to the number of enrollees.

}
\end{figure}

For identification, we exploit variation in the take-up of affirmative action across UERJ's majors (Figure \ref{fig:variation_aa}) in a difference-in-differences (DD) specification:
\begin{align}
	Y_{imc} &= \gamma_m + \gamma_{cf(m)} + \pi[\text{ExposureToAA}_m\times\text{Post}_c] + \varepsilon_{imc}. \label{dd}
\end{align}
$Y_{imc}$ is an outcome for individual $i$ who enrolled in major $m$ and cohort $c$. Our variable of interest is the interaction between a major's exposure to affirmative action and a dummy for post-AA cohorts ($\text{ExposureToAA}_m \times \text{Post}_c$). Our benchmark results use a binary measure of exposure that equals one if the share of affirmative action enrollees in 2004--2011 was above 30 percent (the horizontal line in Figure \ref{fig:variation_aa}). We include major and cohort fixed effects, and cluster standard errors at the major level.

We estimate equation \eqref{dd} in our sample of top enrollees to examine the effects of affirmative action on untargeted students. In this case, the $\pi$ coefficient measures how affirmative action changed top enrollees' outcomes in more-affected majors relative to less-affected majors. We refer to these estimates as ``spillover'' effects because they reflect the impacts of affirmative action students' enrollment on top enrollees' outcomes \citep{arcidiacono2010does}. We also present DD coefficients for the small subset of top enrollees who are from underrepresented minority (URM) groups; this sheds light on whether affirmative action impacted URM students who could have gained admission to UERJ in absence of the policy.

Our DD specification identifies the effects of a 19 percentage point increase in the fraction of enrollees in an individual's program/cohort who entered through affirmative action (Panel A of Table \ref{tab:dd_traits}). This is a large effect on diversity relative to the scale of affirmative action at many US universities, but it is similar to the magnitude of \citet{chetty2020income}'s ``need-affirmative'' counterfactual admission policy. 

\subsection{Identification assumptions}

Our key identification assumption is that the outcomes of enrollees in more- and less-affected majors would have followed parallel trends in the absence of affirmative action. A potential concern is that Brazil experienced a recession in the mid-2010s, which may have had heterogeneous impacts across UERJ's majors. To address this, we interact the cohort dummies in equation \eqref{dd}, $\gamma_c$, with fixed effects for five field of study groups, $f(m)$: business, health, humanities, natural sciences, and social sciences (see Appendix B.1). This restricts identification to comparisons between majors in the same field, which were likely to be similarly affected by macroeconomic conditions.

Appendix Figure A6 shows that mean wages evolved similarly in industries that hired UERJ students from majors with more and less exposure to affirmative action. For this figure, we first compute the mean hourly wage in each industry $\times$ year pair using all workers in the RAIS. We then compute a weighted average of these industry $\times$ year means for each UERJ major using the share of pre-AA top enrollees who were employed in each industry as weights. These industry mean wages trended similarly between more- and less-exposed majors across all years of our data. In the years in which post-AA graduates were in the labor market (2009--2019), the change in industry mean wages between more- and less-affected majors is small and statistically insignificant ($-0.02$ log points). This suggests that our results are not driven by divergent industry growth rates or heterogeneous impacts of the mid-2010s recession. Below we also present event study and robustness results to test our identification assumption.


\section{Spillover effects of affirmative action}\label{sec:dd-spillovers}

\subsection{Characteristics of UERJ enrollees}\label{sub:dd_traits}

We begin our DD analysis by asking whether affirmative action affected the composition of UERJ's top enrollees. Research finds that families prefer schools with high-achieving peers \citep{abdulkadirouglu2020}. Thus UERJ's policy may have induced some students to attend other colleges. To test this hypothesis, Table \ref{tab:dd_traits} uses UERJ enrollees' demographic characteristics and entrance exam scores as dependent variables in regression \eqref{dd}. Column (A) shows dependent variable means for top enrollees in pre-AA cohorts (1995--2001). Our main results are the DD coefficients, $\pi$, for top enrollees in column (B). Column (C) shows DD estimates for URM top enrollees, which we define as top enrollees who identify as Black or indigenous in the RAIS. Column (D) shows DD estimates for students who are not top enrollees.

We do not find significant effects of exposure to affirmative action on top enrollees' observable characteristics. The DD coefficients for top enrollees' age, gender, and race are small and statistically insignificant (Table \ref{tab:dd_traits}, Panel B, column B). We find no effects on top enrollees' field exam or admission scores (Panel C). In Panel D, the dependent variables are indices of predicted log wages based on demographic characteristics and admission scores. We find no effect on these predicted wages, and the estimates are similar when we restrict to enrollees who appear in the RAIS. Thus, the composition of top enrollees in more- and less-affected majors did not diverge significantly with the adoption of affirmative action.\footnote{\enskip We also find no evidence of differential changes in the composition of top enrollees using using SES indices based on individuals' first and last names (Appendix Table A13).}

A possible explanation for this finding is that prospective students may not have known that the take-up of affirmative action would differ across UERJ's majors. Students were surely aware of the admission policy, but our DD analysis nets out school-level changes in top enrollees' characteristics. Before enrolling, students may not have known the affirmative action share would be, for example, 15 percentage points lower in Economics than in Business. Thus while affirmative action may have deterred some students from enrolling in \emph{any} UERJ major, compositional changes are unlikely to fully explain our DD results.

By contrast, in majors with high exposure to affirmative action, the population of non-top enrollees became more racially diverse, older, and lower-ability as measured by entrance exam scores (column D of Table \ref{tab:dd_traits}). This reflects the intended effects of affirmative action on diversity.

\subsection{Labor market outcomes}

\begin{figure}[htp]{\scriptsize
\begin{centering}
%
\includegraphics[width=\linewidth]{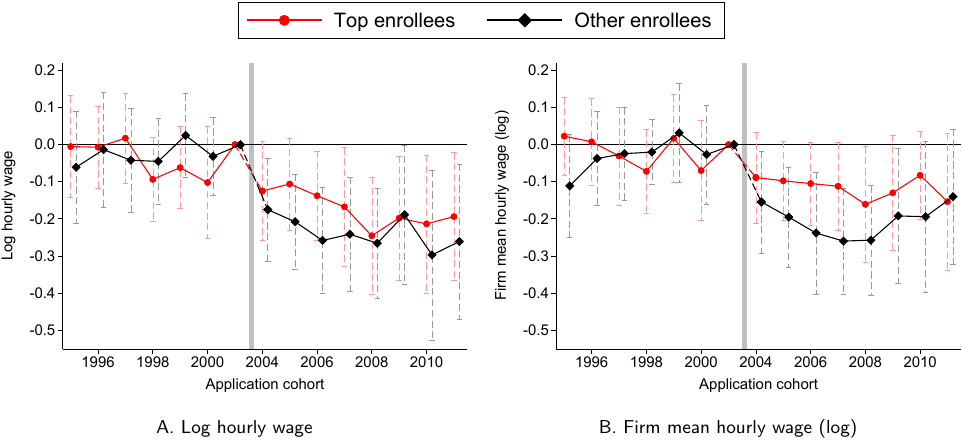}

\par\end{centering}
\protect\caption{Event study estimates for individual and firm mean hourly wages 6--9 years after application} \label{fig:event}
\singlespacing\justify
\emph{Notes}: This figure plots $\pi_c$ coefficients from an event-study version of our DD regression \eqref{dd}, which replaces $\text{Post}_c$ with dummies for each cohort (omitting 2001). Dashed lines are 95\% confidence intervals using standard errors clustered at the program level. The dependent variables are log hourly wage (Panel A) and firm mean log hourly wage (Panel B) measured 6--9 years after application. Red markers show estimates for top enrollees. Blue markers show estimates for other enrollees.

}
\end{figure}

Our main finding is that greater exposure to affirmative action reduced top enrollees' earnings. Table \ref{tab:dd_labor} presents DD estimates for graduation and labor market outcomes measured 6--9 years after application using the same table structure as Table \ref{tab:dd_traits}. We find that UERJ's policy reduced the mean hourly wage of top enrollees by 14 percent in more-affected majors relative to less-affected majors (Panel B, column B). The DD estimate for average monthly earnings is similar in magnitude ($-170$ USD). Panel A of Figure \ref{fig:event} shows an event-study version of this result. The hourly wage coefficient for top enrollees (red line) drops sharply between the last pre-AA cohort (2001) and the first post-AA cohort (2004), and it declines further to $-0.20$ log points by the 2011 cohort. These negative effects persist at a similar magnitude for earnings measured 10--13 years after application (see Appendix Table A14).

The decline in top enrollees' earnings was largely driven by a decline in firm quality as measured by firm average wages. The DD estimate for log firm mean hourly wage is $-0.095$ for top enrollees (Panel B of Table \ref{tab:dd_labor}), which is 70 percent of the individual wage coefficient. The event-study coefficients for firm average wage also decline sharply in the first post-AA cohort (Panel B of Figure \ref{fig:event}). Exposure to affirmative action did not affect top enrollees' graduation rates (Panel A of Table \ref{tab:dd_labor}), suggesting that the earnings effect is not driven by changes in educational attainment. The DD estimate for formal employment is negative and marginally significant ($-0.027$), but it is relatively small compared to the mean formal employment rate (0.74).

Appendix Table A16 shows that our results for top enrollees are robust to multiple specification checks. Our earnings estimates are similar if we restrict to pre-recession years or if we include program-specific linear trends estimated in the pre-AA cohorts. Controlling for student demographics and entrance exam scores only slightly reduces the DD coefficients, consistent with the small compositional effects in Table \ref{tab:dd_traits}. We continue to find negative effects when we compare programs in the same quartile of selectivity (defined by the $x$-axis in Figure \ref{fig:variation_aa}) and when we exclude field of study controls. Lastly, our results are similar when we use a continuous treatment variable, $\text{ExposureToAA}_m$, which is the share of 2004--2011 enrollees who were from an affirmative action track (the $y$-axis in Figure \ref{fig:variation_aa}).

Notably, we also find UERJ's affirmative action policy reduced the earnings of top enrollees from URM groups (column C of Table \ref{tab:dd_labor}). These estimates are imprecise because our top enrollee sample includes only about 1,600 Black and indigenous students. Nonetheless, we find negative and significant point estimates for both individual and firm average wages (Panel B). We also find large earnings declines for non-top enrollees in more- vs.\ less-exposed majors (column D). The DD estimate for non-top enrollees' hourly wages ($-0.212$) is larger in magnitude than the predicted wage effect based on individual characteristics ($-0.154$). Thus spillover effects may have also reduced the wages of affirmative action students, although this evidence is suggestive. 

\subsection{Networking mechanisms}\label{sub:dd_networks}

To shed light on mechanisms for these spillover effects, we first ask whether affirmative action affected the jobs that UERJ students obtained through networking. Affirmative action may have caused some employers to forgo recruiting at UERJ because it reduced the \emph{expected} ability of a UERJ student \citep{macleod2017,weinstein2018employer}. Further, affirmative action students typically obtained lower-paying jobs than general track students (Table \ref{tab:rd_returns}), so the value of referrals from classmates likely declined in majors with high exposure to the policy.

To test for these mechanisms, Panels C--D of Table \ref{tab:dd_labor} use dependent variables that measure employment at firms that hired UERJ alumni from different cohorts and application tracks.\footnote{\enskip These alumni firm variables are similar to those in our RD analysis except we define them to be non-overlapping. For example, in the second row of Panel C, the dependent variable equals one only if the firm did \emph{not} hire a pre-AA alum. As in Table \ref{tab:rd_job}, we require that the applicant and alum are from the same major.} In Panel C, the outcome variables are indicators for employment at firms with pre-AA alumni from the enrollee's program versus firms that hired \emph{only} post-AA alumni. In Panel D, the outcome variables are indicators for employment at firms with general track alumni from the enrollee's own cohort versus firms that hired \emph{only} alumni from other cohorts or from the affirmative action tracks. These variables test whether affirmative action changed the types of firms that hired UERJ students (Panel C) and the peer connections that UERJ students used to obtain jobs (Panel D). Firms with pre-AA and general track alumni paid significantly higher average wages than those that hired only post-AA or affirmative action alumni (Appendix Table A12).

We find that affirmative action reduced top enrollees' employment rates at higher-paying alumni network jobs. Top enrollees' likelihood of employment at firms with pre-AA alumni declined by 5.5 percentage points in more- vs.\ less-affected majors (Table \ref{tab:dd_labor}, Panel C, column B). This decline was offset by a 4.9 percentage point increase in the rate of employment at firms that hired only post-AA alumni. Similarly, the likelihood of employment with same-cohort general track alumni declined by 9.8 percentage points for top enrollees in more- vs.\ less-affected majors (Panel D, column B). Correspondingly, top enrollees in these majors became relatively more likely to work at firms that hired only general track alumni from another cohort (+4.2pp) or only affirmative action alumni (+4.6pp). Thus employment shifted toward firms with lower average wages (Appendix Table A12). This suggests that the negative spillover effects of affirmative action on earnings can partly be explained by a decline in the value of networking. 

\subsection{Learning mechanisms}\label{sub:enade}

Affirmative action may also have reduced top enrollees' earnings through human capital channels. For example, UERJ students became less academically prepared on average under affirmative action. This may have reduced the benefits of peer interactions or caused professors to teach less advanced material.

To test for learning mechanisms, we use data from Brazil's national college exit exam, the Enade \citep{inep2022enade}. The Enade is a field-specific exam that has been administered every year since 2004, although each field is tested every three years on a staggered schedule. The government uses Enade scores to rate higher education programs, so many universities ask students to take the exam when they are close to graduation \citep{pedrosa2013assessing}. The Enade is typically low stakes from the student's perspective; it is not a graduation requirement at most universities.

Table \ref{tab:dd_enade} shows how affirmative action affected the characteristics and performance of UERJ's Enade participants. This table presents DD estimates that compare 2004--2015 exam takers at UERJ to those at other federal and state universities that did not have affirmative action during this period.\footnote{\enskip See Appendix Table A17 for details on our Enade sample and the exam fields.} Column (A) shows the means of each outcome in the 2004--2006 cohorts at UERJ; we define 2004--2006 as the pre-AA period since these Enade cohorts typically enrolled in UERJ prior to 2003. Column (B) displays DD estimates for all exam takers, which are the coefficients on an indicator for UERJ interacted with an indicator for the post-AA cohorts (2007--2015).\footnote{\enskip Our DD specification for Table \ref{tab:dd_enade} is 
\begin{align}
Y_{mjt} &= \gamma_{mj} + \gamma_{mt} + \pi[\text{UERJ}_j\times\text{Post}_t] + \varepsilon_{mjt}. \label{eq:dd_enade}
\end{align}
Regressions are at the exam field ($m$) by institution ($j$) by year ($t$) level, with observations weighted by the number of exam takers. We include field $\times$ institution dummies, $\gamma_{mj}$, and field $\times$ year dummies, $\gamma_{mt}$. Thus identification comes only from within-field comparisons. The coefficient of interest, $\pi$, is on an indicator for UERJ interacted with an indicator for the 2007--2015 cohorts, $\text{UERJ}_j\times\text{Post}_t$.} The Enade data is not linked to our UERJ records at the individual level, so we cannot estimate this regression in our top enrollee sample. As an alternative, column (C) restricts the sample to white students from private high schools, who were not eligible for affirmative action. Column (D) presents results for non-white and/or public high school exam takers.

Panel A of Table \ref{tab:dd_enade} shows that affirmative action increased the diversity of UERJ's Enade exam takers, but we do not find compositional changes within the sample of white private school students. The mean number of UERJ exam takers per program/cohort (36 students) did not change significantly with affirmative action (first row of Panel A, column B), but there were 7.9 fewer white private student students on average (column C), and 9.1 more non-white and/or public school students (column D). Affirmative action significantly increased the racial and socioeconomic diversity of the average UERJ Enade participant (column B). However, we do not find significant changes in gender, age, mother's education, or household income within the sample of white private school students (column C). The relative change in the composition of UERJ's white private school students is close to zero using an index of predicted Enade scores based on demographic characteristics (last row of Panel A).

\begin{figure}[t]{\scriptsize
\begin{centering}
\includegraphics[width=.7\linewidth]{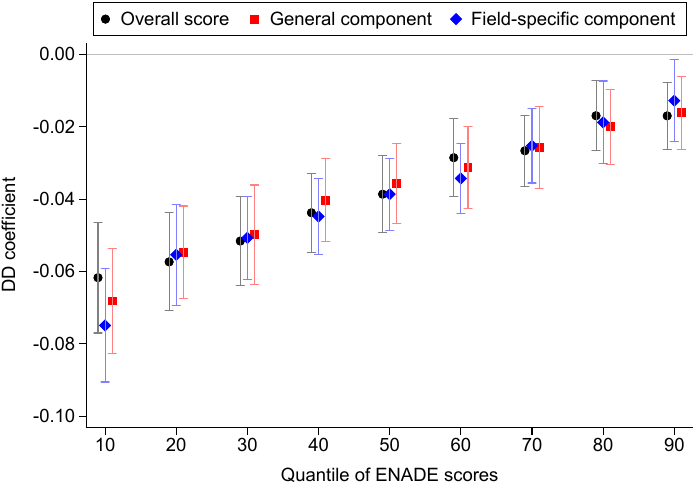}

\par\end{centering}
\protect\caption{Effects of affirmative action at UERJ on quantiles of Enade scores}\label{fig:dd_enade_quantiles}
\singlespacing\justify
\emph{Notes}: This table displays DD estimates of the effect of UERJ's affirmative action policy on quantiles of its graduates Enade exam scores. These estimate are similar to those in Panel B of Table \ref{tab:dd_enade}, but the dependent variables are quantiles of Enade scores within each institution $\times$ program $\times$ cohort cell. Markers depict the DD coefficient ($y$-axis) for each quantile ($x$-axis). Vertical bars are 95 percent confidence intervals using standard errors clustered at the institution level.

}
\end{figure}

Panel B of Table \ref{tab:dd_enade} shows that affirmative action decreased the Enade scores of UERJ students, including \emph{within} the sample of white private school students. Enade scores are expressed as the proportion of correct answers, and the overall score is a weighted average of its field-specific and general components. For the average UERJ exam taker, affirmative action reduced the proportion of correct answers by 3.8 percentage points (column B) from a pre-AA mean of 55 percent (column A). This average effect is likely due in part to the policy's effects on diversity. Yet the overall scores of UERJ's white private school students also declined by 2.2 percentage points (column C). This decline is 15 percent of a standard deviation of the full distribution of Enade scores (14.4pp). Similarly, Figure \ref{fig:dd_enade_quantiles} shows that Enade performance declined by about two percentage points at the highest quantiles of UERJ's score distribution.\footnote{\enskip Figure \ref{fig:dd_enade_quantiles} plots DD coefficients in which the dependent variables are quantiles of Enade scores within each exam field $\times$ institution $\times$ year cell (rather than mean scores, as in Table \ref{tab:dd_enade}).} We also find that Enade scores declined in UERJ majors with more exposure to affirmative action relative to less-exposure majors (Appendix Table A18).

These findings suggest that affirmative action reduced the learning of UERJ's top students. At high quantiles and in the white private school sample, the declines in Enade performance are not likely to be driven by compositional effects. We find no evidence of negative selection in the sample of white private school students, and all else equal, one would expect \emph{positive} selection within this sample because the bar for admission was higher in cohorts with affirmative action. Thus these results suggest that the negative effects of affirmative action on top enrollees' earnings were partly driven by learning spillovers.

\subsection{Discussion}\label{sub:dd_discuss}

Our point estimates imply that a one percentage point increase in the affirmative action share led to a 0.7 percent decrease in the wages of UERJ's highly-ranked students. Thus the negative effects on top enrollees' earnings were large in majors with the highest exposure to affirmative action. These spillover results are consistent with our RD analysis, which found that general applicants in the post-AA cohorts had a negative early-career earnings return to attending UERJ (Table \ref{tab:rd_returns}).

These spillover effects were driven by a combination of compositional, networking, and learning mechanisms. Although the DD estimates for top enrollee composition are not statistically significant, the point estimate for the log wage index in Panel D of Table \ref{tab:dd_traits} ($-0.033$) is 25 percent of our main effect on log wages ($-0.132$). By combining the DD estimates for access to alumni firms (Panels C--D of Table \ref{tab:dd_labor}) with the OLS wage premia for these jobs (Appendix Table A12), network mechanisms can explain 10--17 percent of the overall wage effect. Our Enade dataset is not linked to wages, but \citet{reyes2022endurance} finds that a one percentage point increase in the proportion of correct answers on Brazil's national college entrance exam (ENEM) is associated with a 0.02 log point increase in early-career wages. Assuming that the relationship between correct answers and wages is the same on the Enade exam, the decline in overall scores for white private school students in Panel B of Table \ref{tab:dd_enade} (2.2pp) can explain 32 percent of the overall wage effect. Taken together, these compositional, networking, and learning effects explain two-thirds of the decrease in top enrollees' hourly wages. (See Appendix B.6 for details.)

Further, UERJ's adoption of affirmative action reduced the earnings of highly-ranked URM students. High-scoring URM students were likely affected by networking and learning spillover effects in the same way as other top enrollees. They may also have faced statistical discrimination from employers when the URM share of the student body increased \citep{coate1993}. Thus our results suggest that highly-ranked URM students may have been better off if UERJ's affirmative action policy had been smaller in scale.


\section{Conclusion \label{sec:con}}

This paper documented a tradeoff between the direct and spillover effects of affirmative action at Rio de Janeiro State University (UERJ). On the one hand, marginally-admitted Black and low-income students who attended UERJ as a result of affirmative action experienced a 14 percent increase in early-career earnings. This earnings gain was driven not by educational attainment, but rather by increased access to high-paying firms affiliated with UERJ alumni. This suggests that the primary benefit of affirmative action at elite universities may be to help disadvantaged students gain access to job networks in high-wage sectors of the economy. Yet we found some evidence that affirmative action students' earnings and networking gains decreased as their careers progressed, suggesting that they faced additional barriers to career advancement in the labor market.

On the other hand, UERJ's affirmative action policy had negative impacts on the careers of its other students, including highly-ranked URM students. Our results suggest that a 19 percentage point increase in the share of students admitted through affirmative action led to a 14 percent decrease in the wages of UERJ's top students. This earnings effect may have been due in part to a change in the composition of UERJ's top students, but it was also driven by negative spillover effects on their learning and a decline in the value of peer networking. These results can explain why elite schools around the world use admission policies that favor high-achieving and wealthy students \citep{arcidiacono2022legacy}, and why they may be hesitant to unilaterally adopt affirmative action at a large scale. 

Our paper shows that elite universities face a tradeoff between serving as engines of upward mobility for disadvantaged students and maintaining sources of their value added that stem from admitting high-achieving and wealthy students. An important caveat is that we do not examine non-pecuniary benefits of interacting with classmates from diverse backgrounds \citep[e.g.,][]{carrell2019impact}, which can further justify the adoption of large-scale affirmative action.


\clearpage
\begin{singlespace}
\bibliographystyle{chicago}
\bibliography{bib_er,add_gr}
\end{singlespace}


\clearpage

\section*{Figures and tables}

\clearpage 

\clearpage
\begin{landscape}
\begin{figure}[htp]{\scriptsize
\begin{centering}
%
\includegraphics[width=\linewidth]{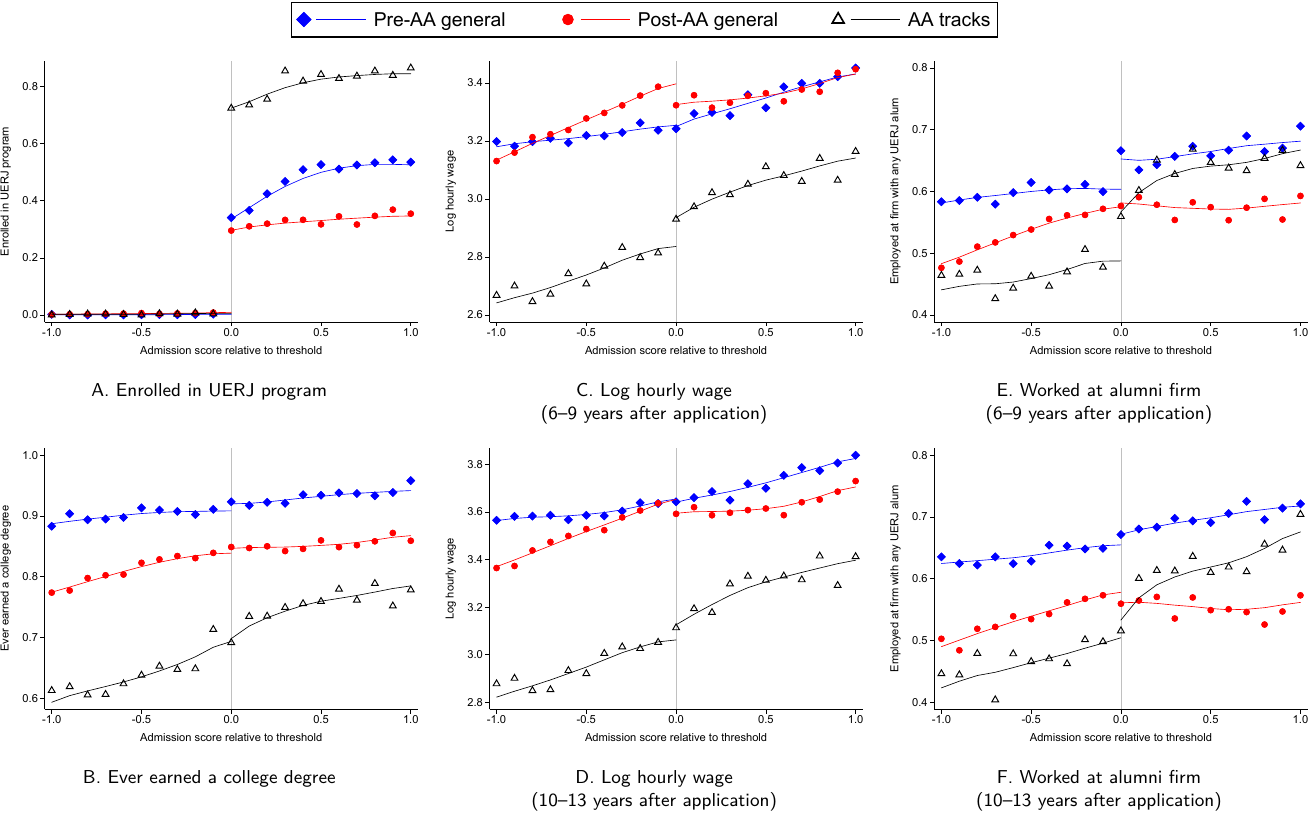}

\par\end{centering}
\protect\caption{RD graphs for UERJ enrollment, earnings, and employment at alumni firms} \label{fig:rd}
\singlespacing\justify
\emph{Notes}: This figure presents RD graphs for pre-AA general applicants (blue diamonds), post-AA general applicants (red circles), and Black/public school applicants (black triangles). The $x$-axis in each panel is an applicant's standardized admission score normalized to zero at the cutoff. The $y$-axis plots means of each outcome in 0.1 SD bins of the standardized score. Outcomes are measured 6--9 years after UERJ application in Panels C and E, and 10--13 years after application in Panels D and F. Lines are predicted values from local linear regressions estimated with a triangular kernel.

}
\end{figure}
\end{landscape}

\clearpage

\clearpage

\clearpage

\clearpage

\thispagestyle{empty}

\begin{table}[htp]{\scriptsize\begin{centering}
\protect\caption{Summary statistics for RD and DD samples} \label{tab:summ_stats}
\newcommand\w{1.6}
\begin{tabular}{lR{\w cm}l@{}R{\w cm}R{\w cm}R{\w cm}R{\w cm}}
\midrule
&(A)&&(B)&(C)&(D)&(E)\\\addlinespace
&\multicolumn{1}{c}{1995--2001}    &&\multicolumn{4}{c}{2004--2011 cohorts} \\\cmidrule{2-2}\cmidrule{4-7}
Sample sizes and & General&& General & Public &  &Other  \\
characteristics of all applicants&track &&track& high school&Black&AA\\\midrule

\addlinespace\addlinespace
\multicolumn{7}{l}{Panel A.\ Programs in both RD and DD samples (24 programs)} \\\midrule
Total applicants &95,659 & &159,408 &10,996 &7,263 &318 \\
Applicants in RD sample &93,930 & &159,383 &9,624 &5,600 &0 \\
Enrollees in DD sample &15,512 & &11,588 &4,465 &3,241 &211 \\
Top enrollees in DD sample &7,932 & &8,922 &362 &178 &2 \\\addlinespace
Female &0.50 & &0.55 &0.60 &0.60 &0.48 \\
Age at applicantion &20.75 & &20.28 &21.88 &23.04 &24.30 \\
White (UERJ data) & & &0.64 &0.49 &0.03 &0.35 \\
White (RAIS data) &0.78 & &0.67 &0.57 &0.15 &0.48 \\
Mother has a high school degree & & &0.85 &0.49 &0.56 &0.54 \\
HH income $>$ 1.5$\times$ min. wage & & &0.82 &0.35 &0.35 &0.45 \\

\addlinespace \addlinespace
\multicolumn{7}{l}{Panel B.\  Programs in DD sample only (19 programs)} \\\midrule
Total applicants &47,633 & &50,553 &4,374 &2,118 &58 \\
Applicants in RD sample &0 & &0 &0 &0 &0 \\
Enrollees in DD sample &13,765 & &14,105 &2,469 &1,326 &38 \\
Top enrollees in DD sample &8,534 & &9,179 &495 &253 &9 \\\addlinespace
Female &0.56 & &0.53 &0.62 &0.63 &0.57 \\
Age at applicantion &22.34 & &21.62 &22.54 &24.09 &26.24 \\
White (UERJ data) & & &0.59 &0.49 &0.03 &0.32 \\
White (RAIS data) &0.75 & &0.65 &0.60 &0.20 &0.47 \\
Mother has a high school degree & & &0.78 &0.45 &0.52 &0.43 \\
HH income $>$ 1.5$\times$ min. wage & & &0.74 &0.28 &0.30 &0.25 \\
\midrule

\end{tabular}
\par\end{centering}

\singlespacing\justify
\emph{Notes}: This table reports summary statistics for UERJ applicants in our sample.

Panel A includes 24 programs that are in our RD and DD samples: Accounting, Biological Sciences, Business Administration, Chemical Engineering, Chemistry, Computer Science, Dentistry, General Engineering, Geography, Geology, Greek/Latin/Literature, History, History Ed.\ (SGO), Industrial Design, Journalism, Law, Mechanical Engineering, Medicine, Nursing, Nutrition, Production Engineering, Psychology, Social Science, Social Work.

Panel B includes 19 programs that are in our DD sample only: Art, Biological Sciences (SGO), Cartographic Engineering, Economics, English/German/Japanese, Geography Ed.\ (SGO), Language (SGO), Math, Math Ed.\ (SGO), Mechanical Engineering (NF), Oceanography, Philosophy, Physical Ed., Physics, Production Engineering (RES), Spanish/French/Italian, Statistics, Teaching, Teaching (DDC).

Programs are at UERJ's main campus in Rio unless denoted with parentheses. Column (A) includes applicants in the pre-AA cohorts. Columns (B)--(E) include applicants to the four admission tracks in the post-AA cohorts. See Appendix B.1 for variable definitions and our grouping of programs into fields of study. See Appendix B.4 for our sample definition.

}
\end{table}

\clearpage

\begin{table}[htp]{\scriptsize\begin{centering}
\protect\caption{RD estimates of the effects of UERJ enrollment on graduation and earnings}\label{tab:rd_returns}
\newcommand\w{1.33}
\begin{tabular}{l@{}lR{\w cm}@{}L{0cm}R{\w cm}@{}L{0.43cm}R{\w cm}@{}L{0cm}R{\w cm}@{}L{0.43cm}R{\w cm}@{}L{0cm}R{\w cm}@{}L{0.43cm}}
\midrule
&&(A)&&(B)&&(C)&&(D)&&(E)&&(F)\\\addlinespace
&&\multicolumn{3}{c}{1995--2001} &&\multicolumn{3}{c}{2004--2011}&&\multicolumn{3}{c}{2004--2011}\\
&&\multicolumn{3}{c}{general track}&&\multicolumn{3}{c}{ general track}&&\multicolumn{3}{c}{AA tracks} \\\cmidrule{3-5}\cmidrule{7-9}\cmidrule{11-13}
&& Mean&&RD  && Mean&&RD && Mean&&RD  \\
Dependent variable &&below &&coef&&below &&coef&&below &&coef\\\midrule

\addlinespace\addlinespace
\multicolumn{13}{l}{Panel A.\ First stage} \\\midrule
Enrolled in UERJ program&&0.003&&0.313&\sym{***}&0.008&&0.292&\sym{***}&0.004&&0.689&\sym{***}\\
&&&&(0.010&)&&&(0.006&)&&&(0.014&)\\\addlinespace
\(N\)&&3,234&&17,519&&4,012&&47,838&&543&&6,121&\\\addlinespace

\addlinespace\addlinespace
\multicolumn{13}{l}{Panel B.\ Graduation and earnings 6--9 years after application (2SLS)} \\\midrule
Graduated from UERJ program&&0.002&&0.711&\sym{***}&0.003&&0.677&\sym{***}&0.004&&0.640&\sym{***}\\
&&&&(0.017&)&&&(0.013&)&&&(0.018&)\\\addlinespace
Formal employment&&0.627&&0.064&\sym{**}&0.672&&$-$0.031&&0.729&&$-$0.002&\\
&&&&(0.029&)&&&(0.027&)&&&(0.026&)\\\addlinespace
Log hourly wage&&3.237&&$-$0.003&&3.387&&$-$0.079&&2.813&&0.132&\sym{***}\\
&&&&(0.050&)&&&(0.049&)&&&(0.044&)\\\addlinespace
Monthly earnings (2019 USD)&&1,356.069&&0.295&&1,390.819&&$-$153.473&\sym{**}&816.821&&110.230&\sym{**}\\
&&&&(75.313&)&&&(77.290&)&&&(49.546&)\\\addlinespace
\(N\) (formal employment)&&3,234&&37,794&&4,012&&55,030&&543&&8,147&\\
\(N\) (log hourly wage)&&2,027&&24,564&&2,694&&32,972&&394&&6,100&\\\addlinespace

\addlinespace\addlinespace
\multicolumn{13}{l}{Panel C.\ Graduation and earnings 10--13 years after application (2SLS)} \\\midrule
Graduated from UERJ program&&0.002&&0.718&\sym{***}&0.003&&0.693&\sym{***}&0.003&&0.661&\sym{***}\\
&&&&(0.017&)&&&(0.014&)&&&(0.021&)\\\addlinespace
Formal employment&&0.693&&0.032&&0.686&&$-$0.026&&0.714&&0.037&\\
&&&&(0.027&)&&&(0.031&)&&&(0.039&)\\\addlinespace
Log hourly wage&&3.636&&0.005&&3.637&&0.005&&3.052&&0.024&\\
&&&&(0.054&)&&&(0.058&)&&&(0.063&)\\\addlinespace
Monthly earnings (2019 USD)&&2,005.191&&$-$84.946&&1,757.947&&$-$99.418&&1,041.942&&56.577&\\
&&&&(94.587&)&&&(109.084&)&&&(75.202&)\\\addlinespace
\(N\) (formal employment)&&3,234&&39,134&&2,974&&41,138&&388&&4,320&\\
\(N\) (log hourly wage)&&2,237&&24,273&&2,021&&26,407&&273&&3,746&\\

\midrule

\end{tabular}
\par\end{centering}
\singlespacing\justify
\emph{Notes}: This table presents RD estimates for the effects of UERJ enrollment on graduation, formal employment, and earnings. Columns (A), (C), and (E) show means of each dependent variable for applicants in each group who scored $(-0.1,0)$ SDs below the cutoff. Columns (B), (D), and (F) show RD coefficients. Panel A reports reduced-form RD coefficients, $\theta$, from equation \eqref{rd_rf}, which measure the effects of UERJ admission on UERJ enrollment. Panels B--C report 2SLS RD coefficients, $\beta$, from equation \eqref{rd_iv}, which measure the effects of UERJ enrollment on outcomes 6--9 (Panel B) and 10--13 (Panel C) years after application. Parentheses contain standard errors clustered at the individual level. Sample sizes refer to the dependent variables indicated in the parentheses after $N$.

{*} $p<0.10$, {*}{*} $p<0.05$, {*}{*}{*} $p<0.01$.	

}
\end{table}

\clearpage

\begin{table}[htp]{\scriptsize\begin{centering}
\protect\caption{RD estimates for enrollment in other universities and degree attainment} \label{tab:rd-census-enrol}
\newcommand\w{1.15}
\begin{tabular}{l@{}lR{\w cm}@{}L{0cm}R{\w cm}@{}L{0.43cm}R{\w cm}@{}L{0cm}R{\w cm}@{}L{0.43cm}R{\w cm}@{}L{0cm}R{\w cm}@{}L{0.43cm}}
\midrule
&&(A)&&(B)&&(C)&&(D)&&(E)&&(F)\\\addlinespace
&&\multicolumn{3}{c}{1995--2001} &&\multicolumn{3}{c}{2004--2011}&&\multicolumn{3}{c}{2004--2011}\\
&&\multicolumn{3}{c}{general track}&&\multicolumn{3}{c}{ general track}&&\multicolumn{3}{c}{AA tracks} \\\cmidrule{3-5}\cmidrule{7-9}\cmidrule{11-13}
&& Mean&&RD  && Mean&&RD && Mean&&RD  \\
Dependent variable &&below &&coef&&below &&coef&&below &&coef\\\midrule

\addlinespace\addlinespace
\multicolumn{13}{l}{Panel A.\ Enrollment in Rio de Janeiro universities (reduced form, 2009--2011 cohorts only)} \\\midrule
\# enrolled in UERJ&&&&&&1.465&&0.271&\sym{***}&1.051&&0.880&\sym{***}\\
&&&&&&&&(0.037&)&&&(0.088&)\\\addlinespace
\# enrolled in UFRJ&&&&&&3.369&&$-$0.147&\sym{***}&2.381&&$-$0.111&\\
&&&&&&&&(0.057&)&&&(0.137&)\\\addlinespace
\# enrolled in other federal universities&&&&&&4.407&&$-$0.165&\sym{**}&3.181&&0.041&\\
&&&&&&&&(0.083&)&&&(0.168&)\\\addlinespace
\# enrolled in a top-100 private university&&&&&&5.154&&$-$0.176&\sym{**}&4.312&&0.147&\\
&&&&&&&&(0.077&)&&&(0.164&)\\\addlinespace
\# enrolled in other private universities&&&&&&5.110&&$-$0.041&&5.181&&$-$0.457&\sym{**}\\
&&&&&&&&(0.062&)&&&(0.229&)\\\addlinespace
\# enrolled in same progam area (2-dig)&&&&&&3.448&&0.120&\sym{**}&2.647&&0.351&\sym{**}\\
&&&&&&&&(0.059&)&&&(0.138&)\\\addlinespace
\# enrolled in same progam area (3-dig)&&&&&&1.661&&0.192&\sym{***}&1.367&&0.459&\sym{***}\\
&&&&&&&&(0.040&)&&&(0.082&)\\\addlinespace
\(N\) (\# enrolled in UERJ)&&&&&&1,553&&19,895&&215&&2,757&\\\addlinespace

\addlinespace\addlinespace
\multicolumn{13}{l}{Panel B.\ Educational attainment measured in RAIS (2SLS)} \\\midrule
Any college degree, 6-9 years later&&0.731&&0.044&&0.785&&0.006&&0.636&&$-$0.002&\\
&&&&(0.032&)&&&(0.029&)&&&(0.038&)\\\addlinespace
Ever earned a college degree&&0.911&&0.012&&0.839&&0.026&&0.713&&0.010&\\
&&&&(0.017&)&&&(0.025&)&&&(0.033&)\\\addlinespace
Ever earned a graduate degree&&0.107&&$-$0.004&&0.069&&$-$0.017&&0.051&&$-$0.006&\\
&&&&(0.020&)&&&(0.017&)&&&(0.015&)\\\addlinespace
\(N\) (ever college degree)&&2,417&&32,718&&2,925&&36,617&&415&&5,978&\\
\midrule

\end{tabular}
\par\end{centering}
\singlespacing\justify
\emph{Notes}: This table presents RD estimates for enrollment in Rio de Janeiro universities and educational attainment.

Panel A reports reduced-form RD coefficients, $\theta$, from equation \eqref{rd_rf}. The dependent variables are the \emph{total} number of enrollees in a given group of universities or field of study who share the applicant's birthdate, gender, and enrollment year as measured in Brazil's higher education census (see Appendix B.5). We categorize universities into four groups by ownership and selectivity: 1) The federal university in Rio de Janeiro municipality (UFRJ); 2) Federal universities in the Rio de Janeiro suburbs (UFF, UFRRJ, UNIRIO); 3) Private universities in Rio de Janeiro municipality that ranked in the top 100 of the 2012 \emph{Folha} ranking (PUC-Rio, UNESA); and 4) Other private universities in Rio de Janeiro municipality (UGF, UVA, UCAM, Universo, UCB). Fields of study are defined by 2- and 3-digit census major codes. The sample is 2009--2011 UERJ applicants. Regressions include gender and age dummies to increase precision.

Panel B reports 2SLS RD coefficients, $\beta$, from equation \eqref{rd_iv}. The dependent variables are indicators for educational attainment measured in the RAIS. Regressions include all UERJ applicants.

The columns are defined in the same way as Table \ref{tab:rd_returns}. Parentheses contain standard errors clustered at the individual level. Sample sizes refer to the dependent variables indicated in the parentheses after $N$.

{*} $p<0.10$, {*}{*} $p<0.05$, {*}{*}{*} $p<0.01$.

}
\end{table}

\clearpage

\begin{table}[htp!]{\scriptsize\begin{centering}\vspace{-1em}
\protect\caption{RD estimates for employment at alumni firms}\label{tab:rd_job}
\newcommand\w{1.15}
\begin{tabular}{l@{}lR{\w cm}@{}L{0cm}R{\w cm}@{}L{0.43cm}R{\w cm}@{}L{0cm}R{\w cm}@{}L{0.43cm}R{\w cm}@{}L{0cm}R{\w cm}@{}L{0.43cm}}
\midrule
&&(A)&&(B)&&(C)&&(D)&&(E)&&(F)\\\addlinespace
&&\multicolumn{3}{c}{1995--2001} &&\multicolumn{3}{c}{2004--2011}&&\multicolumn{3}{c}{2004--2011}\\
&&\multicolumn{3}{c}{general track}&&\multicolumn{3}{c}{ general track}&&\multicolumn{3}{c}{AA tracks} \\\cmidrule{3-5}\cmidrule{7-9}\cmidrule{11-13}
&& Mean&&RD  && Mean&&RD && Mean&&RD  \\
Dependent variable && below&&coef&& below&&coef&&below &&coef\\\midrule

\addlinespace\addlinespace
\multicolumn{13}{l}{Panel A.\ Employment 6--9 years after application (2SLS)} \\\midrule
Employed at firm with any  &&0.600&&0.118&\sym{***}&0.572&&0.070&\sym{**}&0.477&&0.137&\sym{***}\\
\quad UERJ alum&&&&(0.033&)&&&(0.034&)&&&(0.038&)\\\addlinespace
Employed at firm with any &&0.579&&0.129&\sym{***}&0.540&&0.076&\sym{**}&0.437&&0.106&\sym{***}\\
\quad general track alum&&&&(0.035&)&&&(0.034&)&&&(0.035&)\\\addlinespace
Employed at firm with any &&0.375&&0.036&&0.402&&0.076&\sym{**}&0.386&&0.129&\sym{***}\\
\quad AA track alum&&&&(0.031&)&&&(0.033&)&&&(0.037&)\\\addlinespace
\# UERJ alumni per 1000&&6.561&&4.652&\sym{***}&7.120&&9.738&\sym{***}&3.334&&8.523&\sym{**}\\
\quad workers at firm &&&&(1.624&)&&&(2.583&)&&&(3.580&)\\\addlinespace
Firm mean wage (log)&&3.303&&0.018&&3.475&&$-$0.095&\sym{*}&3.073&&0.106&\sym{*}\\
&&&&(0.043&)&&&(0.053&)&&&(0.062&)\\\addlinespace
\(N\) (firm mean wage)&&2,024&&30,345&&2,681&&31,087&&394&&4,306&\\\addlinespace

\addlinespace\addlinespace
\multicolumn{13}{l}{Panel B.\ Employment 10--13 years after application (2SLS)} \\\midrule
Employed at firm with any &&0.649&&0.059&\sym{*}&0.573&&0.028&&0.498&&0.080&\sym{*}\\
\quad UERJ alum&&&&(0.033&)&&&(0.037&)&&&(0.044&)\\\addlinespace
Employed at firm with any &&0.629&&0.060&&0.550&&0.014&&0.451&&0.086&\sym{**}\\
\quad general track alum&&&&(0.037&)&&&(0.038&)&&&(0.039&)\\\addlinespace
Employed at firm with any &&0.411&&0.032&&0.384&&0.056&&0.383&&0.082&\sym{*}\\
\quad AA track alum&&&&(0.034&)&&&(0.036&)&&&(0.044&)\\\addlinespace
\# UERJ alumni per 1000 &&5.873&&3.648&\sym{***}&5.620&&5.954&\sym{**}&3.224&&$-$0.092&\\
\quad workers at firm &&&&(1.078&)&&&(3.025&)&&&(2.767&)\\\addlinespace
Firm mean wage (log)&&3.572&&0.093&\sym{*}&3.581&&$-$0.053&&3.223&&0.049&\\
&&&&(0.053&)&&&(0.062&)&&&(0.071&)\\\addlinespace
\(N\) (firm mean wage)&&2,236&&24,701&&2,010&&21,071&&275&&3,133&\\
\midrule

\end{tabular}
\par\end{centering}
\singlespacing\justify
\emph{Notes}: This table presents 2SLS RD coefficients, $\beta$, from equation \eqref{rd_iv}, which measure the effects of UERJ enrollment on employment at alumni firms 6--9 (Panel A) and 10--13 (Panel B) years after application. The first three outcomes are indicators for employment at any firm during each time period with: 1) any UERJ alum; 2) any general track alum; and 3) any AA track alum. The fourth outcome is the average number of alumni per 1000 workers at the applicant's firms over the time period. The fifth outcome is the natural log of the mean hourly wage at the applicant's firms averaged over the time period.

The columns are defined in the same way as Table \ref{tab:rd_returns}. Parentheses contain standard errors clustered at the individual level. Sample sizes refer to the dependent variables indicated in the parentheses after $N$. 

{*} $p<0.10$, {*}{*} $p<0.05$, {*}{*}{*} $p<0.01$.

}
\end{table}

\clearpage

\begin{table}[htp!]{\scriptsize\begin{centering}
\protect\caption{DD estimates of the effects of AA exposure on student body composition}\label{tab:dd_traits}
\newcommand\w{1.4}
\begin{tabular}{l@{}lR{2 cm}@{}L{0.43cm}R{\w cm}@{}L{0.43cm}R{\w cm}@{}L{0.43cm}R{\w cm}@{}L{0.43cm}R{\w cm}@{}L{0.43cm}R{\w cm}@{}L{0.43cm}R{\w cm}@{}L{0.43cm}}
\midrule
&&(A)&&(B)&&(C)&&(D)\\\addlinespace
&&Pre-AA mean&&\multicolumn{5}{c}{DD coefficients}\\\cmidrule{3-3}\cmidrule{5-9}
&& All top && All top && URM top && Non-top\\
Dependent variable &&enrollees&&enrollees&&enrollees&&enrollees\\\midrule

\addlinespace\addlinespace
\multicolumn{9}{l}{Panel A.\ Exposure to affirmative action} \\\midrule
Prop.\ of classmates from AA tracks&&0.000&&0.189&\sym{***}&0.208&\sym{***}&0.192&\sym{***}\\
&&&&(0.017&)&(0.021&)&(0.018&)\\

\addlinespace\addlinespace
\multicolumn{9}{l}{Panel B.\ Demographic characteristics} \\\midrule
Age at application&&21.921&&0.191&&0.794&&0.666&\sym{***}\\
&&&&(0.312&)&(0.899&)&(0.229&)\\\addlinespace
Female&&0.501&&0.032&&0.102&&0.038&\sym{*}\\
&&&&(0.022&)&(0.078&)&(0.021&)\\\addlinespace
White&&0.810&&0.013&&&&$-$0.121&\sym{***}\\
&&&&(0.018&)&&&(0.025&)\\\addlinespace
Brown&&0.156&&0.000&&&&0.043&\sym{**}\\
&&&&(0.012&)&&&(0.017&)\\\addlinespace
Black&&0.025&&$-$0.005&&$-$0.009&&0.077&\sym{***}\\
&&&&(0.010&)&(0.037&)&(0.012&)\\

\addlinespace\addlinespace
\multicolumn{9}{l}{Panel C.\ Admission exam scores (standardized in population of all enrollees)} \\\midrule
Field exam writing score&&0.178&&$-$0.045&&0.031&&$-$0.246&\sym{***}\\
&&&&(0.043&)&(0.129&)&(0.046&)\\\addlinespace
Mean field exam subject score&&0.151&&$-$0.029&&0.039&&$-$0.182&\sym{**}\\
&&&&(0.064&)&(0.121&)&(0.084&)\\\addlinespace
Admission score&&0.270&&$-$0.080&&0.034&&$-$0.498&\sym{***}\\
&&&&(0.112&)&(0.160&)&(0.143&)\\

\addlinespace\addlinespace
\multicolumn{9}{l}{Panel D.\ Predicted log wage based on characteristics and scores} \\\midrule
Predicted log wage&&3.298&&$-$0.023&&$-$0.011&&$-$0.161&\sym{***}\\
&&&&(0.029&)&(0.049&)&(0.043&)\\\addlinespace
Predicted log wage (if in RAIS)&&3.251&&$-$0.033&&$-$0.009&&$-$0.154&\sym{***}\\
&&&&(0.028&)&(0.053&)&(0.043&)\\\addlinespace
\(N\) (enrollees)&&16,466&&35,866&&1,631&&30,854&\\
\midrule

\end{tabular}
\par\end{centering}
\singlespacing\justify
\emph{Notes}: This table displays DD estimates of the effect of affirmative action exposure on student characteristics. Column (A) shows the mean of each dependent variable for all top enrollees in the 1995--2001 cohorts. Columns (B)--(D) display estimates of $\pi$ from equation \eqref{dd} for all top enrollees, URM top enrollees, and non-top enrollees. The dependent variables are:
\begin{itemize}[leftmargin=1.5em]
\item Panel A. The proportion of enrollees in an individual's program/cohort who were from an affirmative action track. 

\item Panel B. Demographic characteristics of enrollees.

\item Panel C. Applicants' field exam and overall admission scores, normalized to be mean 0/SD 1 in the population of all UERJ enrollees in a given cohort. Field exam score regressions include dummies for cohorts $\times$ the applicant's set of subject tests (which vary by major).

\item Panel D. The predicted value from a regression of log hourly wage on all variables in Panels B--C. 
\end{itemize}
Parentheses contain standard errors clustered at the program level. 

{*} $p<0.10$, {*}{*} $p<0.05$, {*}{*}{*} $p<0.01$.

}
\end{table}

\clearpage

\begin{table}[htp]{\scriptsize\begin{centering}
\protect\caption{DD estimates for graduation, employment, and earnings 6--9 years after application}\label{tab:dd_labor}
\newcommand\w{1.4}
\begin{tabular}{l@{}lR{2 cm}@{}L{0.43cm}R{\w cm}@{}L{0.43cm}R{\w cm}@{}L{0.43cm}R{\w cm}@{}L{0.43cm}R{\w cm}@{}L{0.43cm}R{\w cm}@{}L{0.43cm}R{\w cm}@{}L{0.43cm}}
\midrule
&&(A)&&(B)&&(C)&&(D)\\\addlinespace
&&Pre-AA mean&&\multicolumn{5}{c}{DD coefficients}\\\cmidrule{3-3}\cmidrule{5-9}
&& All top && All top && URM top && Non-top\\
Dependent variable &&enrollees&&enrollees&&enrollees&&enrollees\\\midrule
	
\addlinespace\addlinespace
\multicolumn{9}{l}{Panel A.\ Graduation and formal employment} \\\midrule
Graduated from UERJ program&&0.556&&0.013&&0.013&&0.006&\\
&&&&(0.021&)&(0.063&)&(0.028&)\\\addlinespace
Formal employment&&0.734&&$-$0.027&\sym{*}&0.076&&$-$0.012&\\
&&&&(0.015&)&(0.054&)&(0.015&)\\

\addlinespace\addlinespace
\multicolumn{9}{l}{Panel B.\ Earnings} \\\midrule
Log hourly wage&&3.245&&$-$0.132&\sym{***}&$-$0.220&\sym{**}&$-$0.212&\sym{***}\\
&&&&(0.045&)&(0.107&)&(0.062&)\\\addlinespace
Monthly earnings (2019 USD)&&1,380.558&&$-$169.838&\sym{***}&$-$28.036&&$-$272.989&\sym{***}\\
&&&&(53.057&)&(100.006&)&(89.500&)\\\addlinespace
Firm mean hourly wage (log)&&3.316&&$-$0.095&\sym{**}&$-$0.334&\sym{***}&$-$0.183&\sym{***}\\
&&&&(0.035&)&(0.113&)&(0.051&)\\

\addlinespace\addlinespace
\multicolumn{9}{l}{Panel C.\ Employment at firms with pre- and post-AA alumni} \\\midrule
Pre-AA alumni&&0.602&&$-$0.055&\sym{**}&$-$0.009&&$-$0.044&\\
&&&&(0.023&)&(0.081&)&(0.033&)\\\addlinespace
Only post-AA alumni&&0.067&&0.049&\sym{**}&0.019&&0.036&\\
&&&&(0.023&)&(0.056&)&(0.023&)\\

\addlinespace\addlinespace
\multicolumn{9}{l}{Panel D.\ Alumni firm employment by application track and cohort} \\\midrule
General track alumni from same cohort&&0.451&&$-$0.098&\sym{***}&$-$0.118&\sym{*}&$-$0.072&\sym{**}\\
&&&&(0.021&)&(0.068&)&(0.028&)\\\addlinespace
General track alumni from diff.\ cohort&&0.233&&0.042&\sym{**}&0.138&&0.004&\\
&&&&(0.016&)&(0.086&)&(0.017&)\\\addlinespace
Only AA alumni from same cohort&&0.000&&0.036&\sym{***}&0.035&\sym{**}&0.051&\sym{***}\\
&&&&(0.009&)&(0.014&)&(0.007&)\\\addlinespace
Only AA alumni from diff.\ cohort&&0.012&&0.010&\sym{**}&$-$0.017&&0.014&\sym{***}\\
&&&&(0.005&)&(0.022&)&(0.004&)\\\addlinespace

\(N\) (enrollees)&&16,466&&35,866&&1,631&&30,854&\\
\(N\) (wage observations)&&12,062&&26,445&&1,323&&22,975&\\
\midrule

\end{tabular}
\par\end{centering}
\singlespacing\justify
\emph{Notes}: This table displays DD estimates of the effect of affirmative action exposure on graduation, earnings, and employment at alumni firms measured 6--9 years after application. The columns are defined in the same way as Table \ref{tab:dd_traits}. The dependent variables are defined similarly to those in Tables \ref{tab:rd_returns} and \ref{tab:rd_job}. In Panel C, we categorize firms using alumni from the pre- and post-AA cohorts. In Panel D, we categorize firms using the alum's cohort (same or different than the applicant's cohort) and application track (general or AA). The outcomes in Panels C--D are non-overlapping, i.e., variables in the lower rows equal one only if the firm did not hire alumni who meet the criteria for the higher rows. Parentheses contain standard errors clustered at the program level. 

{*} $p<0.10$, {*}{*} $p<0.05$, {*}{*}{*} $p<0.01$	

}
\end{table}

\clearpage

\begin{table}[htp]{\scriptsize\begin{centering}
\protect\caption{DD estimates for UERJ's Enade exam taker characteristics and scores}\label{tab:dd_enade}
\newcommand\w{1.6}
\begin{tabular}{l@{}lR{2 cm}@{}L{0.43cm}R{\w cm}@{}L{0.43cm}R{\w cm}@{}L{0.43cm}R{\w cm}@{}L{0.43cm}R{\w cm}@{}L{0.43cm}R{\w cm}@{}L{0.43cm}R{\w cm}@{}L{0.43cm}}
\midrule
&&(A)&&(B)&&(C)&&(D)\\\addlinespace
&&UERJ&&\\
&&pre-AA mean&&\multicolumn{5}{c}{DD coefficients}\\\cmidrule{3-3}\cmidrule{5-9}
&& &&  && White \\
&& All && All && private HS  && Other \\
Dependent variable &&students&&students&&students&&students\\\midrule
	
\addlinespace\addlinespace
\multicolumn{9}{l}{Panel A.\ Characteristics of Enade exam takers} \\\midrule
\# exam takers&&36.086&&4.322&&$-$7.926&\sym{***}&9.142&\sym{***}\\
&&&&(2.664&)&(1.398&)&(1.777&)\\\addlinespace
White&&0.716&&$-$0.132&\sym{***}&&&$-$0.066&\sym{***}\\
&&&&(0.009&)&&&(0.009&)\\\addlinespace
Private HS student&&0.570&&$-$0.131&\sym{***}&&&$-$0.040&\sym{***}\\
&&&&(0.015&)&&&(0.009&)\\\addlinespace
Female&&0.526&&$-$0.005&&$-$0.009&&$-$0.016&\\
&&&&(0.011&)&(0.012&)&(0.011&)\\\addlinespace
Age&&26.520&&0.661&\sym{**}&0.042&&0.700&\sym{*}\\
&&&&(0.261&)&(0.209&)&(0.361&)\\\addlinespace
Mother has a HS degree&&0.692&&$-$0.086&\sym{***}&0.016&&$-$0.081&\sym{***}\\
&&&&(0.012&)&(0.012&)&(0.012&)\\\addlinespace
HH income / minimum wage&&7.724&&$-$1.577&\sym{***}&0.292&&$-$1.873&\sym{***}\\
&&&&(0.098&)&(0.210&)&(0.117&)\\\addlinespace
Predicted general score &&0.563&&$-$0.008&\sym{***}&0.001&\sym{**}&$-$0.007&\sym{***}\\
&&&&(0.001&)&(0.001&)&(0.001&)\\

\addlinespace\addlinespace
\multicolumn{9}{l}{Panel B.\ Enade scores (proportion correct answers)} \\\midrule
Overall score&&0.553&&$-$0.038&\sym{***}&$-$0.022&\sym{**}&$-$0.051&\sym{***}\\
&&&&(0.005&)&(0.008&)&(0.006&)\\\addlinespace
Field-specific component&&0.519&&$-$0.037&\sym{***}&$-$0.021&\sym{**}&$-$0.048&\sym{***}\\
&&&&(0.005&)&(0.009&)&(0.006&)\\\addlinespace
General component&&0.657&&$-$0.041&\sym{***}&$-$0.026&\sym{***}&$-$0.059&\sym{***}\\
&&&&(0.005&)&(0.007&)&(0.007&)\\\addlinespace

\(N\) (programs $\times$ years)&&36&&1,664&&1,664&&1,664&\\
\(N\) (exam takers)&&1,059&&61,112&&16,851&&37,992&\\

\midrule
	
\end{tabular}
\par\end{centering}
\singlespacing\justify
\emph{Notes}: This table displays DD estimates of the effect of affirmative action on the characteristics (Panel A) and scores (Panel B) of UERJ's Enade exam takers. The sample is 2004--2015 Enade participants from UERJ and other federal/state universities that did not implement affirmative action through 2012 (see Appendix Table A17). Column (A) shows means for UERJ exam takers in 2004--2006. Other columns show $\pi$ coefficients from the DD regression:
\begin{align*}
	Y_{mjt} &= \gamma_{mj} + \gamma_{mt} + \pi[\text{UERJ}_j\times\text{Post}_t] + \varepsilon_{mjt}.
\end{align*}
Regressions are at the exam field ($m$) by institution ($j$) by year ($t$) level, with observations weighted by the number of exam takers. (In the first row of Panel A, we weight by the number of 2004--2006 exam takers in each $mj$ cell.) We include dummies for field $\times$ institution, field $\times$ year, and UERJ $\times$ the 2007--2015 cohorts ($\text{UERJ}_j\times\text{Post}_t$). Columns (B)--(D) include all students, white private high school students, and non-white and/or public school students. ``Predicted general score'' is the predicted value from a regression of general component scores on age and dummies for gender, race, private high school, mother's education, father's education, and family income. Parentheses contain standard errors clustered at the institution level. 

{*} $p<0.10$, {*}{*} $p<0.05$, {*}{*}{*} $p<0.01$	

}
\end{table}


\clearpage

\renewcommand \thesection{\Alph{section}}\setcounter{section}{0}
\renewcommand \thesubsection{\thesection.\arabic{subsection}}\setcounter{subsection}{0}
\renewcommand{\thetable}{\thesection\arabic{table}}\setcounter{table}{0}
\renewcommand{\thefigure}{\thesection\arabic{figure}}\setcounter{figure}{0}
\renewcommand{\theequation}{\thesection\arabic{equation}}\setcounter{equation}{0}

\clearpage
\begin{center}\noindent {\large \textbf{Online Appendix for ``The direct and spillover effects of large-scale affirmative action at an elite Brazilian university'' \\ by Cecilia Machado, Germ\'an Reyes, and Evan Riehl}}\end{center}
\vspace{1em}
\noindent Outline:
\begin{enumerate}[label=\Alph*.]
\item Appendix figures and tables
\item Empirical appendix
\end{enumerate}

\clearpage

\section{Appendix figures and tables}
\vspace{2em}

\begin{figure}[htp]{\scriptsize
\begin{centering}
\includegraphics[width=.7\linewidth]{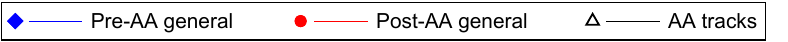}\vspace{1em}

\begin{minipage}{.48\textwidth}
\includegraphics[width=\linewidth]{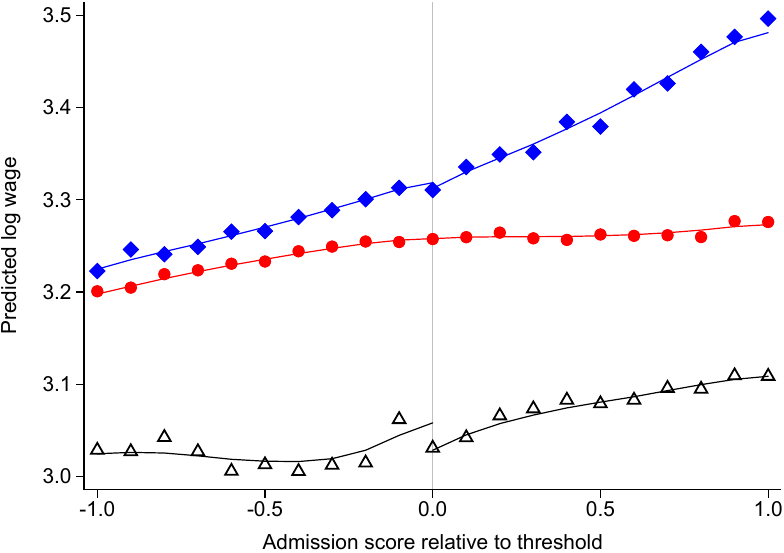}
\captionof*{figure}{Panel A. All applicants}
\end{minipage}\hspace{1em}
\begin{minipage}{.48\textwidth}
\includegraphics[width=\linewidth]{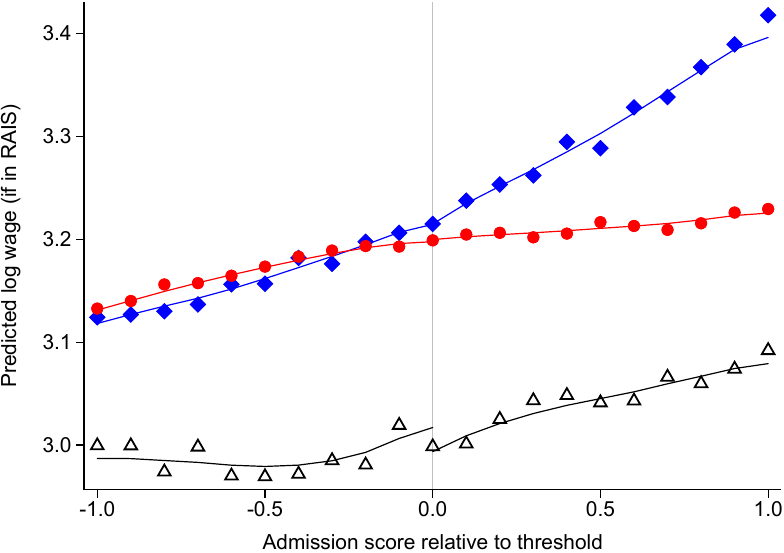}
\captionof*{figure}{Panel B. Applicants who appear in the RAIS}
\end{minipage}
\par\end{centering}
\protect\caption{Predicted log wage based on applicant characteristics} \label{fig:bal}
\singlespacing\justify
\emph{Notes}: This figure presents RD graphs for pre-AA general applicants (blue diamonds), post-AA general applicants (red circles), and Black/public school applicants (black triangles). The $x$-axis in each panel is an applicant's standardized admission score normalized to zero at the cutoff. The dependent variable on the $y$-axis is the predicted value from a regression of log hourly wage (6--9 years after application) on student characteristics (age, gender, race, mother's education, family income, score on the writing component of the field exam, and qualifying exam score). Markers depict means in 0.1 SD bins of the standardized score. Lines are predicted values from local linear regressions estimated separately above and below the threshold with a triangular kernel.

}
\end{figure}

\clearpage
\begin{figure}[hpb!]{\scriptsize
\begin{centering}
\begin{minipage}{.48\textwidth}
\includegraphics[width=\linewidth]{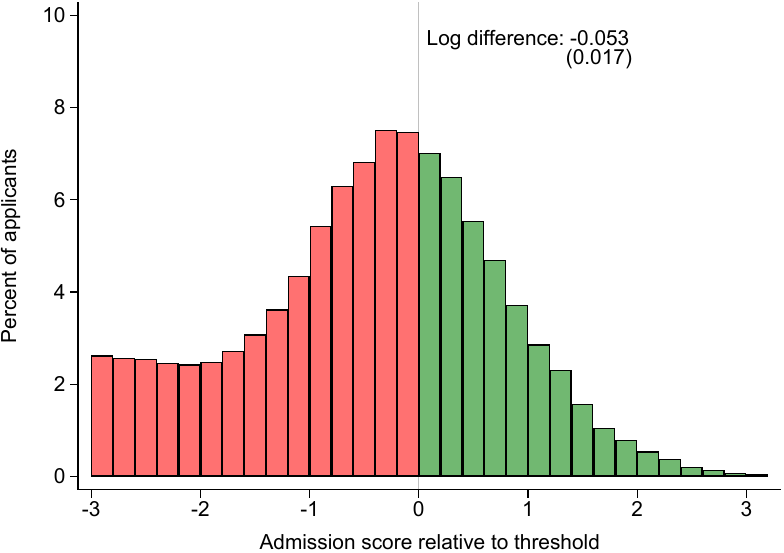}
\captionof*{figure}{Panel A. General track (1995--2001)}
\end{minipage}\hspace{1em}
\begin{minipage}{.48\textwidth}
\includegraphics[width=\linewidth]{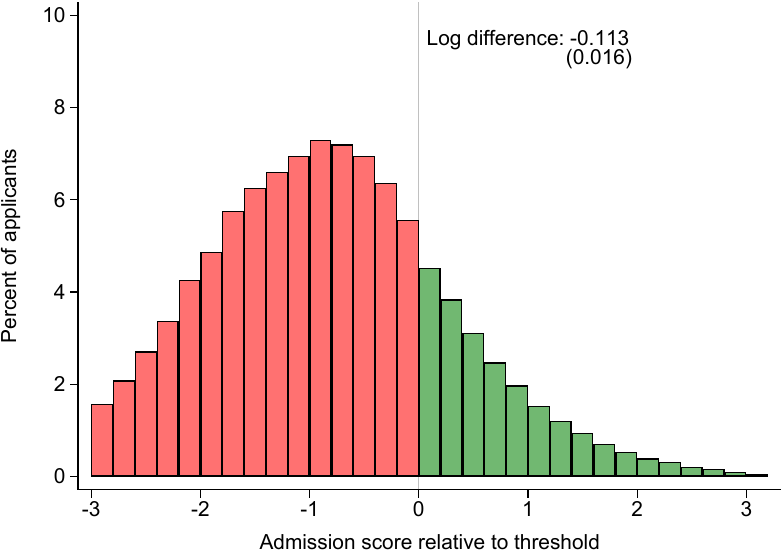}
\captionof*{figure}{Panel B. General track (2004--2011)}
\end{minipage}\vspace{1em}

\begin{minipage}{.48\textwidth}
\includegraphics[width=\linewidth]{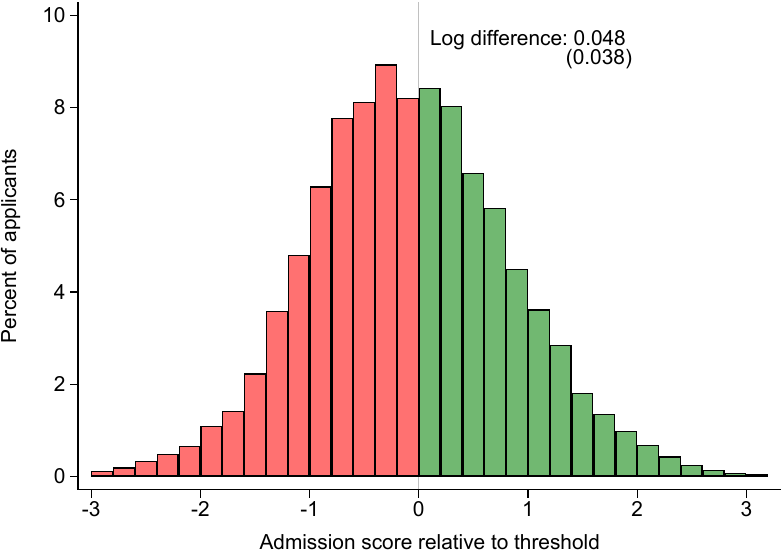}
\captionof*{figure}{Panel C. AA tracks (2004--2011)}
\end{minipage}\hspace{1em}
\par\end{centering}
\protect\caption{Density of admission scores relative to the threshold} \label{fig:hist}
\singlespacing\justify
\emph{Notes}: This figure shows the density of admission scores relative to the cutoff. The $x$-axis is a student's admission score normalized to zero at the cutoff of the relevant application pool. The $y$-axis shows the percent of applicants within 0.20 SD unit bins of the admission score. We restrict the figure to only display normalized scores within three SD of the cutoff. We also exclude applicants whose score defines the cutoff.

Panel A shows the distribution of admission scores for pre-AA general applicants, Panel B for post-AA general applicants, and Panel C for Black/public school applicants. 

Each figure displays the estimated the log difference in height at the threshold using the \citet{mccrary2008} density test. The standard error is shown in parentheses.

}
\end{figure}

\clearpage
\begin{figure}[htpb!]{\scriptsize
\begin{centering}
\includegraphics[width=.5\linewidth]{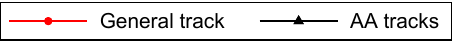}\vspace{1em}

\includegraphics[width=.75\linewidth]{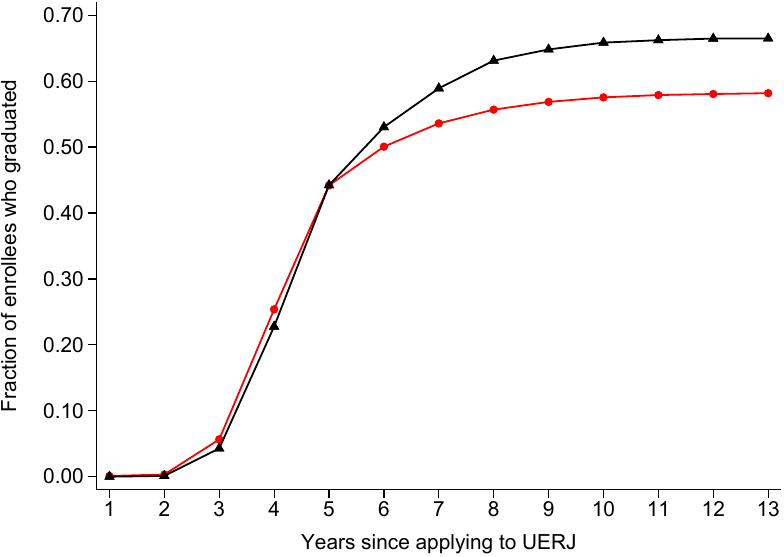}
\protect\caption{UERJ graduation rate by year since application} \label{fig:cdf_grad}
\par\end{centering}
\singlespacing \justify\emph{Notes}: This figure show the empirical cumulative distribution function of the graduation rate of students in programs in our RD sample (Panel A of Table \ref{tab:summ_stats}). We plot separately the graduation rate of general track enrollees (red line) and Black/public school enrollees (black line).

}
\end{figure}

\clearpage

\begin{figure}[htp]{\scriptsize
\begin{centering}
\includegraphics[width=.8\linewidth]{results/rd_legend}\vspace{1em}

\begin{minipage}{.48\textwidth}
\includegraphics[width=\linewidth]{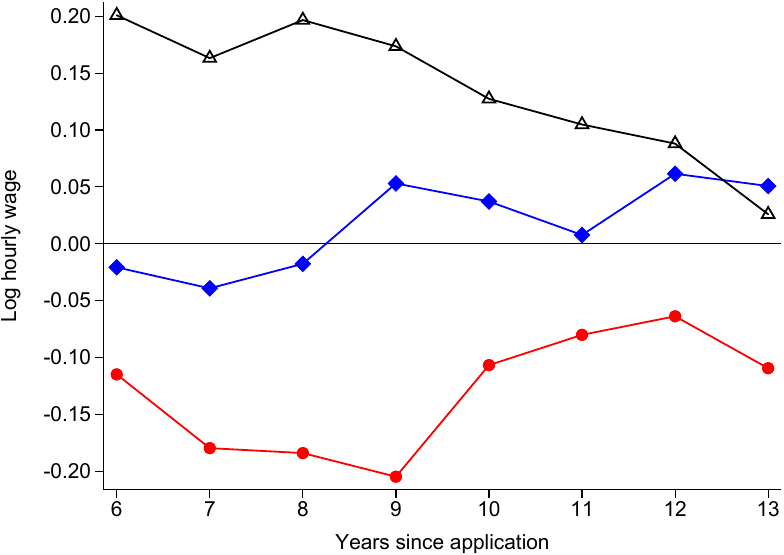}
\captionof*{figure}{Panel A. By years since application \\ (1997--2006 cohorts)}
\end{minipage}\hspace{1em}
\begin{minipage}{.48\textwidth}
\includegraphics[width=\linewidth]{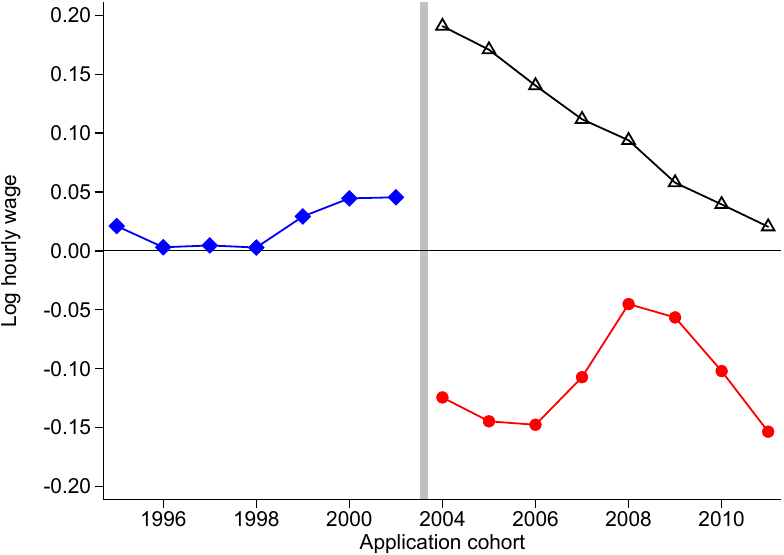}
\captionof*{figure}{Panel B. By cohort \\ (Wages measured 6--9 years later) }
\end{minipage}
\par\end{centering}
\protect\caption{RD estimates for log hourly wages by years since application and cohort} \label{fig:fade}
\singlespacing\justify
\emph{Notes}: This figure displays 2SLS RD coefficients, $\beta$, for pre-AA general applicants (blue diamonds), post-AA general applicants (red circles), and Black/public school applicants (black triangles).

Panel A plots $\beta$ coefficients for log hourly wages measured at different years since individuals applied to UERJ. To smooth estimates, we use the applicant's three-year average wage as the dependent variable in each regression  (years $t-1$, $t$, and $t+1$). We include only 1997--2006 cohorts since we observe their outcomes in each of 6--13 years later.

Panel B plots $\beta$ coefficients for log hourly wages estimated in different application cohorts. To smooth estimates, we include three adjacent cohorts for each regression (cohorts $t-1$, $t$, and $t+1$). All regressions use mean log hourly wage measured 6--9 years after application as the dependent variable.

}
\end{figure}
	
\clearpage

\begin{figure}[t]{\scriptsize
\begin{centering}
\begin{minipage}{.65\textwidth}
\includegraphics[width=\linewidth]{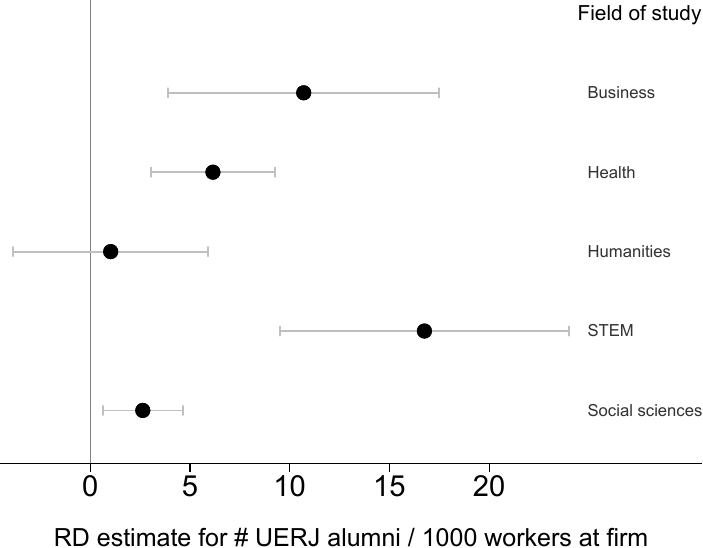}
\end{minipage}

\par\end{centering}
\protect\caption{Field of study heterogeneity in RD estimates for alumni per 1000 workers at the firm} \label{fig:network_het_field}
\singlespacing\justify
\emph{Notes}: This figure examines heterogeneity by field of study in the effects of UERJ enrollment on employment at alumni firms. The figure displays RD estimates for the effects of UERJ enrollment on the mean number of alumni per 1000 workers at the applicants' firms measured 6--9 years after application. These estimates are analogous to the 2SLS RD coefficients in the fourth row of Table \ref{tab:rd_job} (Panel A), except we estimate equation \eqref{rd_iv} pooling across all applicant groups. We estimate these regressions separately for the five field of study groups listed in Appendix \ref{app:var_def}. Markers depict the RD coefficients, $\beta$, coefficients from these regressions, and horizontal bars are 95 percent confidence intervals using standard errors clustered at the individual level.

}
\end{figure}

\clearpage

\begin{figure}[htp]{\scriptsize
\begin{centering}

\begin{minipage}{.48\textwidth}
\includegraphics[width=\linewidth]{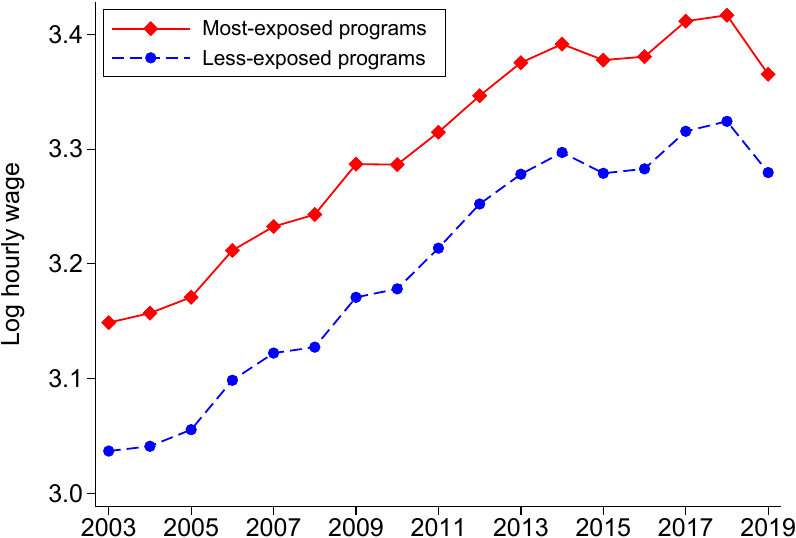}
\captionof*{figure}{Panel A. Industry mean wage in majors w/ more and less AA exposure}
\end{minipage}\hspace{1em}
\begin{minipage}{.48\textwidth}
\includegraphics[width=\linewidth]{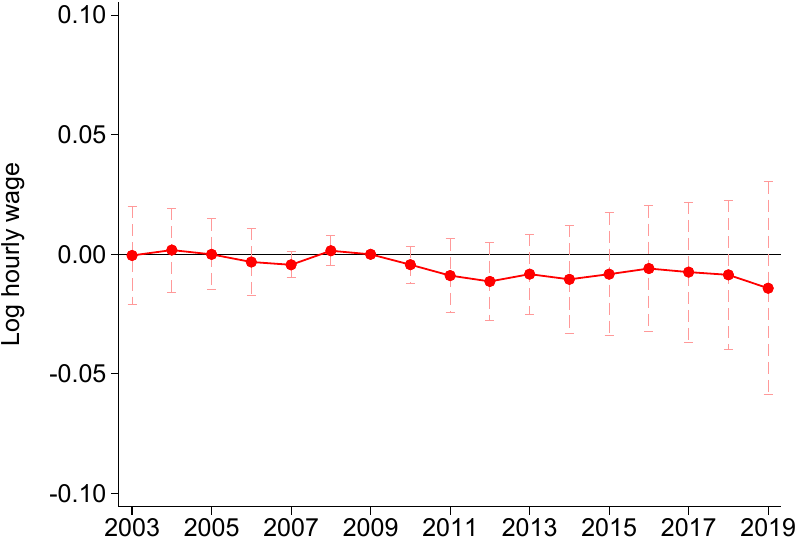}
\captionof*{figure}{Panel B. Event study of industry mean wage \\ \text{} }
\end{minipage}

\par\end{centering}
\protect\caption{Wage trends in industries that employed pre-AA top enrollees}\label{fig:wage_trends}
\singlespacing\justify

\emph{Notes}:  

This figure shows how hourly wages trended over time in industries that employed top enrollees from the pre-AA cohorts. We begin by computing the mean hourly in each (5-digit) industry $j$ and year $t$ using all workers in the RAIS data; we denote this industry $\times$ year mean wage by $\bar{w}_{jt}$. We then take our sample of top UERJ enrollees in the pre-AA cohorts (1995--2001), and compute the share of individuals in each major $m$ who were employed in industry $j$ measured 6--9 years after UERJ application; we denote these shares $s_{mj}$, with $\sum_j s_{mj} = 1$ for each $m$. Lastly, we define $\bar{w}_{mt} = \sum_j s_{mj} * \bar{w}_{jt}$ as the industry mean wage for UERJ major $m$ in year $t$, where this mean is computed using the pre-AA employment shares in each industry.

In Panel A, we plot the average value of $\bar{w}_{mt}$ in majors with more- (red line) and less- (blue line) exposure to affirmative action in each year from $t = 2003$ to $2019$. More-exposed programs are those in which the share of affirmative action enrollees in the 2004--2011 cohorts was 30 percent or higher (Panel A of Table \ref{tab:summ_stats}). Less-exposed programs are those in which the share of affirmative action enrollees in the 2004--2011 cohorts was below 30 percent (Panel B of Table \ref{tab:summ_stats}).

Panel B plots $\pi_t$ coefficients from the following event study regression:
\begin{align*}
         \bar{w}_{mt} = \gamma_m + \gamma_{tf(m)} + \pi_{t}\text{ExposureToAA}_m + \varepsilon_{mt},
\end{align*}
where $\gamma_m$ are program fixed effects, $\gamma_{tf(m)}$ are year $\times$ field of study fixed effects, and $\pi_t$ are coefficients on the interaction between year dummies (omitting 2009) and an indicator for more-exposed programs, $\text{ExposureToAA}_m$. Observations in this event study are weighted by the number of pre-AA top enrollees in each major $m$. Vertical dashed lines are 95 percent confidence intervals with standard errors clustered at the program level.

}
\end{figure}

\clearpage

\begin{table}[htp]{\scriptsize\begin{centering}
\protect\caption{RD balance tests}\label{tab:bal}
\newcommand\w{1.1}
\begin{tabular}{l@{}lR{\w cm}@{}L{0cm}R{\w cm}@{}L{0.43cm}R{\w cm}@{}L{0cm}R{\w cm}@{}L{0.43cm}R{\w cm}@{}L{0cm}R{\w cm}@{}L{0.43cm}}
&&(A)&&(B)&&(C)&&(D)&&(E)&&(F)\\\addlinespace
&&\multicolumn{3}{c}{1995--2001} &&\multicolumn{3}{c}{2004--2011}&&\multicolumn{3}{c}{2004--2011}\\
&&\multicolumn{3}{c}{general track}&&\multicolumn{3}{c}{ general track}&&\multicolumn{3}{c}{AA tracks} \\\cmidrule{3-5}\cmidrule{7-9}\cmidrule{11-13}
Dependent variable&& Mean&&RD  && Mean&&RD && Mean&&RD  \\
\quad (cohorts observed) &&below &&coef&&below &&coef&&below &&coef\\\midrule\addlinespace

\addlinespace
\multicolumn{13}{l}{\hspace{-1em} \textbf{Panel A.\ Applicant characteristics}} \\\midrule

Female&&&&&&0.530&&0.001&&0.602&&0.001&\\
\quad (2004--2011 cohorts) &&&&&&&&(0.008&)&&&(0.021&)\\
\addlinespace
Female (measured in RAIS)&&0.486&&0.008&&0.518&&0.012&&0.600&&0.001&\\
\quad (1995--2011 cohorts) &&&&(0.010&)&&&(0.009&)&&&(0.023&)\\
\addlinespace

White&&&&&&0.696&&0.002&&0.343&&$-$0.022&\\
\quad (2007--2011 cohorts) &&&&&&&&(0.010&)&&&(0.028&)\\
\addlinespace
White (measured in RAIS)&&0.790&&0.011&&0.714&&0.003&&0.436&&$-$0.027&\\
\quad (1995--2011 cohorts) &&&&(0.009&)&&&(0.008&)&&&(0.020&)\\
\addlinespace

Brown&&&&&&0.212&&$-$0.009&&0.313&&0.023&\\
\quad (2007--2011 cohorts) &&&&&&&&(0.009&)&&&(0.030&)\\
\addlinespace
Brown (measured in RAIS)&&0.173&&$-$0.008&&0.216&&$-$0.010&&0.321&&0.028&\\
\quad (1995--2011 cohorts) &&&&(0.010&)&&&(0.007&)&&&(0.022&)\\
\addlinespace

Age at application&&20.608&&0.181&\sym{**}&20.043&&0.037&&22.306&&$-$0.446&\sym{*}\\
\quad (1995--2011 cohorts) &&&&(0.081&)&&&(0.072&)&&&(0.262&)\\
\addlinespace

Mother has HS degree&&&&&&0.901&&$-$0.002&&0.534&&0.001&\\
\quad (2007--2011 cohorts) &&&&&&&&(0.007&)&&&(0.033&)\\
\addlinespace
HH income $ > $ 1.5x min.\ wage&&&&&&0.886&&$-$0.007&&0.341&&0.008&\\
\quad (2007--2011 cohorts) &&&&&&&&(0.007&)&&&(0.026&)\\
\addlinespace

Writing score (SD units)&&0.174&&0.020&&0.477&&0.011&&$-$0.202&&0.011&\\
\quad (1995--2001, 2007--2011 cohorts) &&&&(0.015&)&&&(0.016&)&&&(0.052&)\\
\addlinespace
Qualifying exam score (SD units)&&$-$0.148&&$-$0.009&\\
\quad (1995--2001 cohorts) &&&&(0.008&)\\

\midrule
Joint balance test ($ p$ value)&&&&0.110&&&&0.411&&&&0.875&\\
\addlinespace

\addlinespace
\multicolumn{13}{l}{\hspace{-1em} \textbf{Panel B.\ Predicted log wage based on applicant characteristics}} \\\midrule
Predicted log wage&&3.313&&$-$0.004&&3.254&&0.001&&3.062&&$-$0.003&\\
\quad (1995--2011 cohorts) &&&&(0.006&)&&&(0.003&)&&&(0.010&)\\
\addlinespace
Predicted log wage (if in RAIS)&&3.206&&0.004&&3.193&&0.003&&3.019&&$-$0.006&\\
\quad (1995--2011 cohorts) &&&&(0.006&)&&&(0.004&)&&&(0.011&)\\
\addlinespace

\midrule
$N$ (predicted wage) &&3,234&&27,610&&4,012&&45,731&&543&&6,410&\\

\(N\) (predicted wage if in RAIS)&&2,027&&17,027&&2,694&&30,315&&394&&4,303&\\
\addlinespace

\end{tabular}
\par\end{centering}
\singlespacing\justify
\emph{Notes}: This table presents RD balance tests. Columns (A), (C), and (E) show means of each dependent variable for applicants in each group who scored $(-0.1,0)$ SDs below the cutoff. Columns (B), (D), and (F) display reduced-form RD coefficients, $\theta$, from equation \eqref{rd_rf}, using the dependent variable listed in the row header. 

The last row in Panel A reports the $p$ values from $F$ tests that the coefficients on all covariates are jointly equal to zero. Parentheses contain standard errors clustered at the individual level. Sample sizes refer to the dependent variables indicated in the parentheses after $N$. 

{*} $p<0.10$, {*}{*} $p<0.05$, {*}{*}{*} $p<0.01$.

}
\end{table}

\clearpage

\begin{table}[htp]{\scriptsize\begin{centering}
\protect\caption{RD complier characteristics}\label{tab:compliers}
\newcommand\w{1.24}
\begin{tabular}{l@{}lR{\w cm}@{}L{0.43cm}R{\w cm}@{}L{0.43cm}R{\w cm}@{}L{0.43cm}R{\w cm}@{}L{0.43cm}R{\w cm}@{}L{0.43cm}R{\w cm}@{}L{0.43cm}}
&&(A)&&(B)&&(C)&&(D)&&(E)&&(F)\\\addlinespace
&&\multicolumn{5}{c}{} &&\multicolumn{5}{c}{Differences in}\\
&&\multicolumn{5}{c}{Complier characteristics}&&\multicolumn{5}{c}{complier characteristics} \\\cmidrule{3-7}\cmidrule{9-13}
&&Pre-AA&&Post-AA \\
Dependent variable&& general&&general  && AA&&  \\
\quad (cohorts observed) &&track &&track&&tracks &&(B)$-$(A)&&(C)$-$(A) &&(C)$-$(B)\\\midrule\addlinespace

\addlinespace

\multicolumn{13}{l}{\hspace{-1em} \textbf{Panel A.\ Applicant characteristics}} \\\midrule

Female (measured in RAIS)&&0.507&\sym{***}&0.523&\sym{***}&0.616&\sym{***}&0.016&&0.109&\sym{***}&0.093&\sym{***}\\
\quad (1995--2011 cohorts) &&(0.014&)&(0.014&)&(0.019&)&(0.020&)&(0.023&)&(0.024&)\\
\addlinespace
White (measured in RAIS)&&0.795&\sym{***}&0.727&\sym{***}&0.382&\sym{***}&$-$0.068&\sym{***}&$-$0.412&\sym{***}&$-$0.344&\sym{***}\\
\quad (1995--2011 cohorts) &&(0.012&)&(0.012&)&(0.017&)&(0.017&)&(0.021&)&(0.021&)\\
\addlinespace
Brown (measured in RAIS)&&0.180&\sym{***}&0.204&\sym{***}&0.338&\sym{***}&0.023&&0.158&\sym{***}&0.134&\sym{***}\\
\quad (1995--2011 cohorts) &&(0.013&)&(0.010&)&(0.017&)&(0.017&)&(0.022&)&(0.020&)\\
\addlinespace
Age at application&&21.485&\sym{***}&20.595&\sym{***}&21.879&\sym{***}&$-$0.891&\sym{***}&0.394&\sym{*}&1.285&\sym{***}\\
\quad (1995--2011 cohorts) &&(0.129&)&(0.118&)&(0.191&)&(0.175&)&(0.230&)&(0.224&)\\
\addlinespace
Writing score (SD units)&&0.359&\sym{***}&0.678&\sym{***}&$-$0.257&\sym{***}&0.319&\sym{***}&$-$0.616&\sym{***}&$-$0.935&\sym{***}\\
\quad (1995--2001, 2007--2011 cohorts) &&(0.024&)&(0.028&)&(0.043&)&(0.037&)&(0.049&)&(0.051&)\\
\addlinespace

\addlinespace
\multicolumn{13}{l}{\hspace{-1em} \textbf{Panel B.\ Predicted log wage based on applicant characteristics}} \\\midrule
Predicted log wage&&3.327&\sym{***}&3.291&\sym{***}&3.027&\sym{***}&$-$0.036&\sym{***}&$-$0.300&\sym{***}&$-$0.264&\sym{***}\\
\quad (1995--2011 cohorts) &&(0.012&)&(0.005&)&(0.010&)&(0.013&)&(0.015&)&(0.011&)\\
\addlinespace
Predicted log wage (if in RAIS)&&3.231&\sym{***}&3.228&\sym{***}&2.985&\sym{***}&$-$0.003&&$-$0.245&\sym{***}&$-$0.242&\sym{***}\\
\quad (1995--2011 cohorts) &&(0.013&)&(0.006&)&(0.010&)&(0.014&)&(0.017&)&(0.012&)\\
\addlinespace

\midrule
\(N\) (predicted wage)&&27,707&&45,674&&6,111&&73,381&&33,818&&51,785&\\

\(N\) (predicted wage if in RAIS)&&18,093&&30,796&&4,277&&48,889&&22,370&&35,073&\\
\addlinespace

\end{tabular}
\par\end{centering}
\singlespacing\justify
\emph{Notes}: This table displays complier characteristics for our 2SLS RD regressions. Columns (A)--(C) display complier characteristics for pre-AA applicants, post-AA general track applicants, and AA applicants, respectively. These complier characteristics are the $\beta$ coefficients from our 2SLS RD specification \eqref{rd_iv} in which the dependent variables are the interaction between an indicator for enrolling in UERJ and the applicant characteristic listed in the row header. In Panel B, the dependent variables are the interaction between an indicator for enrolling in UERJ and the applicant's predicted log wage based on their demographic characteristics (as in Panel B of Table \ref{tab:bal}). Columns (D)--(F) report the differences between the complier characteristics for: (D) post-AA general and pre-AA applicants; (E) AA and pre-AA applicants; and (F) AA and post-AA general applicants.

Parentheses contain standard errors clustered at the individual level. Sample sizes refer to the dependent variables indicated in the parentheses after $N$. 

{*} $p<0.10$, {*}{*} $p<0.05$, {*}{*}{*} $p<0.01$.

}
\end{table}

\clearpage

\begin{landscape}
\begin{table}[h]{\scriptsize\begin{centering}
\protect\caption{Effects of UERJ enrollment on job outcomes measured 10--13 years later}\label{tab:rd_later}
\newcommand\w{1.4}
\begin{tabular}{l@{}lR{\w cm}@{}L{0.43cm}R{\w cm}@{}L{0.43cm}R{\w cm}@{}L{0.43cm}R{\w cm}@{}L{0.43cm}R{\w cm}@{}L{0.43cm}R{\w cm}@{}L{0.43cm}}
&&(A)&&(B)&&(C)&&(D)&&(E)&&(F)\\\addlinespace
&&\multicolumn{5}{c}{RD coefficient}&&\multicolumn{5}{c}{\emph{Change} in RD coefficients}\\
&&\multicolumn{5}{c}{measured 10--13 years later}&&\multicolumn{5}{c}{from 6--9 to 10--13 years later} \\\cmidrule{3-7}\cmidrule{9-13}
&&1995--01&&2004--11&&AA&&1995--01&&2004--11&&AA\\
Dependent variable && general&&general&&tracks&& general&&general&&tracks\\\midrule\addlinespace

\addlinespace
\multicolumn{13}{l}{\hspace{-1em} \textbf{Panel A.\ Formal employment and earnings (2SLS)}} \\\midrule
Formal employment &&0.032&&$-$0.026&&0.037&&$-$0.032&&0.005&&0.038&\\
&&(0.027&)&(0.031&)&(0.039&)&(0.021&)&(0.028&)&(0.036&)\\
\addlinespace
Log hourly wage&&0.005&&0.005&&0.024&&0.008&&0.084&&$-$0.108&\sym{**}\\
&&(0.054&)&(0.058&)&(0.063&)&(0.046&)&(0.052&)&(0.055&)\\
\addlinespace
Monthly earnings (2019 USD)&&$-$84.946&&$-$99.418&&56.577&&$-$85.241&&54.056&&$-$53.652&\\
&&(94.587&)&(109.084&)&(75.202&)&(71.026&)&(87.121&)&(60.324&)\\
\midrule
\(N\) (log hourly wage)&&24,273&&26,407&&3,746&&48,837&&59,379&&9,846&\\
\addlinespace

\addlinespace\addlinespace
\multicolumn{13}{l}{\hspace{-1em} \textbf{Panel B.\ Employment at firms that hired UERJ alumni (2SLS)}} \\\midrule
Employed at firm with any UERJ alum&&0.059&\sym{*}&0.028&&0.080&\sym{*}&$-$0.059&\sym{*}&$-$0.042&&$-$0.057&\\
&&(0.033&)&(0.037&)&(0.044&)&(0.032&)&(0.036&)&(0.042&)\\
\addlinespace
\# UERJ alumni / 1000 workers at firm&&3.648&\sym{***}&5.954&\sym{**}&$-$0.092&&$-$1.004&&$-$3.784&&$-$8.615&\sym{**}\\
&&(1.078&)&(3.025&)&(2.767&)&(1.497&)&(2.742&)&(3.514&)\\
\midrule
\(N\) (employed at firm with any alum)&&25,535&&28,217&&4,179&&50,859&&62,344&&9,239&\\

\end{tabular}
\par\end{centering}
\singlespacing\justify
\emph{Notes}: This table presents RD estimates for employment and earnings measured 10--13 years after application. Columns (A)--(C) show 2SLS RD coefficients, $\beta$, from equation \eqref{rd_iv} for each applicant group. Columns (D)--(F) show the difference in the 2SLS RD coefficients between the periods of 6--9 and 10--13 years after application. Parentheses contain standard errors clustered at the individual level. Sample sizes refer to the dependent variables indicated in the parentheses after $N$. 

{*} $p<0.10$, {*}{*} $p<0.05$, {*}{*}{*} $p<0.01$.

}
\end{table}
\end{landscape}

\clearpage
\begin{table}[htpb]{\scriptsize
\begin{center}
\caption{Robustness of RD estimates --- General track (1995--2001)} \label{tab:rd_bw_1}
\newcommand\w{1.4}
\newcommand\s{1}
\begin{tabular}{l@{}lR{\w cm}@{}L{0.43cm}R{\w cm}@{}L{0.43cm}R{\w cm}@{}L{0.43cm}R{\w cm}@{}L{0.43cm}R{\w cm}@{}L{0.43cm}R{\w cm}@{}L{0.43cm}}
&& (A) && (B) && (C) && (D) && (E)  \\\addlinespace
&&  \multicolumn{6}{c}{RD coefficients by bandwidth, $h^Y$}  \\\cmidrule{3-8}    
&& $1.0 \times$  && $0.5 \times$  && $1.5 \times$    && Donut  && Adding \\
&& CCT   &&  CCT  && CCT && RD && controls \\
\midrule\addlinespace

\multicolumn{8}{l}{\hspace{-1em} \textbf{Panel A.\ First stage}} \\\midrule
Enrolled in UERJ program&&0.689&\sym{***}&0.708&\sym{***}&0.713&\sym{***}&0.626&\sym{***}&0.689&\sym{***}\\
&&(0.014&)&(0.019&)&(0.012&)&(0.030&)&(0.014&)\\
\midrule
\(N\)&&6,121&&3,262&&8,606&&3,308&&6,121&\\
\addlinespace\addlinespace

\multicolumn{8}{l}{\hspace{-1em} \textbf{Panel B.\ Returns to UERJ enrollment 6--9 years later (2SLS)}} \\\midrule
Graduated from UERJ program&&0.640&\sym{***}&0.642&\sym{***}&0.643&\sym{***}&0.660&\sym{***}&0.638&\sym{***}\\
&&(0.018&)&(0.023&)&(0.015&)&(0.034&)&(0.018&)\\
\addlinespace
Formal employment &&$-$0.002&&$-$0.008&&$-$0.013&&0.022&&$-$0.010&\\
&&(0.026&)&(0.037&)&(0.021&)&(0.038&)&(0.017&)\\
\addlinespace
Log hourly wage&&0.132&\sym{***}&0.123&\sym{**}&0.130&\sym{***}&0.161&\sym{***}&0.125&\sym{***}\\
&&(0.044&)&(0.062&)&(0.036&)&(0.058&)&(0.043&)\\
\addlinespace
Monthly earnings (2019 USD)&&110.230&\sym{**}&114.289&\sym{*}&112.040&\sym{***}&202.213&\sym{***}&108.147&\sym{**}\\
&&(49.523&)&(66.820&)&(40.360&)&(76.211&)&(48.984&)\\
\midrule
\(N\) (employment regression)&&8,147&&4,459&&11,011&&6,276&&8,147&\\

\(N\) (wage regression)&&6,100&&3,311&&8,203&&5,405&&6,100&\\
\addlinespace\addlinespace

\multicolumn{8}{l}{\hspace{-1em} \textbf{Panel C.\ Returns to UERJ enrollment 10--13 years later (2SLS)}} \\\midrule
Graduated from UERJ program&&0.661&\sym{***}&0.654&\sym{***}&0.670&\sym{***}&0.653&\sym{***}&0.660&\sym{***}\\
&&(0.021&)&(0.028&)&(0.018&)&(0.041&)&(0.021&)\\
\addlinespace
Formal employment &&0.037&&0.060&&0.011&&0.057&&0.025&\\
&&(0.039&)&(0.055&)&(0.031&)&(0.054&)&(0.031&)\\
\addlinespace
Log hourly wage&&0.025&&0.101&&0.052&&0.019&&0.014&\\
&&(0.063&)&(0.087&)&(0.050&)&(0.080&)&(0.062&)\\
\addlinespace
Monthly earnings (2019 USD)&&56.577&&69.183&&66.656&&120.434&&45.488&\\
&&(75.149&)&(102.962&)&(59.072&)&(104.238&)&(74.087&)\\
\midrule
\(N\) (employment regression)&&4,320&&2,280&&6,109&&3,958&&4,320&\\

\(N\) (wage regression)&&3,748&&2,024&&5,240&&3,693&&3,748&\\

\end{tabular}
\par\end{center}
\singlespacing\justify
\emph{Notes}: This table display RD coefficients using different specifications of our estimating equation. The coefficients are estimated on the sample of general track applicants in the pre-AA cohorts (1995--2001).

Columns (A)--(C) display the estimated RD coefficients using different sample bandwidths. Column (A) reproduces our baseline specification, which uses the \citet{calonico2014} (CCT) optimal bandwidth for each outcome. In Column (B), we use a bandwidth half the size of the optimal CCT bandwidth. In Column (C), we use a bandwidth twice as large as the CCT bandwidth. In Column (D), we exclude applicants with an admission score within $0.05$ SD of the cutoff. In Column (E), we include controls for age, gender, race, mother's educational attainment, family income, score on the writing component of the field exam, and qualifying exam score.

Panel A displays the first-stage effect, which the estimated $\theta$ from equation \eqref{rd_rf}. Panels B--C display 2SLS RD coefficients, $\beta$, from equation \eqref{rd_iv}. Parentheses contain standard errors clustered at the individual level.

{*} $p<0.10$, {*}{*} $p<0.05$, {*}{*}{*} $p<0.01$.

}
\end{table}

\clearpage
\begin{table}[htpb]{\scriptsize
\begin{center}
\caption{Robustness of RD estimates --- General track (2004--2011)} \label{tab:rd_bw_2}
\newcommand\w{1.4}
\newcommand\s{2}
\begin{tabular}{l@{}lR{\w cm}@{}L{0.43cm}R{\w cm}@{}L{0.43cm}R{\w cm}@{}L{0.43cm}R{\w cm}@{}L{0.43cm}R{\w cm}@{}L{0.43cm}R{\w cm}@{}L{0.43cm}}
&& (A) && (B) && (C) && (D) && (E)  \\\addlinespace
&&  \multicolumn{6}{c}{RD coefficients by bandwidth, $h^Y$}  \\\cmidrule{3-8}    
&& $1.0 \times$  && $0.5 \times$  && $1.5 \times$    && Donut  && Adding \\
&& CCT   &&  CCT  && CCT && RD && controls \\\midrule\addlinespace
\multicolumn{8}{l}{\hspace{-1em} \textbf{Panel A.\ First stage}} \\\midrule
\midrule
\addlinespace\addlinespace

\multicolumn{8}{l}{\hspace{-1em} \textbf{Panel B.\ Returns to UERJ enrollment 6--9 years later (2SLS)}} \\\midrule
\addlinespace
\addlinespace
\addlinespace
\midrule

\addlinespace\addlinespace

\multicolumn{8}{l}{\hspace{-1em} \textbf{Panel C.\ Returns to UERJ enrollment 10--13 years later (2SLS)}} \\\midrule
\addlinespace
\addlinespace
\addlinespace
\midrule

\end{tabular}
\par\end{center}
\singlespacing\justify
\emph{Notes}: This table display RD coefficients using different specifications of our estimating equation. The table is structured similarly to Table \ref{tab:rd_bw_1}, but the coefficients are estimated on the sample of general track applicants in the post-AA cohorts (2004--2011). See notes to Table \ref{tab:rd_bw_1} for details.

{*} $p<0.10$, {*}{*} $p<0.05$, {*}{*}{*} $p<0.01$.

}
\end{table}

\clearpage
\begin{table}[htpb]{\scriptsize
\begin{center}
\caption{Robustness of RD estimates --- Affirmative action tracks} \label{tab:rd_bw_3}
\newcommand\w{1.4}
\newcommand\s{3}
\begin{tabular}{l@{}lR{\w cm}@{}L{0.43cm}R{\w cm}@{}L{0.43cm}R{\w cm}@{}L{0.43cm}R{\w cm}@{}L{0.43cm}R{\w cm}@{}L{0.43cm}R{\w cm}@{}L{0.43cm}}
&& (A) && (B) && (C) && (D) && (E)  \\\addlinespace
&&  \multicolumn{6}{c}{RD coefficients by bandwidth, $h^Y$}  \\\cmidrule{3-8}    
&& $1.0 \times$  && $0.5 \times$  && $1.5 \times$    && Donut  && Adding \\
&& CCT   &&  CCT  && CCT && RD && controls \\\midrule\addlinespace

\multicolumn{8}{l}{\hspace{-1em} \textbf{Panel A.\ First stage}} \\\midrule
\midrule
\addlinespace\addlinespace

\multicolumn{8}{l}{\hspace{-1em} \textbf{Panel B.\ Returns to UERJ enrollment 6--9 years later (2SLS)}} \\\midrule
\addlinespace
\addlinespace
\addlinespace
\midrule

\addlinespace\addlinespace

\multicolumn{8}{l}{\hspace{-1em} \textbf{Panel C.\ Returns to UERJ enrollment 10--13 years later (2SLS)}} \\\midrule
\addlinespace
\addlinespace
\addlinespace
\midrule

\end{tabular}
\par\end{center}
\singlespacing\justify
\emph{Notes}: This table display RD coefficients using different specifications of our estimating equation. The table is structured similarly to Table \ref{tab:rd_bw_1}, but the coefficients are estimated on the sample of Black/public school applicants in the post-AA cohorts (2004--2011). See notes to Table \ref{tab:rd_bw_1} for details.

{*} $p<0.10$, {*}{*} $p<0.05$, {*}{*}{*} $p<0.01$.

}
\end{table}

\clearpage

\begin{table}[htp]{\scriptsize\begin{centering}
\protect\caption{Heterogeneity in RD estimates by field of study --- Affirmative action tracks}\label{tab:rd_het_field}
\newcommand\w{1.5}
\begin{tabular}{l@{}lR{\w cm}@{}L{0.43cm}R{\w cm}@{}L{0.43cm}R{\w cm}@{}L{0.43cm}R{\w cm}@{}L{0.43cm}R{\w cm}@{}L{0.43cm}}
&&(A)&&(B)&&(C)&&(D)&&(E)\\\addlinespace
&& \multicolumn{10}{c}{Field of study} \\\cmidrule{3-12}
&&&&            &&        Human              && Natural     && Social  \\
Dependent variable &&Business&&Health && -ities    &&sciences &&sciences  \\\midrule\addlinespace

\addlinespace
\multicolumn{9}{l}{\hspace{-1em} \textbf{Panel A.\ Returns to UERJ enrollment 6--9 years after applying (2SLS)}} \\\midrule
Graduated from UERJ program&&0.697&\sym{***}&0.801&\sym{***}&0.528&\sym{***}&0.411&\sym{***}&0.722&\sym{***}\\
&&(0.059&)&(0.029&)&(0.045&)&(0.038&)&(0.033&)\\
\addlinespace
Formal employment&&0.071&&$-$0.055&&0.075&&0.037&&$-$0.053&\\
&&(0.077&)&(0.053&)&(0.058&)&(0.055&)&(0.051&)\\
\addlinespace
Log hourly wage&&0.224&\sym{*}&0.204&\sym{**}&0.079&&0.090&&0.095&\\
&&(0.131&)&(0.087&)&(0.103&)&(0.094&)&(0.091&)\\
\addlinespace
Monthly earnings (2019 USD)&&270.006&\sym{**}&217.408&\sym{*}&97.492&&50.829&&3.713&\\
&&(128.029&)&(116.354&)&(87.611&)&(113.661&)&(96.297&)\\
\midrule
\(N\) (formal employment)&&784&&1,895&&1,295&&1,681&&2,492&\\

\(N\) (log hourly wage)&&698&&1,382&&1,009&&1,269&&1,719&\\
\addlinespace

\addlinespace\addlinespace
\multicolumn{9}{l}{\hspace{-1em} \textbf{Panel B.\ Returns to UERJ enrollment 10--13 years after applying (2SLS)}} \\\midrule
Graduated from UERJ program&&0.724&\sym{***}&0.803&\sym{***}&0.556&\sym{***}&0.482&\sym{***}&0.715&\sym{***}\\
&&(0.072&)&(0.036&)&(0.053&)&(0.048&)&(0.041&)\\
\addlinespace
Formal employment&&0.080&&0.018&&$-$0.024&&0.105&&0.026&\\
&&(0.146&)&(0.075&)&(0.093&)&(0.083&)&(0.077&)\\
\addlinespace
Log hourly wage&&0.240&&0.172&&0.025&&$-$0.055&&$-$0.141&\\
&&(0.220&)&(0.127&)&(0.145&)&(0.133&)&(0.122&)\\
\addlinespace
Monthly earnings (2019 USD)&&327.743&&243.571&&17.347&&$-$153.081&&$-$45.814&\\
&&(239.378&)&(159.063&)&(157.375&)&(163.135&)&(151.932&)\\
\midrule
\(N\) (formal employment)&&423&&1,022&&709&&830&&1,336&\\

\(N\) (log hourly wage)&&403&&903&&624&&727&&1,089&\\

\end{tabular}
\par\end{centering}
\singlespacing\justify
\emph{Notes}: This table displays RD coefficients estimated on the sample of Black/public school applicants. Each column shows the result for applicants to different fields of study. Column (A) shows the results for applicants to business programs; column (B) for health programs; column (C) for humanities programs; column (D) for natural sciences programs, and column (E) for social sciences programs. See Appendix \ref{app:var_def} for the programs included in each field of study and Appendix Tables \ref{tab:samp-gen}--\ref{tab:samp-blk} for the number of applicants by program/cohort.

Panels A--B display 2SLS RD coefficients, $\beta$, from equation \eqref{rd_iv}. The dependent variables are program completion, formal employment, and earnings, each measured 6--9 years after applying (Panel A) and 10--13 years after applying (Panel B). See Appendix \ref{app:var_def} for variable definitions.

Parentheses contain standard errors clustered at the individual level. Sample sizes refer to the dependent variables indicated in the parentheses after $N$. 

{*} $p<0.10$, {*}{*} $p<0.05$, {*}{*}{*} $p<0.01$.

}
\end{table}

\clearpage

\clearpage
\begin{landscape}
\begin{table}[!htb]{\scriptsize\begin{centering}
\protect\caption{Summary statistics for Rio de Janeiro universities in 2010}\label{tab:uerj_competitors}
\newcommand\w{1.3}
\begin{tabular}{lllrrrrrrrr}
(A)&(B)&(C)&(D)&(E)&(F)&(G)&(H)&(I)\\\addlinespace
&&&&&& \multicolumn{3}{c}{in 2010 US dollars}\\\cmidrule{7-9}
&&&\emph{Folha}&&& Annual& Annual & Expenses\\
&&& national & Undergrad  & Graduate & revenue & expenses & per\\
University name & Abbr.& Ownership & ranking & enrollment & enrollment & (millions)& (millions) & student \\\midrule\addlinespace\addlinespace
Univ. Federal do Rio de Janeiro &UFRJ &Federal &3 &50,342 &12,453 &1,254.4 &1,254.4 &19,976 \\
\textbf{Univ. do Estado do Rio de Janeiro} &\textbf{UERJ} &\textbf{State} &\textbf{11} &\textbf{30,144} &\textbf{5,767} &\textbf{463.9} &\textbf{465.5} &\textbf{12,962} \\
Pont. Univ. Cat\'olica do Rio de Janeiro &PUC-Rio &Private &13 &17,061 &3,352 &291.3 &265.8 &13,022 \\
Univ. Federal Fluminense &UFF &Federal &15 &48,809 &5,720 &767.9 &1,268.3 &23,259 \\
Univ. Federal Rural do Rio de Janeiro &UFRRJ &Federal &48 &14,826 &2,116 &308.8 &250.0 &14,759 \\
Univ. Federal do Estado do Rio de Janeiro &UNIRIO &Federal &67 &14,418 &934 &140.5 &109.2 &7,112 \\
Univ. Est\'acio de S\'a &UNESA &Private &89 &181,832 &492 &341.0 &237.6 &1,303 \\
Univ. Gama Filho &UGF &Private &110 &21,020 &243 &94.8 &97.1 &4,568 \\
Univ. Veiga de Almeida &UVA &Private &147 &21,983 &184 &62.4 &55.7 &2,512 \\
Univ. Salgado de Oliveira &Universo &Private &155 &48,130 &173 &246.5 &114.8 &2,377 \\
Univ. Castelo Branco &UCB &Private &160 &71,524 &0 &33.8 &33.1 &463 \\
Univ. C\^andido Mendes &UCAM &Private &168 &21,454 &458 &69.8 &83.9 &3,827 \\

\end{tabular}
\par\end{centering}
\singlespacing \justify
\emph{Notes}: This table displays summary statistics for universities in Rio de Janeiro. The sample includes private universities in the municipality of Rio, federal universities in the state of Rio, and UERJ. These are the universities we use to define enrollment outcomes in Table \ref{tab:rd-census-enrol}.

Columns (A)--(C) show the university's name, abbreviation, and ownership type. Column (D) reports the university's rank in the 2012 national ranking by the newspaper \emph{Folha}. Column (E) shows the number of undergraduate students enrolled in each institution in 2010, which we compute from the individual-level dataset of Brazil's higher education census (\emph{Censo da Educa\c{c}\~ao Superior}). Column (F) shows the number of graduate students at each institution in 2010, which we compute from the CAPES census of graduate programs (\emph{Discentes dos Programas de P\'os-Gradua\c{c}\~ao stricto sensu no Brasil}). Columns (G)--(H) report annual revenue and expenses in 2010 (converted to US dollars) from the school-level dataset of the \emph{Censo da Educa\c{c}\~ao Superior}. Column (I) shows annual expenses (column H) divided by total enrollment (columns E $+$ F). 

}
\end{table}
\end{landscape}

\clearpage
\begin{landscape}
\begin{table}[htp]{\scriptsize\begin{centering}
\caption{Top employers of UERJ alumni}\label{tab:firms_top10}
\begin{tabular}{llrrrrrrrrrr}
&                        &(A)              &(B)                  &(C)    &(D)       &(E)     & (F)      & (G) \\\addlinespace
&                         & No.\ UERJ & No. \ alumni   & Firm  &          &       & Prop. of  & Firm mean  \\
&                         & graduates &  hired per        & size   & Located & Public & employees & hourly wage \\
\# & Firm  & hired         & 1000 workers  &(mean)& in Rio   & firm   &  w/ college   & (2019 USD)\\\midrule

\multicolumn{9}{l}{\hspace{-1em} \textbf{Panel A.\ Top 10 firms by total number of UERJ alumni employees}} \\
1& City Hall of Rio de Janeiro &1,161 &13.30 &87,274 &Yes &Yes &0.461 &6.891 \\
2& State Secretary of Education &1,093 &11.97 &91,309 &Yes &Yes &0.398 &3.959 \\
3& State University of Rio de Janeiro (UERJ) &409 &56.29 &7,266 &Yes &Yes &0.690 &13.062 \\
4& Brazilian Petroleum (Petrobras - HQ) &384 &62.91 &6,104 &Yes &No &0.780 &27.690 \\
5& State Secretary of Health &377 &15.35 &24,563 &Yes &Yes &0.330 &3.132 \\
6& State Court of Law &321 &21.27 &15,093 &Yes &Yes &0.718 &16.466 \\
7& Center for Payment of the Army &307 &1.91 &161,115 &No &Yes &0.172 &6.048 \\
8& Federal University of Rio de Janeiro (UFRJ) &238 &22.97 &10,362 &Yes &Yes &0.715 &13.051 \\
9& State Public Ministry &227 &71.22 &3,187 &Yes &Yes &0.740 &25.150 \\
10& City Hall of Duque de Caxias &221 &17.83 &12,395 &Yes &Yes &0.874 &6.914 \\
--& All other firms &-- &-- &510 &0.777 &0.084 &0.384 &6.971 \\
 \addlinespace

\multicolumn{9}{l}{\hspace{-1em} \textbf{Panel B.\ Top 10 firms by number of UERJ alumni hired per 1000 workers}} \\
1& National Bank of Econ. \& Social Dev. &217 &109.23 &1,987 &Yes &No &0.875 &38.010 \\
2& Accenture &184 &107.13 &1,718 &Yes &No &0.810 &13.268 \\
3& Petrobras - EDIHB &176 &102.91 &1,710 &Yes &No &0.841 &24.554 \\
4& General Public Defender of the State &142 &80.84 &1,757 &Yes &Yes &0.536 &24.971 \\
5& Petrobras - Research Center &137 &72.20 &1,898 &Yes &No &0.693 &22.740 \\
6& State Public Ministry &227 &71.22 &3,187 &Yes &Yes &0.740 &25.150 \\
7& Petrobras - Vibra Energy &86 &68.34 &1,258 &Yes &No &0.781 &20.840 \\
8& TIM Cellular &112 &67.67 &1,655 &Yes &No &0.813 &13.542 \\
9& Pedro II Federal Public School &139 &63.35 &2,194 &Yes &Yes &0.828 &10.560 \\
10& Petrobras - EDISE &384 &62.91 &6,104 &Yes &No &0.780 &27.690 \\
--& All other firms &-- &-- &537 &0.777 &0.085 &0.384 &6.964 \\
 \addlinespace

\end{tabular}
\par\end{centering}
\singlespacing\justify 
\emph{Notes}: This table displays summary statistics for top employers of UERJ alumni from the programs in our RD sample (Panel A of Table \ref{tab:summ_stats}). Panel A lists the top ten firms ranked according to column (A), which is the number of UERJ graduates hired across all cohorts in our data. Panel B lists the top ten firms ranked according to column (B), which is the number of UERJ graduates (column A) divided by the firm size (column C) and multiplied by 1000. Column (C) shows the average firm size (number of employees). Column (D) indicates whether the firm is located in the state of Rio. Column (E) indicates whether the firm is public. Column (F) shows the proportion of the firm's employees with a college degree (from any school). Column (G) shows the firm mean hourly wage, measured in 2019 USD. The last row of each Panel shows the average of all other firms that hired at least one UERJ graduate in our sample.

}
\end{table}
\end{landscape}

\clearpage

\begin{table}[htp!]{\scriptsize\begin{centering}\vspace{-1em}
\protect\caption{RD estimates for number of UERJ alumni in an applicant's firm and labor market}\label{tab:rd_labor_market}
\newcommand\w{1.15}
\begin{tabular}{l@{}lR{\w cm}@{}L{0cm}R{\w cm}@{}L{0.43cm}R{\w cm}@{}L{0cm}R{\w cm}@{}L{0.43cm}R{\w cm}@{}L{0cm}R{\w cm}@{}L{0.43cm}}
&&(A)&&(B)&&(C)&&(D)&&(E)&&(F)\\\addlinespace
&&\multicolumn{3}{c}{1995--2001} &&\multicolumn{3}{c}{2004--2011}&&\multicolumn{3}{c}{2004--2011}\\
&&\multicolumn{3}{c}{general track}&&\multicolumn{3}{c}{ general track}&&\multicolumn{3}{c}{AA tracks} \\\cmidrule{3-5}\cmidrule{7-9}\cmidrule{11-13}
&& Mean&&RD  && Mean&&RD && Mean&&RD  \\
Dependent variable && below&&coef&& below&&coef&&below &&coef\\\midrule\addlinespace

\addlinespace
\multicolumn{13}{l}{\hspace{-1em} \textbf{Panel A.\ Employment 6--9 years after application (2SLS)}} \\\midrule
\# UERJ alumni per 1000&&6.561&&4.652&\sym{***}&7.120&&9.738&\sym{***}&3.334&&8.523&\sym{**}\\
\quad workers at firm &&&&(1.624&)&&&(2.583&)&&&(3.580&)\\
\addlinespace
\# UERJ alumni per 1000 workers &&5.021&&1.646&\sym{**}&5.333&&0.718&&3.817&&4.639&\sym{***}\\
\quad in labor market&&&&(0.774&)&&&(1.000&)&&&(1.353&)\\
\addlinespace
\# UERJ alumni per 1000 workers &&1.540&&3.505&\sym{**}&1.786&&7.279&\sym{**}&$-$0.483&&5.186&\sym{*}\\
\quad at firm (net of labor market)&&&&(1.390&)&&&(2.989&)&&&(2.867&)\\
\midrule
\(N\) (\# UERJ alumni at firm)&&2,029&&25,324&&2,698&&34,127&&396&&5,060&\\
\addlinespace

\addlinespace\addlinespace
\multicolumn{13}{l}{\hspace{-1em} \textbf{Panel B.\ Employment 10--13 years after application (2SLS)}} \\\midrule
\# UERJ alumni per 1000 &&5.873&&3.648&\sym{***}&5.620&&5.954&\sym{**}&3.224&&$-$0.092&\\
\quad workers at firm &&&&(1.078&)&&&(3.025&)&&&(2.767&)\\
\addlinespace
\# UERJ alumni per 1000 workers&&4.421&&0.433&&4.515&&0.316&&2.960&&2.382&\sym{***}\\
\quad in labor market &&&&(0.693&)&&&(0.940&)&&&(0.894&)\\
\addlinespace
\# UERJ alumni per 1000 workers&&1.452&&3.179&\sym{***}&1.105&&4.552&&0.264&&$-$0.821&\\
\quad at firm (net of labor market)&&&&(1.159&)&&&(2.819&)&&&(1.601&)\\
\midrule
\(N\) (\# UERJ alumni at firm)&&2,242&&25,535&&2,039&&28,217&&277&&4,179&\\
\addlinespace

\end{tabular}
\par\end{centering}
\singlespacing\justify
\emph{Notes}: This table presents RD estimates for the effects of UERJ enrollment on the number of UERJ alumni in an applicant's firm and labor market. Columns (A), (C), and (E) show means of each dependent variable for applicants in each group who scored $(-0.1,0)$ SDs below the cutoff. Columns (B), (D), and (F) show 2SLS RD coefficients, $\beta$, from equation \eqref{rd_iv}, which measure the effects of UERJ enrollment on employment outcomes. Panel A measures outcomes 6--9 years after application, and Panel B measures outcomes 10--13 years after application. In the first row of each panel, the dependent variable is the number of alumni per 1000 workers at the applicant's firms (averaged over the time period); this replicates results from Table \ref{tab:rd_job}. In the second row, the dependent variable is the number of alumni per 1000 workers in the applicant's labor markets (averaged over the time period). We define labor markets as a municipality $\times$ 5-digit industry code, and we compute the number of workers in each firm and labor market using its mean size over all years of our data. In the last row of each panel, the dependent variable is the \emph{difference} between the dependent variables in the first and second rows. Parentheses contain standard errors clustered at the individual level. Sample sizes refer to the dependent variables indicated in the parentheses after $N$. 

{*} $p<0.10$, {*}{*} $p<0.05$, {*}{*}{*} $p<0.01$.

}
\end{table}

\clearpage

\begin{table}[htp!]{\scriptsize\begin{centering}
\protect\caption{Effects of UERJ enrollment on firm, occupation, industry, and municipality mean wages}\label{tab:rd_indices}
\newcommand\w{1.15}
\begin{tabular}{l@{}lR{\w cm}@{}L{0cm}R{\w cm}@{}L{0.43cm}R{\w cm}@{}L{0cm}R{\w cm}@{}L{0.43cm}R{\w cm}@{}L{0cm}R{\w cm}@{}L{0.43cm}}
&&(A)&&(B)&&(C)&&(D)&&(E)&&(F)\\\addlinespace
&&\multicolumn{3}{c}{1995--2001} &&\multicolumn{3}{c}{2004--2011}&&\multicolumn{3}{c}{2004--2011}\\
&&\multicolumn{3}{c}{general track}&&\multicolumn{3}{c}{ general track}&&\multicolumn{3}{c}{AA tracks} \\\cmidrule{3-5}\cmidrule{7-9}\cmidrule{11-13}
&& Mean&&RD  && Mean&&RD && Mean&&RD  \\
Dependent variable &&below &&coef&&below &&coef&&below &&coef\\\midrule\addlinespace

\addlinespace\addlinespace
\multicolumn{13}{l}{\hspace{-1em} \textbf{Panel A.\ Outcomes 6--9 years after application (2SLS)}} \\\midrule
Firm mean wage (log)&&3.303&&0.018&&3.475&&$-$0.095&\sym{*}&3.073&&0.106&\sym{*}\\
&&&&(0.043&)&&&(0.053&)&&&(0.062&)\\
\addlinespace
Occupation mean wage (log)&&3.271&&0.028&&3.388&&$-$0.062&&3.017&&0.053&\\
&&&&(0.034&)&&&(0.039&)&&&(0.041&)\\
\addlinespace
Industry mean wage (log)&&3.219&&0.009&&3.201&&$-$0.024&&3.000&&0.044&\\
&&&&(0.031&)&&&(0.037&)&&&(0.036&)\\
\addlinespace
Municipality mean wage (log)&&3.186&&0.009&&3.175&&$-$0.013&&3.137&&$-$0.005&\\
&&&&(0.017&)&&&(0.018&)&&&(0.020&)\\
\midrule
\(N\) (firm mean wage)&&2,024&&30,345&&2,681&&31,087&&394&&4,306&\\
\addlinespace

\addlinespace\addlinespace
\multicolumn{13}{l}{\hspace{-1em} \textbf{Panel B.\ Outcomes 10--13 years after application (2SLS)}} \\\midrule
Firm mean wage (log)&&3.572&&0.093&\sym{*}&3.581&&$-$0.053&&3.223&&0.049&\\
&&&&(0.053&)&&&(0.062&)&&&(0.071&)\\
\addlinespace
Occupation mean wage (log)&&3.428&&$-$0.009&&3.509&&0.025&&3.157&&$-$0.071&\\
&&&&(0.033&)&&&(0.041&)&&&(0.053&)\\
\addlinespace
Industry mean wage (log)&&3.314&&0.039&&3.271&&0.029&&3.094&&0.018&\\
&&&&(0.034&)&&&(0.040&)&&&(0.044&)\\
\addlinespace
Municipality mean wage (log)&&3.209&&$-$0.025&&3.206&&$-$0.027&&3.131&&0.017&\\
&&&&(0.021&)&&&(0.020&)&&&(0.023&)\\
\midrule
\(N\) (firm mean wage)&&2,236&&24,701&&2,010&&21,071&&275&&3,133&\\
\addlinespace

\end{tabular}
\par\end{centering}
\singlespacing\justify
\emph{Notes}: This table presents RD estimates for the effects of UERJ enrollment on mean wages at the firm, occupation, industry, and municipality levels. Columns (A), (C), and (E) show means of each dependent variable for applicants in each group who scored $(-0.1,0)$ SDs below the cutoff. Columns (B), (D), and (F) show 2SLS RD coefficients, $\beta$, from equation \eqref{rd_iv}, which measure the effects of UERJ enrollment on mean hourly wages associated with four different characteristics of individuals' jobs: 1) firm; 2) occupation (using 4-digit CBO codes); 3) industry (using 4-digit CNAE codes); and 4) municipality. Panel A measures outcomes 6--9 years after application, and Panel B measures outcomes 10--13 years after application. Parentheses contain standard errors clustered at the individual level. Sample sizes refer to the dependent variables indicated in the parentheses after $N$. 

}
\end{table}

\clearpage
\begin{table}[htp]{\scriptsize\begin{centering}
\protect\caption{OLS regressions on alumni firm variables}\label{tab:alumni_ols}
\newcommand\w{1.1}
\begin{tabular}{l@{}lR{\w cm}@{}L{0.13cm}R{\w cm}@{}L{0.13cm}R{\w cm}@{}L{0.13cm}R{\w cm}@{}L{0.13cm}R{\w cm}@{}L{0.13cm}R{\w cm}@{}L{0.13cm}R{\w cm}@{}L{0.13cm}}

&&(A)&&(B)&&(C)&&(D)\\\addlinespace
&&\multicolumn{7}{c}{Dependent variable:}\\
Covariate&&\multicolumn{7}{c}{Log firm mean hourly wage} \\\midrule

Any alumni firm&&0.444&&&&&&&\\
&&(0.004&)&&&&&&\\
\addlinespace
Firm w/ 50+ alumni per 1000 workers&&&&0.398&&&&&\\
&&&&(0.011&)&&&&\\
\addlinespace
Firm w/ 25$-$50 alumni per 1000 workers&&&&0.589&&&&&\\
&&&&(0.010&)&&&&\\
\addlinespace
Firm w/ 10$-$25 alumni per 1000 workers&&&&0.639&&&&&\\
&&&&(0.006&)&&&&\\
\addlinespace
Firm w/ 5$-$10 alumni per 1000 workers&&&&0.488&&&&&\\
&&&&(0.007&)&&&&\\
\addlinespace
Firm w/ 1$-$5 alumni per 1000 workers&&&&0.439&&&&&\\
&&&&(0.005&)&&&&\\
\addlinespace
Firm w/ 0$-$1 alumni per 1000 workers&&&&0.303&&&&&\\
&&&&(0.005&)&&&&\\
\addlinespace
Pre-AA alumni&&&&&&0.481&&&\\
&&&&&&(0.004&)&&\\
\addlinespace
Only post-AA alumni&&&&&&0.274&&&\\
&&&&&&(0.005&)&&\\
\addlinespace
General track alumni from same cohort&&&&&&&&0.533&\\
&&&&&&&&(0.005&)\\
\addlinespace
General track alumni from diff.\ cohort&&&&&&&&0.455&\\
&&&&&&&&(0.004&)\\
\addlinespace
Only AA alumni from same cohort&&&&&&&&0.293&\\
&&&&&&&&(0.008&)\\
\addlinespace
Only AA alumni from diff.\ cohort&&&&&&&&0.070&\\
&&&&&&&&(0.010&)\\
\addlinespace
Admission score&&0.170&&0.164&&0.173&&0.164&\\
&&(0.002&)&(0.002&)&(0.002&)&(0.002&)\\
\midrule
\(N\)&&549,675&&549,675&&549,675&&549,675&\\
\addlinespace

\end{tabular}
\par\end{centering}
\singlespacing\justify
\emph{Notes}: This table shows OLS estimates of the wage premia associated with employment at UERJ alumni firms. The sample includes all UERJ applicants. Regressions are at the applicant $\times$ year level and include observations 6--9 years after UERJ application. The dependent variable is log firm mean hourly wage. We use four types of our alumni firm variables as covariates:
\begin{itemize}
\item Column (A): Any alumni firm;
\item Column (B): Firms categorized by the number of UERJ alumni they hired relative to their mean size;
\item Column (C): Firms that hired any alumni from the pre-AA cohorts (1995--2001) vs.\ firms that hired alumni \emph{only} from the post-AA cohorts (2004--2011);
\item Column (D): Firms defined by the alum's cohort (same or different than the applicant's cohort) and application track (general or AA). We define these outcomes to be non-overlapping, i.e., the variables in the lower rows equal one only if the firm did not hire alumni who meet the criteria for the higher rows.

\end{itemize}
All regressions control for the applicant's standardized admission score and application pool $\times$ calendar year dummies. Parentheses contain standard errors clustered at the individual level. 

{*} $p<0.10$, {*}{*} $p<0.05$, {*}{*}{*} $p<0.01$.

}
\end{table}

\clearpage

\begin{table}[htp!]{\scriptsize\begin{centering}
\protect\caption{DD estimates of the effects of AA exposure on student SES \\ as measured by first and last names}\label{tab:dd_wage_index}
\newcommand\w{1.4}
\begin{tabular}{l@{}lR{2 cm}@{}L{0.43cm}R{\w cm}@{}L{0.43cm}R{\w cm}@{}L{0.43cm}R{\w cm}@{}L{0.43cm}R{\w cm}@{}L{0.43cm}R{\w cm}@{}L{0.43cm}R{\w cm}@{}L{0.43cm}}
&&(A)&&(B)&&(C)&&(D)\\\addlinespace
&&Pre-AA mean&&\multicolumn{5}{c}{DD coefficients}\\\cmidrule{3-3}\cmidrule{5-9}
&& All top && All top && URM top && Non-top\\
Dependent variable &&enrollees&&enrollees&&enrollees&&enrollees\\\midrule\addlinespace

\addlinespace
\multicolumn{10}{l}{\hspace{-1em} \textbf{Panel A.\ SES indices based on first and last names}} \\\midrule
First-name-based log wage&&2.424&&0.004&&$-$0.004&&$-$0.008&\sym{**}\\
&&&&(0.003&)&(0.016&)&(0.004&)\\
\addlinespace 
First-name-based wage&&4.464&&0.037&&0.006&&$-$0.080&\sym{***}\\
&&&&(0.022&)&(0.140&)&(0.029&)\\
\midrule
Last-name-based log wage&&2.400&&$-$0.006&&$-$0.007&&$-$0.026&\sym{***}\\
&&&&(0.007&)&(0.019&)&(0.007&)\\
\addlinespace 
Last-name-based wage&&4.406&&$-$0.071&&$-$0.052&&$-$0.191&\sym{***}\\
&&&&(0.052&)&(0.142&)&(0.063&)\\
\addlinespace 

\addlinespace
\multicolumn{10}{l}{\hspace{-1em} \textbf{Panel B.\ Predicted log wage based on characteristics, scores, and SES name indices}} \\\midrule
Predicted log wage&&3.303&&$-$0.023&&$-$0.012&&$-$0.163&\sym{***}\\
&&&&(0.028&)&(0.048&)&(0.043&)\\
\addlinespace
Predicted log wage (if in RAIS)&&3.255&&$-$0.034&&$-$0.011&&$-$0.156&\sym{***}\\
&&&&(0.028&)&(0.052&)&(0.043&)\\
\addlinespace

\(N\) (enrollees)&&16,466&&35,866&&1,631&&30,854&\\

\end{tabular}
\par\end{centering}
\singlespacing\justify
\emph{Notes}: This table displays DD estimates of the effect of affirmative action exposure on student SES as measured by first and last names. Column (A) shows the mean of each dependent variable for all top enrollees in the 1995--2001 cohorts. Columns (B)--(D) display estimates of $\pi$ from equation \eqref{dd} for all top enrollees, URM top enrollees, and non-top enrollees.

In Panel A, the dependent variables are SES indices defined based on the applicant's first or last name. To define these indices, we take the nationwide RAIS data for the year 2019 and compute the leave-individual-out mean hourly wage associated with each first name and each last name. We exclude first and last names that appear fewer than ten times. The dependent variables in Panel A are these first and last name mean wages in both logs and levels.

In Panel B, the dependent variables are the predicted values from a regression of log hourly wage (6--9 years after application) on each of the variables in Panels B--C of Table \ref{tab:dd_traits} plus the SES name indices from Panel A of this table. These are similar to the dependent variables in Panel D of Table \ref{tab:dd_traits}, except these predicted wages are also based on the SES name indices.

Parentheses contain standard errors clustered at the program level. 

{*} $p<0.10$, {*}{*} $p<0.05$, {*}{*}{*} $p<0.01$.

}
\end{table}

\clearpage

\begin{table}[htp]{\scriptsize\begin{centering}
\protect\caption{DD estimates for graduation, employment, and earnings 10--13 years after application}\label{tab:dd_labor10}
\newcommand\w{1.35}
\begin{tabular}{l@{}lR{2 cm}@{}L{0.43cm}R{\w cm}@{}L{0.43cm}R{\w cm}@{}L{0.43cm}R{\w cm}@{}L{0.43cm}R{\w cm}@{}L{0.43cm}R{\w cm}@{}L{0.43cm}R{\w cm}@{}L{0.43cm}}
&&(A)&&(B)&&(C)&&(D)\\\addlinespace
&&Pre-AA mean&&\multicolumn{5}{c}{DD coefficients}\\\cmidrule{3-3}\cmidrule{5-9}
&& All top && All top && URM top && Non-top\\
Dependent variable &&enrollees&&enrollees&&enrollees&&enrollees\\\midrule\addlinespace

\multicolumn{9}{l}{\hspace{-1em} \textbf{Panel A.\ Graduation and formal employment}} \\\midrule
Graduated from UERJ program&&0.568&&0.011&&$-$0.003&&0.010&\\
&&&&(0.018&)&(0.071&)&(0.028&)\\
\addlinespace
Formal employment&&0.768&&$-$0.013&&$-$0.053&&$-$0.010&\\
&&&&(0.011&)&(0.044&)&(0.013&)\\

\addlinespace\addlinespace
\multicolumn{9}{l}{\hspace{-1em} \textbf{Panel B.\ Earnings}} \\\midrule
Log hourly wage&&3.600&&$-$0.115&\sym{**}&$-$0.129&&$-$0.252&\sym{***}\\
&&&&(0.053&)&(0.138&)&(0.072&)\\
\addlinespace
Monthly earnings (2019 USD)&&2,005.914&&$-$224.443&\sym{**}&$-$83.051&&$-$469.037&\sym{***}\\
&&&&(90.068&)&(189.155&)&(133.630&)\\
\addlinespace
Firm mean hourly wage (log)&&3.565&&$-$0.114&\sym{**}&$-$0.185&&$-$0.191&\sym{***}\\
&&&&(0.044&)&(0.142&)&(0.055&)\\

\addlinespace\addlinespace
\multicolumn{9}{l}{\hspace{-1em} \textbf{Panel C.\ Employment at firms with pre- and post-AA alumni}} \\\midrule
Pre-AA alumni&&0.595&&$-$0.043&&0.011&&$-$0.023&\\
&&&&(0.027&)&(0.084&)&(0.034&)\\
\addlinespace
Only post-AA alumni&&0.089&&0.034&\sym{**}&$-$0.070&&0.034&\\
&&&&(0.016&)&(0.059&)&(0.021&)\\

\addlinespace\addlinespace
\multicolumn{9}{l}{\hspace{-1em} \textbf{Panel D.\ Alumni firm employment by application track and cohort}} \\\midrule
General track alumni from same cohort&&0.463&&$-$0.093&\sym{***}&$-$0.121&&$-$0.046&\\
&&&&(0.026&)&(0.076&)&(0.030&)\\
\addlinespace
General track alumni from diff.\ cohort&&0.248&&0.046&\sym{**}&0.031&&0.023&\\
&&&&(0.018&)&(0.078&)&(0.017&)\\
\addlinespace
Only AA alumni from same cohort&&0.000&&0.031&\sym{***}&0.024&&0.036&\sym{***}\\
&&&&(0.006&)&(0.018&)&(0.009&)\\
\addlinespace
Only AA alumni from diff.\ cohort&&0.012&&0.005&&$-$0.016&&0.003&\\
&&&&(0.004&)&(0.015&)&(0.004&)\\
\addlinespace

\(N\) (enrollees)&&16,466&&31,016&&1,157&&26,484&\\

\(N\) (wage observations)&&12,614&&23,381&&928&&20,091&\\

\end{tabular}
\par\end{centering}
\singlespacing\justify
\emph{Notes}: This table displays DD estimates of the effect of affirmative action exposure on graduation, earnings, and employment at alumni firms measured 10--13 years after application. Column (A) shows the mean of each dependent variable for all top enrollees in the 1995--2001 cohorts. Columns (B)--(D) display estimates of $\pi$ from equation \eqref{dd} for all top enrollees, URM top enrollees, and non-top enrollees. The dependent variables are defined similarly to those in Tables \ref{tab:rd_returns} and \ref{tab:rd_job}. In Panel C, we define alumni firms using graduates from the pre- and post-AA cohorts. In Panel D, we define alumni firms using the alum's cohort (same or different than the applicant's cohort) and application track (general or AA). We define the outcomes in Panels C--D to be non-overlapping, i.e., the variables in the lower rows of each panel equal one only if the firm did not hire alumni who meet the criteria for the higher rows. Parentheses contain standard errors clustered at the program level.

{*} $p<0.10$, {*}{*} $p<0.05$, {*}{*}{*} $p<0.01$.

}
\end{table}

\clearpage

\begin{table}[htp]{\scriptsize\begin{centering}
\protect\caption{DD estimates by field of study --- Top enrollees}\label{tab:dd_field}
\newcommand\w{1.48}
\begin{tabular}{l@{}lR{\w cm}@{}L{0.43cm}R{\w cm}@{}L{0.43cm}R{\w cm}@{}L{0.43cm}R{\w cm}@{}L{0.43cm}R{\w cm}@{}L{0.43cm}R{\w cm}@{}L{0.43cm}R{\w cm}@{}L{0.43cm}R{\w cm}@{}L{0.43cm}R{\w cm}@{}L{0.43cm}R{\w cm}@{}L{0.43cm}}
&& (A)      && (B)        && (C)     && (D)         \\ \addlinespace
&& && Human&& Natural && Social \\
Dependent variable && Business && -ities &&sciences &&sciences    \\\midrule\addlinespace

\multicolumn{9}{l}{\hspace{-1em} \textbf{Panel A.\ Graduation and formal employment}} \\\midrule
Graduated from UERJ program&&$-$0.011&&0.046&&$-$0.020&&0.052&\\
&&(0.043&)&(0.034&)&(0.039&)&(0.025&)\\
\addlinespace
Formal employment&&0.022&&$-$0.021&&$-$0.029&&$-$0.030&\\
&&(0.032&)&(0.029&)&(0.018&)&(0.026&)\\

\addlinespace\addlinespace
\multicolumn{9}{l}{\hspace{-1em} \textbf{Panel B.\ Earnings}} \\\midrule
Log hourly wage&&$-$0.035&&$-$0.212&\sym{**}&$-$0.082&&$-$0.181&\sym{*}\\
&&(0.040&)&(0.094&)&(0.071&)&(0.070&)\\
\addlinespace
Monthly earnings (2019 USD)&&$-$118.732&&$-$211.522&\sym{***}&$-$86.695&&$-$301.923&\\
&&(91.622&)&(60.937&)&(77.045&)&(185.385&)\\
\addlinespace
Firm mean hourly wage (log)&&$-$0.100&\sym{**}&$-$0.137&\sym{**}&$-$0.066&&$-$0.079&\\
&&(0.018&)&(0.055&)&(0.072&)&(0.058&)\\

\addlinespace\addlinespace
\multicolumn{9}{l}{\hspace{-1em} \textbf{Panel C.\ Employment at firms with pre- and post-AA alumni}} \\\midrule
Pre-AA alumni&&$-$0.072&&$-$0.026&&$-$0.062&&$-$0.158&\sym{***}\\
&&(0.034&)&(0.032&)&(0.037&)&(0.023&)\\
\addlinespace
Only post-AA alumni&&0.035&&0.047&&0.059&&0.088&\sym{**}\\
&&(0.054&)&(0.038&)&(0.034&)&(0.026&)\\

\addlinespace\addlinespace
\multicolumn{9}{l}{\hspace{-1em} \textbf{Panel D.\ Alumni firm employment by application track and cohort}} \\\midrule
General track alumni from same cohort&&$-$0.057&\sym{**}&$-$0.095&\sym{**}&$-$0.103&\sym{**}&$-$0.151&\sym{**}\\
&&(0.008&)&(0.036&)&(0.036&)&(0.034&)\\
\addlinespace
General track alumni from diff.\ cohort&&$-$0.049&\sym{**}&0.077&\sym{***}&0.036&&0.071&\sym{**}\\
&&(0.007&)&(0.021&)&(0.024&)&(0.016&)\\
\addlinespace
Only AA alumni from same cohort&&0.043&\sym{*}&0.023&&0.046&\sym{***}&0.044&\sym{**}\\
&&(0.011&)&(0.020&)&(0.008&)&(0.010&)\\
\addlinespace
Only AA alumni from diff.\ cohort&&0.019&&0.011&&0.016&\sym{***}&$-$0.017&\sym{**}\\
&&(0.012&)&(0.011&)&(0.004&)&(0.006&)\\

\midrule
\(N\) (enrollees)&&2,895&&13,410&&11,266&&5,445&\\

\(N\) (wage observations)&&2,343&&10,312&&8,409&&3,515&\\

\end{tabular}
\par\end{centering}
\singlespacing\justify
\emph{Notes}: This table examines heterogeneity by field of study in our DD estimates for top enrollees. The sample and dependent variables are the same as in column (B) of Table \ref{tab:dd_labor}, but we estimate regressions for programs in four field of study groups: (A) business, (B) humanities, (C) natural sciences, and (D) social sciences. See Appendix \ref{app:var_def} for the programs included in each field of study group. We cannot estimate our DD specification for health programs because there is no variation in our binary measure of exposure to affirmative action. All outcomes are measured 6--9 years after UERJ application.

{*} $p<0.10$, {*}{*} $p<0.05$, {*}{*}{*} $p<0.01$.
 
}
\end{table}

\clearpage

\begin{landscape}
\begin{table}[htp]{\scriptsize\begin{centering}\vspace{-4em}
\protect\caption{Robustness of DD estimates --- Top enrollees}\label{tab:dd_rob}
\newcommand\w{1.48}
\begin{tabular}{l@{}lR{\w cm}@{}L{0.43cm}R{\w cm}@{}L{0.43cm}R{\w cm}@{}L{0.43cm}R{\w cm}@{}L{0.43cm}R{\w cm}@{}L{0.43cm}R{\w cm}@{}L{0.43cm}R{\w cm}@{}L{0.43cm}R{\w cm}@{}L{0.43cm}R{\w cm}@{}L{0.43cm}R{\w cm}@{}L{0.43cm}}
&& (A)      && (B)        && (C)     && (D)       && (E)        && (F)   && (G)    \\ \addlinespace
&& Bench-         && Pre-crisis     && Linear    && Demo- &&Selectivity&&No field  && Actual  \\
Dependent variable && mark && years  && trends && graphics && controls && of study        && AA share    \\\midrule\addlinespace

\multicolumn{9}{l}{\hspace{-1em} \textbf{Panel A.\ Graduation and formal employment}} \\\midrule
Graduated from UERJ program&&0.013&&0.004&&0.037&\sym{*}&0.013&&0.011&&0.039&\sym{*}&0.006&\\
&&(0.021&)&(0.019&)&(0.021&)&(0.021&)&(0.020&)&(0.021&)&(0.020&)\\
\addlinespace
Formal employment&&$-$0.027&\sym{*}&$-$0.022&&$-$0.039&\sym{**}&0.005&&$-$0.015&&$-$0.044&\sym{**}&$-$0.023&\\
&&(0.015&)&(0.014&)&(0.019&)&(0.008&)&(0.021&)&(0.020&)&(0.015&)\\

\addlinespace
\multicolumn{9}{l}{\hspace{-1em} \textbf{Panel B.\ Earnings}} \\\midrule
Log hourly wage&&$-$0.132&\sym{***}&$-$0.118&\sym{**}&$-$0.133&\sym{**}&$-$0.110&\sym{**}&$-$0.067&\sym{**}&$-$0.152&\sym{**}&$-$0.125&\sym{**}\\
&&(0.045&)&(0.047&)&(0.061&)&(0.043&)&(0.033&)&(0.061&)&(0.052&)\\
\addlinespace
Firm mean hourly wage (log)&&$-$0.095&\sym{**}&$-$0.094&\sym{***}&$-$0.120&\sym{***}&$-$0.075&\sym{**}&$-$0.059&\sym{**}&$-$0.125&\sym{***}&$-$0.082&\sym{*}\\
&&(0.035&)&(0.032&)&(0.043&)&(0.033&)&(0.028&)&(0.044&)&(0.043&)\\

\addlinespace
\multicolumn{9}{l}{\hspace{-1em} \textbf{Panel C.\ Employment at firms with pre- and post-AA alumni}} \\\midrule
Pre-AA alumni&&$-$0.055&\sym{**}&$-$0.061&\sym{***}&$-$0.072&\sym{***}&$-$0.050&\sym{*}&$-$0.035&\sym{*}&$-$0.064&\sym{**}&$-$0.064&\sym{***}\\
&&(0.023&)&(0.020&)&(0.026&)&(0.025&)&(0.021&)&(0.024&)&(0.021&)\\
\addlinespace
Only post-AA alumni&&0.049&\sym{**}&0.044&\sym{**}&0.080&\sym{***}&0.048&\sym{**}&0.017&&0.073&\sym{***}&0.055&\sym{**}\\
&&(0.023&)&(0.020&)&(0.025&)&(0.023&)&(0.018&)&(0.024&)&(0.024&)\\

\addlinespace
\multicolumn{9}{l}{\hspace{-1em} \textbf{Panel D.\ Alumni firm employment by application track and cohort}} \\\midrule
General track alumni from same cohort&&$-$0.098&\sym{***}&$-$0.094&\sym{***}&$-$0.093&\sym{***}&$-$0.096&\sym{***}&$-$0.090&\sym{***}&$-$0.087&\sym{***}&$-$0.091&\sym{***}\\
&&(0.021&)&(0.016&)&(0.024&)&(0.022&)&(0.022&)&(0.025&)&(0.023&)\\
\addlinespace
General track alumni from diff.\ cohort&&0.042&\sym{**}&0.037&\sym{**}&0.046&\sym{***}&0.045&\sym{**}&0.034&&0.044&\sym{**}&0.029&\\
&&(0.016&)&(0.017&)&(0.016&)&(0.017&)&(0.021&)&(0.017&)&(0.018&)\\
\addlinespace
Only AA alumni from same cohort&&0.036&\sym{***}&0.029&\sym{***}&0.041&\sym{***}&0.036&\sym{***}&0.029&\sym{***}&0.041&\sym{***}&0.035&\sym{***}\\
&&(0.009&)&(0.009&)&(0.007&)&(0.008&)&(0.008&)&(0.007&)&(0.008&)\\
\addlinespace
Only AA alumni from diff.\ cohort&&0.010&\sym{**}&0.007&&0.007&&0.010&\sym{**}&0.006&\sym{**}&0.008&\sym{*}&0.011&\sym{**}\\
&&(0.005&)&(0.005&)&(0.004&)&(0.005&)&(0.003&)&(0.004&)&(0.005&)\\

\midrule
\(N\) (enrollees)&&35,866&&28,591&&35,866&&35,866&&35,866&&35,866&&35,866&\\

\(N\) (wage observations)&&26,445&&20,889&&26,445&&26,445&&26,445&&26,445&&26,445&\\

\end{tabular}
\par\end{centering}
\singlespacing\justify
\emph{Notes}: Column (A) reproduces our benchmark DD results for top enrollees (column B in Table \ref{tab:dd_labor}). Column (B) includes only outcomes measured in 2003--2014. Column (C) includes program-specific linear trends estimated in the 1995--2001 cohorts. Column (D) includes controls for age, gender, race, qualifying exam score, and writing field exam score. Column (E) includes cohort dummies interacted with dummies for quartiles of program selectivity ($x$-axis of Figure \ref{fig:variation_aa}). Column (F) excludes the field of study group interactions, $f(m)$.  Column (G) defines $\text{ExposureToAA}_m$ as each major's affirmative action share in the 2004--2011 cohorts ($y$-axis of Figure \ref{fig:variation_aa}), scaled to represent a 20 percentage point increase. Parentheses contain standard errors clustered at the program level. 

{*} $p<0.10$, {*}{*} $p<0.05$, {*}{*}{*} $p<0.01$.
 
}
\end{table}
\end{landscape}

\clearpage

\begin{landscape}
\begin{table}[!htb]{\scriptsize\begin{centering}\vspace{-2em}
\protect\caption{Number of students in Enade sample by exam year, field, and university}\label{tab:summ_enade}
\newcommand\w{1.3}
\begin{tabular}{lrrrrrrrrrrrrr}
&\multicolumn{13}{c}{Exam year}\\\cmidrule{2-14} 
Exam field & 2004 & 2005 & 2006 & 2007 & 2008 & 2009 & 2010 & 2011 & 2012 & 2013 & 2014 & 2015 & Total \\\midrule
Dentistry &270 & & &156 & & &415 & & &475 & & &1,316 \\
Medicine &342 & & &228 & & &636 & & &738 & & &1,944 \\
Nursing &385 & & &258 & & &479 & & &426 & & &1,548 \\
Nutrition &158 & & &117 & & &255 & & &317 & & &847 \\
Physical education &251 & & &162 & & &141 &176 & &224 &288 & &1,242 \\
Social work &113 & & &116 & & &138 & & &283 & & &650 \\
Biology & &647 & & &474 & & &1,014 & & &1,263 & &3,398 \\
Chemistry & &213 & & &262 & & &364 & & &416 & &1,255 \\
Computation & &400 & & &316 & & &373 & & &633 & &1,722 \\
Engineering I (Civil) & &342 & & &454 & & &441 & & &1,036 & &2,273 \\
Engineering II (Electrical) & &333 & & &340 & & &522 & & &730 & &1,925 \\
Engineering III (Mechanical) & &185 & & &205 & & &381 & & &465 & &1,236 \\
Engineering IV (Chemical) & &173 & & &120 & & &180 & & &291 & &764 \\
Engineering VI (Production) & &108 & & &164 & & &270 & & &365 & &907 \\
Geography & &506 & & &531 & & &708 & & &1,098 & &2,843 \\
History & &526 & & &509 & & &600 & & &1,163 & &2,798 \\
Language & &861 & & &905 & & &1,247 & & &1,336 & &4,349 \\
Math & &388 & & &451 & & &420 & & &602 & &1,861 \\
Philosophy & &96 & & &108 & & &140 & & &119 & &463 \\
Physics & &192 & & &185 & & &139 & & &325 & &841 \\
Social science & &286 & & &193 & & &316 & & &399 & &1,194 \\
Teaching & &835 & & &1,396 & & &4,584 & & &3,807 & &10,622 \\
Accounting & & &194 & & &614 & & &428 & & &610 &1,846 \\
Business & & &710 & & &1,029 & & &1,018 & & &1,128 &3,885 \\
Design & & &198 & & &288 & & &211 & & &320 &1,017 \\
Economics & & &392 & & &410 & & &305 & & &574 &1,681 \\
Law & & &537 & & &1,226 & & &1,229 & & &1,574 &4,566 \\
Psychology & & &210 & & &266 & & &159 & & &311 &946 \\
Social communication & & &288 & & &316 & & &184 & & &385 &1,173 \\\midrule
UERJ &141 &676 &242 &124 &718 &294 &192 &1,086 &253 &304 &1,643 &512 &6,185 \\
Other federal \& state universities &1,378 &5,415 &2,287 &913 &5,895 &3,855 &1,872 &10,789 &3,281 &2,159 &12,693 &4,390 &54,927 \\\midrule
Full sample &1,519 &6,091 &2,529 &1,037 &6,613 &4,149 &2,064 &11,875 &3,534 &2,463 &14,336 &4,902 &61,112 \\

\end{tabular}
\par\end{centering}
\singlespacing \justify
\emph{Notes}: This table shows the number of students in our Enade sample for Table \ref{tab:dd_enade}. The sample includes 2004--2015 Enade exam takers at UERJ and other federal and state universities that did not implement affirmative action until 2012 or later.  We define each university's year of affirmative action adoption as the first year in which more than 10 percent of new enrollees entered through a reserved quota in the \emph{Censo da Educa\c{c}\~ao Superior} data, or, if it is earlier, the year of affirmative action adoption cited in \citet{junior2014} or \citet{vieira2019}. The federal universities in our sample are: UFAC, UFAM, UFC, UFCG, UFCSPA, UFERSA, UFLA, UFMS, UFMT, UFPEL, UFRR, UFV, UFVJM, UNIFAL-MG, UNIFAP, UNIFEI, UNIR, and UNIRIO. The state universities in our sample are: UECE, UERJ, UERR, UNESP, UNITINS, and URCA. We exclude fields with no UERJ exam takers, and we drop any institution $\times$ major pair that does not have exam takers in every year in which the exam was offered.

}
\end{table}
\end{landscape}

\clearpage
\thispagestyle{empty}

\begin{table}[htp]{\scriptsize\begin{centering}
\protect\caption{DDD estimates for UERJ's Enade exam taker characteristics and scores}\label{tab:dd_prog_enade}
\newcommand\w{1.6}
\begin{tabular}{l@{}lR{2 cm}@{}L{0.43cm}R{\w cm}@{}L{0.43cm}R{\w cm}@{}L{0.43cm}R{\w cm}@{}L{0.43cm}R{\w cm}@{}L{0.43cm}R{\w cm}@{}L{0.43cm}R{\w cm}@{}L{0.43cm}}
&&(A)&&(B)&&(C)&&(D)\\\addlinespace
&&UERJ&&\\
&&pre-AA mean&&\multicolumn{5}{c}{DDD coefficients}\\\cmidrule{3-3}\cmidrule{5-9}
&& &&  && White \\
&& All && All && private HS  && Other \\
Dependent variable &&students&&students&&students&&students\\\midrule
	
\addlinespace
\multicolumn{9}{l}{\hspace{-1em} \textbf{Panel A.\ Characteristics of Enade exam takers}} \\\midrule
\# exam takers&&38.250&&$-$5.024&&$-$8.316&\sym{***}&$-$4.539&\\
&&&&(3.766&)&(1.346&)&(3.958&)\\
\midrule
White&&0.750&&$-$0.180&\sym{***}&&&$-$0.138&\sym{***}\\
&&&&(0.017&)&&&(0.014&)\\
\addlinespace
Private HS student&&0.588&&$-$0.133&\sym{***}&&&$-$0.061&\sym{***}\\
&&&&(0.019&)&&&(0.011&)\\
\addlinespace
Female&&0.511&&0.075&\sym{***}&0.026&&0.122&\sym{***}\\
&&&&(0.009&)&(0.019&)&(0.016&)\\
\addlinespace
Age&&26.028&&0.397&&1.018&\sym{**}&0.101&\\
&&&&(0.280&)&(0.389&)&(0.374&)\\
\addlinespace
Mother has a HS degree&&0.726&&$-$0.069&\sym{***}&$-$0.008&&$-$0.051&\sym{**}\\
&&&&(0.016&)&(0.030&)&(0.020&)\\
\addlinespace
HH income / minimum wage&&8.619&&$-$1.706&\sym{***}&$-$0.082&&$-$2.102&\sym{***}\\
&&&&(0.152&)&(0.205&)&(0.121&)\\
\midrule
Predicted general score&&0.567&&$-$0.012&\sym{***}&$-$0.002&&$-$0.012&\sym{***}\\
&&&&(0.001&)&(0.002&)&(0.001&)\\

\addlinespace\addlinespace
\multicolumn{9}{l}{\hspace{-1em} \textbf{Panel B.\ Enade scores (proportion correct answers)}} \\\midrule
Overall score&&0.566&&$-$0.024&\sym{***}&$-$0.030&\sym{**}&$-$0.027&\sym{***}\\
&&&&(0.008&)&(0.011&)&(0.009&)\\
\addlinespace
Field-specific component&&0.535&&$-$0.020&\sym{**}&$-$0.035&\sym{***}&$-$0.020&\sym{**}\\
&&&&(0.009&)&(0.011&)&(0.009&)\\
\addlinespace
General component&&0.661&&$-$0.035&\sym{***}&$-$0.014&&$-$0.048&\sym{***}\\
&&&&(0.008&)&(0.014&)&(0.007&)\\
\addlinespace

\midrule
\(N\) (programs $\times$ years)&&36&&1,664&&1,664&&1,664&\\

\(N\) (exam takers)&&747&&61,112&&16,851&&37,992&\\

\end{tabular}
\par\end{centering}
\singlespacing\justify
\emph{Notes}: This table displays triple-difference (DDD) estimates of the effect of affirmative action on the characteristics (Panel A) and scores (Panel B) of UERJ's Enade exam takers. The sample includes 2004--2015 Enade exam takers from UERJ and other federal and state universities that did not implement affirmative action until 2012 or later. See Appendix Table \ref{tab:summ_enade} for details on our Enade sample and the exam fields. Column (A) shows dependent variable means for UERJ exam takers in 2004--2006. The other columns show $\theta$ coefficients from the DDD regression:
\begin{align*}
	Y_{mjt} &= \gamma_{mj} + \gamma_{mt} + \pi_{f(m)}[\text{UERJ}_j\times\text{Post}_t] + \theta[\text{UERJ}_j\times\text{Post}_t\times\text{ExposureToAA}_m] + \varepsilon_{mjt}.
\end{align*}
Regressions are at the exam field ($m$) by institution ($j$) by year ($t$) level, with observations weighted by the number of exam takers. (In the first two rows of Panel A, we weight by the number of 2004--2006 exam takers in each $mj$ cell.) We include field $\times$ institution dummies, $\gamma_{mj}$, field $\times$ year dummies, $\gamma_{mt}$, and an indicator for UERJ interacted with an indicator for the 2007--2015 cohorts, $\text{UERJ}_j\times\text{Post}_t$. We interact $\text{UERJ}_j\times\text{Post}_t$ with dummies for five field of study groups, $f(m)$ (see Appendix \ref{app:var_def}). The variable of interest is $\text{UERJ}_j\times\text{Post}_t\times\text{ExposureToAA}$, where $\text{ExposureToAA}$ is an indicator for UERJ programs in which the share of affirmative action enrollees in the 2004--2011 cohorts was 30 percent or higher. Columns (B)--(D) include all students, white private high school students, and non-white and/or public school students. ``Predicted general score'' is the predicted value from a regression of general component scores on age and dummies for gender, race, private high school, mother's education, father's education, and family income. Parentheses contain standard errors clustered at the institution level. 

{*} $p<0.10$, {*}{*} $p<0.05$, {*}{*}{*} $p<0.01$	

}
\end{table}

\clearpage

\clearpage

\begin{table}[htp]{\scriptsize\begin{centering}
\protect\caption{DD estimates for Enade exam taker characteristics of UERJ's competitors}\label{tab:dd_enade_comp}
\newcommand\w{1.6}
\begin{tabular}{l@{}lR{2 cm}@{}L{0.43cm}R{\w cm}@{}L{0.43cm}R{\w cm}@{}L{0.43cm}R{\w cm}@{}L{0.43cm}R{\w cm}@{}L{0.43cm}R{\w cm}@{}L{0.43cm}R{\w cm}@{}L{0.43cm}}
&&(A)&&(B)&&(C)&&(D)\\\addlinespace
&&2004--06&&\\
&&mean&&\multicolumn{5}{c}{DD coefficients}\\\cmidrule{3-3}\cmidrule{5-9}
&& UFRJ && UFRJ &&  \\
Dependent variable && and UFF&&and UFF&&UFRJ&&UFF\\\midrule

White&&0.750&&0.009&&0.017&\sym{**}&$-$0.004&\\
&&&&(0.011&)&(0.007&)&(0.009&)\\
\addlinespace
Private HS student&&0.662&&$-$0.019&&$-$0.030&\sym{*}&$-$0.005&\\
&&&&(0.018&)&(0.016&)&(0.015&)\\
\addlinespace
White private HS student&&0.516&&$-$0.008&&$-$0.011&&$-$0.004&\\
&&&&(0.014&)&(0.014&)&(0.013&)\\
\addlinespace
Female&&0.532&&0.008&&0.007&&0.010&\\
&&&&(0.011&)&(0.011&)&(0.011&)\\
\addlinespace
Age&&25.746&&0.190&&0.288&&0.000&\\
&&&&(0.250&)&(0.200&)&(0.262&)\\
\addlinespace
Mother has a HS degree&&0.758&&$-$0.013&&$-$0.026&\sym{**}&0.007&\\
&&&&(0.017&)&(0.012&)&(0.012&)\\
\addlinespace
Father has a HS degree&&0.776&&$-$0.026&\sym{**}&$-$0.021&\sym{*}&$-$0.036&\sym{***}\\
&&&&(0.011&)&(0.011&)&(0.011&)\\
\addlinespace
HH income / minimum wage&&8.510&&$-$0.446&\sym{***}&$-$0.413&\sym{***}&$-$0.506&\sym{***}\\
&&&&(0.093&)&(0.124&)&(0.077&)\\
\midrule
Predicted general score&&0.568&&0.002&\sym{*}&0.002&\sym{*}&0.002&\sym{*}\\
&&&&(0.001&)&(0.001&)&(0.001&)\\

\midrule
\(N\) (programs $\times$ years)&&61&&1,729&&1,601&&1,611&\\

\(N\) (exam takers)&&2,372&&68,267&&62,398&&58,270&\\

\end{tabular}
\par\end{centering}
\singlespacing\justify
\emph{Notes}: This table displays DD estimates of the effect of UERJ's adoption of affirmative action on the characteristics of Enade exam takers at two of its competitor schools: UFRJ and UFF. The sample includes 2004--2015 Enade exam takers from UFRJ, UFF, and other federal and state universities nationwide that did not implement affirmative action until 2012 or later. See Appendix Table \ref{tab:summ_enade} for details on our Enade sample and the exam fields. Column (A) shows dependent variable means for UFRJ and UFF exam takers in 2004--2006. The other columns show $\pi$ coefficients from the DD regression:
\begin{align*}
	Y_{mjt} &= \gamma_{mj} + \gamma_{mt} + \pi[\text{Competitor}_j\times\text{Post}_t] + \varepsilon_{mjt}.
\end{align*}
Regressions are at the exam field ($m$) by institution ($j$) by year ($t$) level, with observations weighted by the number of exam takers (except in the first row of Panel A, where we weight by the number of 2004--2006 exam takers in each $mj$ cell.) We include dummies for field $\times$ institution, field $\times$ year, and the interaction between indicators for UFRJ/UFF and the 2007--2015 cohorts ($\text{Competitor}_j\times\text{Post}_t$). In column (B), the sample includes UFRJ, UFF, and all schools in the control group. Column (C) includes only UFRJ plus control schools, and column (D) includes only UFF plus control schools. ``Predicted general score'' is the predicted value from a regression of general component scores on age and dummies for gender, race, private high school, mother's education, father's education, and family income. Parentheses contain standard errors clustered at the institution level. 

{*} $p<0.10$, {*}{*} $p<0.05$, {*}{*}{*} $p<0.01$	

}
\end{table}


\clearpage

\renewcommand \thesubsection{\thesection.\arabic{subsection}}\setcounter{subsection}{0}
\renewcommand{\thetable}{\thesection\arabic{table}}\setcounter{table}{0}
\renewcommand{\thefigure}{\thesection\arabic{figure}}\setcounter{figure}{0}
\renewcommand{\theequation}{\thesection\arabic{equation}}\setcounter{equation}{0}

\section{Empirical appendix}

\subsection{Variable definitions\label{app:var_def}}

This section describes the main variables in our paper. 

\begin{itemize}[leftmargin=*]

\item
\textbf{Admission score.} Raw admission scores are based on applicants' subject scores on different subjects of the field exam (\emph{Exame discursivo}), plus bonus points from their qualifying exam performance (\emph{exame de qualifica\c{c}\~ao}).  We standardize raw admission scores to represent an individual's distance from the admission cutoff in their application pool in SD units. For this, we subtract the score of the last admitted student in the application pool, and divide by the SD of scores for all applicants to the same program/cohort. We adjust these SDs to be comparable across cohorts because the number of field exam takers varies significantly over time due to changes in UERJ's standards for the qualification exam.

\item
\textbf{Alumni firm employment.} We define UERJ applicant $i$ to major $m$ as obtaining a job at an alumni firm if their firm ever employed another individual $j$ who graduated from major $m$ (the ``alum''). We define different versions of this variable based on the alum's cohort, application track, and year of employment. We also define versions that classify firms based on the number of alumni they hired as a proportion of their mean size over all years of our data.

\item
\textbf{Demographic characteristics.} From the UERJ admission data, we observe age at application, gender, race, mother's education, and household income. Age is available in all cohorts; other characteristics are available only in certain cohorts (see Appendix Table \ref{tab:bal}). These variables come from a survey that applicants completed as part of the application process. We also use gender and race from the RAIS data, which we observe for any applicant who appears in this dataset. We use indicators for three racial groups: \emph{branco} (white), \emph{pardo} (brown), and \emph{preto} (Black).\footnote{\enskip In Brazil, race is commonly classified in five groups: \emph{branco} (white), \emph{pardo} (brown), \emph{preto} (Black), \emph{amarelo} (yellow), and indigenous. UERJ's race-based quota was reserved for individuals who self-identified as Black; this occasionally differs from their racial identity reported in the entrance exam or RAIS data.}

\item 
\textbf{Field exam subject scores.} An applicant's scores on subjects of the field exam (\emph{exame discursivo}). We use an applicant's writing exam score (which is common to all applicants), and their mean score across 2--4 other subjects (which vary depending on the cohort and major they are applying to). We observe field exam subject scores in the 1995--2001 and 2007--2011 cohorts.

\item
\textbf{Field of study.} Our benchmark DD regressions (equation \ref{dd}) include controls for five field of study groups: business, health, humanities, natural science, and social science. We define these field of study groups using UERJ's classification of majors into four areas, but we create a fifth group for business given other research showing that networking plays an important role in this field \citep{zimmerman2019}. The programs in each field of study group are listed below. Program names in italics are those that appear in both our RD and DD samples (Panel A of Table \ref{tab:summ_stats}), while programs in regular text are those that appear only in our DD sample (Panel B of Table \ref{tab:summ_stats}).
\begin{itemize}
\item \textbf{Business}: \emph{Accounting}, \emph{Business Administration}, Economics.
\item \textbf{Health}: \emph{Biological sciences}, \emph{Dentistry}, \emph{Medicine}, \emph{Nursing}, \emph{Nutrition}.
\item \textbf{Humanities}: \emph{Greek/Latin/Literature}, \emph{History Ed.\ (SGO)}, \emph{Journalism}, \emph{Psychology}, Art, Biological Sciences (SGO), English/German/Japanese, Geography Ed.\ (SGO), Language (SGO), Math Ed.\ (SGO), Teaching, Teaching (DDC), Physical Ed., Spanish/French/Italian.
\item \textbf{Natural sciences}: \emph{Chemical Engineering}, \emph{Chemistry}, \emph{Computer Science}, \emph{General Engineering}, \emph{Geography}, \emph{Geology}, \emph{Industrial Design}, \emph{Mechanical Engineering}, \emph{Production Engineering}, Cartographic Engineering, Math, Mechanical Engineering (NF), Oceanography, Physics, Production engineering (RES), Statistics.
\item \textbf{Social sciences}: \emph{History}, \emph{Law}, \emph{Social Science}, \emph{Social Work}, Philosophy.
\end{itemize}

\item
\textbf{Firm.} We define firms at the establishment level. Establishments are identified by their 14-digit CNPJ (short for \textit{Cadastro Nacional da Pessoa Jur\'idica}, or National Registry of Legal Entities). The CNPJ is a tax identifier for legally incorporated identities. The first eight digits identify the company. The rest of the digits identify the branch or subsidiary of the company.

\item
\textbf{Firm mean wage.} The leave-individual-out mean hourly wage at a given firm. 

\item
\textbf{Firm size.} The total number of workers employed by the firm at the end of each year. Throughout the paper, we define firm size using its mean number of workers over all years of our data.

\item
\textbf{Formal employment.} An indicator that takes the value one if an applicant appears in the employee-employer matched dataset (RAIS).

\item
\textbf{Hourly wage.} We calculate the hourly rate of each worker as the ratio between a worker's inflation-adjusted monthly earnings and the hours worked per month. Hours worked reflects the number of hours per week at which the firm hired the worker according to the worker's contract, which may differ from the hours actually worked in any given week.

\item
\textbf{Industry mean wage.} The leave-individual-out mean hourly wage of all UERJ applicants working in a given industry. We define industries at the 4- or 5-digit level of the Brazilian National Classification of Economic Activities (\textit{Classifica\c{c}\~ao Nacional de Atividades Economicas}) level.

\item
\textbf{Monthly earnings (2019 USD).} This variable represents a worker's average monthly salary in a given year. To report this variable, establishments have to calculate the worker's total earnings for the year and divide them by the number of months the firm employed the worker. We adjust earnings for inflation using the consumer price index. We express earnings in 2019 US dollars using the Brazilian Real/US Dollar exchange rate.

\item
\textbf{Municipality mean wage.} The leave-individual-out mean hourly wage of all UERJ applicants working at a given municipality. Municipalities are defined by the location of the worker's establishment.

\item
\textbf{Occupation mean wage.} The leave-individual-out mean hourly wage of all UERJ applicants with a given occupation. We define occupations at the 4-digit of the Brazilian Occupational Code Classification (\textit{Classifica\c{c}\~ao Brasileira de Ocupa\c{c}\~oes}) level.

\item
\textbf{Qualifying exam score.} An applicant's standardized score from the qualifying exam (\emph{exame de qualifica\c{c}\~ao}). This exam includes eight subject tests common to all applicants: Biology, Chemistry, Geography, Foreign language (French, Spanish, or English), History, Literature/Portuguese, Mathematics, and Physics. Students that fail to achieve a minimum score on the qualifying exam cannot take the second round discursivo exam. We observe qualifying exam scores in the 1995--2001 cohorts.

\end{itemize}

We measure graduation and labor market outcomes in two time periods: 6--9 years after application, and 10--13 years after application. For earnings and wage indices, we use the mean value over each four-year period. For binary variables, we use the maximum value over the four year period.

\subsection{Data and merging} \label{app:merge}

Our base dataset includes a list of all individuals who passed UERJ's first-round admission exam and applied to an undergraduate program in the years 1995--2001 and 2004--2011 \citep{uerj2020applicants}. This dataset includes the program(s)/cohort that each individual applied to, their admission score in the second exam of the admission process, and their admission decisions. The 2004--2011 records include the track each applicant applied through. In addition, we have access to socioeconomic variables for the 2007--2011 application cohorts.

We combine the UERJ admission records with two individual-level administrative datasets. The first dataset is from UERJ, and it includes the graduation outcomes of all the students who enrolled in UERJ in 1995--2011 \citep{uerj2020enrollees}. These records contain the student's program, enrollment date, status as of December 2020 (i.e., graduated, dropped out, or still enrolled), and final year in the program. 

The second administrative dataset is called the \textit{Rela\c{c}\~{a}o Anual de Informa\c{c}\~{o}es Sociais}, or RAIS \citep{rais2021}. It includes employment outcomes collected by the Ministry of Labor. We have access to the RAIS for the 2003--2019 period. This dataset has information on all workers with a formal-sector job. The RAIS contains information about both the worker and the firm. Worker information includes demographic variables (e.g., age, gender, and race), educational attainment, occupation, hours worked, and earnings. Firm-level variables include the number of employees, industry, and geographic location.

We merge the admission data with the graduation records using the university ID of each individual. Most individuals match uniquely on the ID, but in cases with duplicated IDs, we corroborate the quality of the matches using individuals' names and programs. We fix a few cases in which different individuals have the same university ID. We match 94.8 percent of individuals in the graduation records to the admission records using the university IDs. We use the names and application years of the remaining unmatched individuals to match them to the graduation records. Overall, we match 97.8 percent of the individuals in the graduation records to the admission records.

Lastly, we link the combined dataset from the above merge to the RAIS dataset using individuals' national ID numbers (\textit{Cadastro de Pessoas F\'isicas}, or CPF for short), birth dates, and names. For this, we follow a two-step procedure. First, we match individuals for whom we have the CPF available in the UERJ records.\footnote{\enskip The UERJ records contain the CPF nearly all individuals who applied in 2000--2001 and 2004--2011. Before 2000, the CPF is rarely available. Virtually all workers in the 2003--2019 RAIS datasets have a CPF.} Second, for individuals who remain unmatched, we merge them using their names and dates of birth. We define a match from this process as observations that have either: 1) the same CPF number; or 2) the same birth date and an exact name match. We match 77.4 of the individuals in merged UERJ records to at least one year of the RAIS through this process. Out of the matched individuals, 66.1 percent were matched using the CPF, and the remaining 33.9 percent were matched using names and dates of birth.

One way to benchmark the merge rate with the RAIS is to compare it with the share of individuals with similar demographic characteristics who have a formal-sector job in Brazil. To do this, we use data from the 2015 Brazilian household survey (\textit{Pesquisa Nacional por Amostra de Domic\'ilios}, abbreviated PNAD), which includes information on the informal economy. Our proxy of working in the formal sector is having the right to a pension when retired.\footnote{\enskip International organizations define informality in two different ways. Under the \textit{legal} definition, a worker is considered informal if she does not have the right to a pension when retired. An alternative to the legal definition is the \textit{productive} definition, where a worker is considered informal if she is a salaried worker in a small firm (i.e., it employs less than five workers), a non-professional self-employed, or a zero-income worker. We use the legal definition in the main text. The share of workers with a formal job under the productive definition is slightly lower than the one based on the legal definition.} The share of economically active individuals aged 25--37 with at least a high-school degree who have a job in the formal sector is 62.4 percent. This suggests that our merge identified most individuals with formal sector jobs.

\subsection{UERJ's admission process} \label{app:adm}

\begin{table}[tbp!]{\scriptsize
		\begin{center}
			\caption{Timeline of events during the 2010 admission process}\label{tab:adm}
			\begin{tabular}{clr}
				\midrule
				& Event & Date \\
				\midrule
				 & First date for applicants to take the qualifying exam & 06/21/2009 \\
				 & Results of the qualifier exam are published & 07/01/2009 \\
				 & Second date for applicants to take the qualifying exam & 09/13/2009 \\
				 & Results of the qualifier exam are published & 09/23/2009 \\
				 & Applicants who passed the qualifier exam take the field exam & 12/13/2009 \\
				 & Results of the field exam are published & 01/16/2010 \\
				 & Results of the field exam are published & 01/30/2010 \\
				 & First round of admission offers is sent & 01/30/2010 \\
				 & Second round of admission offers is sent & 02/12/2010 \\
				 & Admitted students can enroll in first-semester programs & 03/02/2010 -- 03/03/2010 \\
				 & \textbf{First day of classes - 1st semester} & \textbf{03/10/2010} \\
				 & Third round of admission offers is sent & 03/16/2010 \\
				 & Fourth round of admission offers is sent & 07/02/2010 \\
				 & Fifth round of admission offers is sent & 07/16/2010 \\
				 & Newly admitted applicants can enroll in second-semester programs & 07/28/2010 -- 07/29/2010 \\
				 & \textbf{First day of  classes - 2nd semester} & \textbf{08/10/2010} \\
				\midrule
			\end{tabular}
		\end{center}

\singlespacing\justify
\emph{Notes}: This calendar is summarized from information in these two UERJ documents:
\begin{itemize}
\item \href{http://sistema.vestibular.uerj.br/portal_vestibular_uerj/arquivos/arquivos2010/ed/03_anexo1_WEB.pdf}{http://sistema.vestibular.uerj.br/portal\_vestibular\_uerj/arquivos/arquivos2010/ed/03\_anexo1\_WEB.pdf}
\item \href{http://sistema.vestibular.uerj.br/portal_vestibular_uerj/arquivos/arquivos2010/calendario/calendario_eq.pdf}{http://sistema.vestibular.uerj.br/portal\_vestibular\_uerj/arquivos/arquivos2010/calendario/calendario\_eq.pdf}
\end{itemize}
}
\end{table}

Applicants can gain admission to UERJ at one of several stages. The admission process begins with applicants taking a common qualifying exam. Applicants who pass this exam then take a field exam. UERJ ranks applicants based on their field exam scores and sends admissions offers to accepted applicants up to the capacity of each program. The remaining applicants are either rejected (if their score in the field exam is below a minimum threshold) or waitlisted. The first admission offers are typically sent in January, and admitted students have several weeks to accept or reject their offer. UERJ sends a second round of admission offers to waitlisted applicants based on the number of offers that were declined. This process is repeated up to five times per application year if there are remaining open seats, and the last admission offers may occur as late as July. Appendix Table \ref{tab:adm} provides an example of this process for the 2010 cohort.

The admission thresholds in our RD analysis are given by the admission score of the final student who gained admission in each application pool (after all waitlist offers). Any applicant who scored above this threshold could have been admitted to UERJ, although some of these students chose to enroll in other universities by the time they would have gotten in off the waitlist. Potential for non-random sorting around the admission cutoff arises because applicants have control over whether they accept or reject their admission offer. Students just above the final cutoff may therefore be those who particularly want to attend UERJ. We present tests for non-random sorting around the admission cutoff in Section \ref{sub:ident}.

\subsection{Sample} \label{app:samp}

Our initial dataset includes all applicants to UERJ undergraduate majors who passed the first-round qualifying exam and who have a valid second-round admission score (i.e., non-missing, non-zero). UERJ has several campuses; its main campus is in the municipality of Rio de Janeiro, and it has five smaller campuses in other municipalities in the state: Baixada Duque de Caixas (DDC), Nova Friburgo (NF), Resende (RES), S\~ao Gon\c{c}alo (SGO), and Teres\'opolis (TER). The number of undergraduate programs changes across cohorts of our data because UERJ split some large programs into smaller ``sub-programs'' and added some new majors.

Our raw data includes 71 different sub-programs across all cohorts and campuses.  We group these 71 sub-programs into 43 programs to create a consistent set over time. We create these groups using documentation from UERJ detailing how large programs were divided into sub-programs. We exclude six new majors that UERJ created after the introduction of affirmative action: computing engineering (NF), geography (DDC), math (DDC), pedagogy (SGO), tourism (TER), and actuarial sciences (RIO). Appendix Tables \ref{tab:samp-gen}-\ref{tab:samp-blk} show the 43 programs in our data and the sub-programs that they are derived from.

We use data from these 43 programs to create two different samples to analyze the impacts of UERJ's affirmative action policy. For our RD sample, we exclude programs where fewer than 30 percent of the 2004--2011 students entered through an affirmative action track. The second column in Appendix Tables \ref{tab:samp-gen}-\ref{tab:samp-blk} shows the percent of students that entered through an affirmative action track in each program group during 2004--2011. Bolded figures denote programs where this figure is above 30 percent. 24 programs meet this criteria. Within these programs, we also exclude program-cohort-admission track triplets with fewer than five applicants below the admission threshold. We also exclude all applicants to the disabled/indigenous track since these quotas rarely filled up. In Appendix Tables \ref{tab:samp-gen}-\ref{tab:samp-blk}, we highlight in bold the program-cohort pairs in each admission track that satisfy our sample restrictions and appear in our RD sample.

For our DD sample, we focus on applicants who \textit{enrolled} in UERJ. Our DD sample includes the 24 programs in our RD sample plus 19 other programs with lower take-up rates in the affirmative action tracks. These programs are unbolded in Appendix Tables \ref{tab:samp-gen}-\ref{tab:samp-blk}.

\clearpage

\begin{landscape}
\begin{table}[htp]{\tiny\begin{centering}\vspace{-2em}
\protect\caption{Number of applicants by cohort  --- General track}\label{tab:samp-gen}
\newcommand\s{1}
\begin{tabular}{rlrlrrrrrrrrrrrrrrrrrrrrrrrrrr}
$\#$ & Program & Prop.\  AA & Program name(s)  &  1995 & 1996 & 1997 & 1998 & 1999 & 2000 & 2001 & 2004 & 2005 & 2006 & 2007 & 2008 & 2009 & 2010 & 2011 \\ \midrule
1 & Accounting & \textbf{0.364} & Accounting & \textbf{47} &17 &25 &17 &13 &23 & \textbf{33} & \textbf{42} \\\midrule
2 & Art &0.287 & Artistic education & & & & & & & & \\
 & & & Art &53 &12 &10 & & & & & \\
 & & & Art history & & & &3 & & &6 &7 \\
 & & & Visual arts (bach.) & & & &3 & & & & \\
 & & & Visual arts (license) & & & &2 & & & & \\
 & & & Visual arts & & & & &6 &8 &14 &6 \\\midrule
3 & Biology & \textbf{0.494} & Biology & \textbf{64} & \textbf{28} & \textbf{28} & \textbf{32} &14 & \textbf{39} & \textbf{33} & \textbf{44} \\\midrule
4 & Biology (SGO) &0.260 & Biology  &28 &9 &13 &8 &4 &7 &7 &6 \\\midrule
5 & Business & \textbf{0.428} & Business & \textbf{59} & \textbf{38} & \textbf{42} &17 &25 & \textbf{54} & \textbf{42} & \textbf{47} \\\midrule
6 & Cartographic eng. &0.126 & Cartographic eng. &4 &1 &2 &2 & &2 &9 &5 \\\midrule
7 & Chemical eng. & \textbf{0.465} & Chemical eng. &26 &13 &18 &16 &16 & \textbf{34} & \textbf{43} & \textbf{47} \\\midrule
8 & Chemistry & \textbf{0.352} & Chemistry & \textbf{19} &6 &10 &8 &2 &14 &8 &15 \\\midrule
9 & Computer science & \textbf{0.325} & Information science &63 &19 & \textbf{30} &20 &17 &14 &21 & \\
 & & & Computer science & & & & & & & &17 \\\midrule
10 & Dentistry & \textbf{0.404} & Dentistry & \textbf{42} &5 & \textbf{18} &12 &16 &14 & \textbf{24} & \textbf{24} \\\midrule
11 & Economics &0.286 & Economics &36 &14 &20 &5 &6 &25 &30 &22 \\\midrule
12 & General eng. & \textbf{0.307} & Civil eng. &21 &7 &12 &11 &9 &25 & \textbf{31} & \textbf{70} \\
 & & & Engineering & & & & & & & & \\
 & & & Electrical eng. & \textbf{61} &22 &30 &25 &13 &32 &40 & \textbf{62} \\
 & & & Textile eng. & & & & & & & & \\\midrule
13 & Geography & \textbf{0.468} & Geography & \textbf{49} &23 &18 &14 &19 &21 & \textbf{33} &22 \\\midrule
14 & Geog.\ Ed.\ (SGO) &0.275 & Geography  &52 &21 &18 &12 &11 &7 &15 &11 \\\midrule
15 & Geology & \textbf{0.321} & Geology &7 &4 &4 &2 &6 & \textbf{13} &12 & \textbf{13} \\\midrule
16 & History & \textbf{0.457} & History & \textbf{124} & \textbf{37} & \textbf{39} & \textbf{37} &24 & \textbf{50} & \textbf{36} & \textbf{31} \\\midrule
17 & Hist.\ Ed.\  (SGO) & \textbf{0.333} & History  & \textbf{44} &10 &19 &11 &11 &16 &11 &9 \\\midrule
18 & Industrial design & \textbf{0.456} & Industrial design & \textbf{25} &6 & \textbf{14} &7 &8 & \textbf{21} &8 & \textbf{18} \\\midrule
19 & Journalism & \textbf{0.481} & Social communication & & & & & & & & \\
 & & & Journalism & \textbf{67} & \textbf{13} & \textbf{25} &19 & \textbf{23} & \textbf{29} & \textbf{31} & \textbf{31} \\
 & & & Public relations & \textbf{36} & \textbf{22} & \textbf{22} &11 &8 & \textbf{25} & \textbf{30} & \textbf{32} \\\midrule
20 & Language (SGO) &0.205 & Language  &47 &15 &16 &6 &3 &10 &4 &9 \\\midrule
21 & Language I &0.259 & Language I & & & & & & & & \\
 & & & Literature/English &31 &8 &6 &4 &5 &11 &7 &12 \\
 & & & Port./German &4 &1 & &2 &1 &1 & & \\
 & & & Port./Japanese &6 & &1 &1 & &2 &2 & \\\midrule
  \addlinespace

\end{tabular}
\par\end{centering}
}
\end{table}
\end{landscape}

\clearpage

\begin{landscape}
\begin{table}[htp]{\tiny\begin{centering}\vspace{-2em}
\ContinuedFloat\caption{Number of applicants by cohort  --- General track \textit{(continued)}}
\newcommand\s{1}
\begin{tabular}{rlrlrrrrrrrrrrrrrrrrrrrrrrrrrr}
$\#$ & Program & Prop.\  AA & Program name(s)  &  1995 & 1996 & 1997 & 1998 & 1999 & 2000 & 2001 & 2004 & 2005 & 2006 & 2007 & 2008 & 2009 & 2010 & 2011 \\ \midrule
22 & Language II &0.293 & Language II & & & & & & & & \\
 & & & Port./France &10 &3 &2 &4 &4 &2 &5 &3 \\
 & & & Port./Italian &7 &12 &3 &1 &2 &2 &2 &4 \\
 & & & Port./Spanish &36 &9 &18 &17 &9 &6 &7 &8 \\\midrule
23 & Language III & \textbf{0.379} & Language III & & & & & & & & \\
 & & & Port./Greek &7 & &3 &1 & &1 &2 &1 \\
 & & & Port./Latin &6 &4 &2 &5 & &3 &1 & \\
 & & & Port./Literature & \textbf{71} &18 & \textbf{24} &12 &11 &13 &13 &19 \\\midrule
24 & Law & \textbf{0.460} & Law & \textbf{271} & \textbf{89} & \textbf{138} & \textbf{122} & \textbf{125} & \textbf{174} & \textbf{192} & \textbf{247} \\\midrule
25 & Math &0.158 & Math &54 &16 &9 &15 &10 &5 &7 &6 \\\midrule
26 & Math Ed.\ (SGO) &0.143 & Math  &16 &2 &1 &1 &1 &1 &3 &3 \\\midrule
27 & Mech. eng. & \textbf{0.353} & Mech. eng. & \textbf{28} &7 &10 &9 &11 & \textbf{22} & \textbf{18} & \textbf{27} \\\midrule
28 & Mech. eng. (NF) &0.186 & Mech. eng.  &3 &1 &2 &2 & &6 &7 &8 \\\midrule
29 & Medicine & \textbf{0.454} & Medicine & \textbf{123} & \textbf{38} & \textbf{53} & \textbf{58} & \textbf{48} & \textbf{114} & \textbf{120} & \textbf{187} \\\midrule
30 & Nursing & \textbf{0.431} & Nursing & \textbf{109} &17 & \textbf{35} &19 &20 & \textbf{32} & \textbf{23} & \textbf{26} \\\midrule
31 & Nutrition & \textbf{0.411} & Nutrition & \textbf{60} & \textbf{24} & \textbf{25} &10 &16 &22 &19 &28 \\\midrule
32 & Oceanography &0.229 & Oceanography &11 &6 &9 &2 &4 &2 &6 &2 \\\midrule
33 & Teaching &0.234 & Teaching &157 &40 &57 &25 &16 &27 &23 &34 \\\midrule
34 & Teaching (DDC) &0.208 & Teaching  & & & &12 &12 &15 &8 &11 \\
 & & & Teaching I  &55 &13 &17 & & & & & \\
 & & & Teaching II  &17 &8 &5 & & & & & \\\midrule
35 & Philosophy &0.222 & Philosophy &52 &11 &9 &4 &2 &14 &6 &4 \\\midrule
36 & Physical ed. &0.206 & Physical education &56 &6 &24 &9 &5 &6 &11 &12 \\\midrule
37 & Physics &0.135 & Physics &37 &7 &10 &8 &4 &9 &8 &4 \\\midrule
38 & Prod. eng. & \textbf{0.384} & Prod. eng. &9 &6 &5 &3 &12 & \textbf{23} &10 & \textbf{36} \\\midrule
39 & Prod. eng. (RES) &0.152 & Prod. eng.  &14 &5 &8 &2 &1 &5 &7 &7 \\\midrule
40 & Psychology & \textbf{0.480} & Psychology & \textbf{123} & \textbf{32} & \textbf{37} & \textbf{28} &28 & \textbf{43} & \textbf{42} & \textbf{53} \\\midrule
41 & Social science & \textbf{0.408} & Social sciences & \textbf{56} &19 &23 &22 &22 &15 & \textbf{34} &29 \\\midrule
42 & Social work & \textbf{0.432} & Social work & \textbf{119} & \textbf{36} & \textbf{48} & \textbf{27} &30 & \textbf{41} &22 & \textbf{52} \\\midrule
43 & Statistics &0.087 & Statistics &21 &4 &8 & &1 &2 &3 &6 \\\midrule

\end{tabular}
\par\end{centering}
\singlespacing\justify
\emph{Notes}: This table displays the number of applicants in the general track for each program/cohort in our data. See Table \ref{tab:samp-blk} for details on the table structure and statistics.

}
\end{table}
\end{landscape}

\clearpage

\begin{table}[htp]{\tiny\begin{centering}
\protect\caption{Number of applicants by cohort  --- Public high school track}\label{tab:samp-hs}
\newcommand\s{2}
\begin{tabular}{rlrlrrrrrrrrrrrrrrrrrrrrrrrrrr}
$\#$ & Program & Prop.\  AA & Program name(s)  & 2004 & 2005 & 2006 & 2007 & 2008 & 2009 & 2010 & 2011 \\ \midrule
  \addlinespace

\end{tabular}
\par\end{centering}
}
\end{table}

\clearpage

\begin{table}[htp]{\tiny\begin{centering}
\ContinuedFloat\caption{Number of applicants by cohort  --- Public high school track \textit{(continued)}}
\newcommand\s{2}
\begin{tabular}{rlrlrrrrrrrrrrrrrrrrrrrrrrrrrr}
$\#$ & Program & Prop.\  AA & Program name(s)  & 2004 & 2005 & 2006 & 2007 & 2008 & 2009 & 2010 & 2011 \\ \midrule
  \addlinespace
\end{tabular}
\par\end{centering}
\singlespacing\justify
\emph{Notes}: This table displays the number of applicants in the public high school track for each program/cohort in our data. See Table \ref{tab:samp-blk} for details on the table structure and statistics.

}

\end{table}

\clearpage

\begin{table}[htp]{\tiny\begin{centering}
\protect\caption{Number of applicants by cohort  --- Black track}\label{tab:samp-blk}
\newcommand\s{3}
\begin{tabular}{rlrlrrrrrrrrrrrrrrrrrrrrrrrrrr}
$\#$ & Program & Prop.\  AA & Program name(s)  & 2004 & 2005 & 2006 & 2007 & 2008 & 2009 & 2010 & 2011 \\ \midrule
  \addlinespace

\end{tabular}
\par\end{centering}
}
\end{table}

\clearpage

\begin{table}[htp]{\tiny \begin{centering}
\ContinuedFloat\caption{Number of applicants by cohort --- Black track \textit{(continued)}}
\newcommand\s{3}
\begin{tabular}{rlrlrrrrrrrrrrrrrrrrrrrrrrrrrr}
$\#$ & Program & Prop.\  AA & Program name(s)  & 2004 & 2005 & 2006 & 2007 & 2008 & 2009 & 2010 & 2011 \\ \midrule
  \addlinespace
\end{tabular}
\par\end{centering}
\singlespacing\justify
\emph{Notes}: This table displays the number of applicants in the Black track for each program/cohort in our data. The first column shows the 43 programs in our RD and DD samples. The second column shows the proportion of 2004--2011 enrollees in each who were from any affirmative action track ($y$-axis of Figure \ref{fig:variation_aa}); \textbf{bold} numbers in this column show programs with $\text{ExposureToAA}_m =1$ in our benchmark DD specification \eqref{dd}. The third column shows the subgroups that comprise each program. Remaining columns show the number of applicants to each program/cohort; \textbf{bold} numbers denote program/cohorts that we include in our RD sample.

}
\end{table}

\clearpage

\subsection{Fuzzy merge of UERJ and higher education census data} \label{app:censo-superior}

In Section \ref{sec:rd-returns}, we examine the effects of UERJ enrollment on college selectivity using data from a census of all Brazilian college enrollees, called the \emph{Censo da Educa\c{c}\~ao Superior} \citep{inep2019superior}. This subsection describes the merge between UERJ applicants and the higher education census.

We focus on universities in the state of Rio de Janeiro since most Brazilian college students enroll in a university in their home state. We include only 2009--2011 UERJ applicants in this analysis because the higher education census does not exist at the individual level prior to 2009.

We do not observe individuals' ID numbers in the higher education census, so we link the census to the UERJ records using a fuzzy merge based on exact day of birth, gender, and year of enrollment.\footnote{\enskip Each of these variables is available in the public version of the \emph{Censo da Educa\c{c}\~ao Superior} that we downloaded from the website of a Brazilian Ministry of Education agency called INEP (\emph{Instituto Nacional de Estudos e Pesquisas Educacionais Anísio Teixeira}). Some of these variables are no longer available in the current version of this dataset on INEP's website.} In the census data, we compute the \emph{total} number of students at a particular university with a given birthdate, gender, and enrollment year. We merge these variables into our UERJ sample using birthdate, gender, and year of \emph{application}. We then use these totals as dependent variables in our RD specification.

The resulting dependent variables reflect the total number of enrollees in a particular university in Rio de Janeiro who have the same birthdate/gender/enrollment-year triplet as a UERJ applicant. The ideal dependent variable---if we could uniquely identify individuals in the census---would be an indicator variable that takes the value one if a given UERJ applicant enrolled in a given university and zero otherwise. If no college student at a university has the same birthdate/gender/enrollment-year triplet as the applicant, we know that the applicant did not enroll in that university in that year (barring errors in the merge variables). However, if one or more enrollees at the university share the same combination of those three variables, we cannot tell with certainty whether the applicant ended up enrolling in the university.\footnote{\enskip For the average UERJ applicant in our data, there are 29 students with the same birthdate, gender, and enrollment year across all Rio de Janeiro universities.} Thus our dependent variables contain additional measurement error. 

\subsection{Decomposition of DD estimate for top enrollees' log hourly wages}\label{app:decomp}

Our main result in Section \ref{sec:dd-spillovers} is that top enrollees' hourly wages declined by 0.132 log points in UERJ majors with high exposure to affirmative action relative to less-exposed majors (Table \ref{tab:dd_labor}, Panel B, column B). This section describes back-of-the-envelope calculations on the proportion of this estimate that can be explained by compositional, networking, and learning mechanisms.

First, we find that 25 percent of our main DD estimate can be explained by changes in the observable characteristics of top enrollees. In Panel D of Table \ref{tab:dd_traits} we combine applicants' observable characteristics---both demographics and admission scores---into a log wage index, and use this as an outcome variable in our DD specification \eqref{dd}. For individuals who appear in the RAIS data, the observable characteristics of top enrollees declined by 0.033 log wage points in majors with greater exposure to affirmative action relative to less-exposed majors (last row in Panel D of Table \ref{tab:dd_traits}, column B). Although this point estimate is not statistically significant, it is 25 percent of the magnitude of our main DD estimate for top enrollees' log hourly wages, i.e., $-0.033/(-0.132)=25\%$.

Next, we find that 10--17 percent of our DD coefficient for top enrollees' hourly wages can be explained by networking mechanisms. We estimate the contribution of networking mechanisms by multiplying the DD estimates for employment in alumni firms (Panels C--D of Table \ref{tab:dd_labor}) by the OLS wage premia associated with employment in these firms (Appendix Table \ref{tab:alumni_ols}). In Panel C of Table \ref{tab:dd_labor}, we find that greater exposure to affirmative action reduced top enrollees' likelihood of employment at firms that hired pre-AA alumni by 5.5 percentage points, and it increased their likelihood of employment at firms that hired \emph{only} post-AA alumni by 4.9 percentage points. Appendix Table \ref{tab:alumni_ols} shows that firms with pre-AA alumni had an hourly wage premium of 0.481 log points relative to other firms in our sample, whereas firms with only post-AA alumni had an hourly wage premium of 0.274 log points. Under the assumption that these OLS wage premia reflect causal effects, the change in access to pre-AA and post-AA alumni firms can explain 10 percent of our main DD estimate for hourly wages, i.e., $(-0.055*0.481+0.049*0.274)/(-0.132) \approx 10\%$. If we do a similar calculation using the estimates from Panel D of Table \ref{tab:dd_labor}---which measure employment with alumni from different application tracks and cohorts---we find that these estimates can explain 17 percent of our main DD estimate for hourly wages.\footnote{\enskip Specifically, the DD estimates in Panel D of Table \ref{tab:dd_labor} and their associated OLS firm wage premia are:
\begin{itemize}
\item General track alumni from the same cohort: DD coef = $-0.098$; OLS wage premium = 0.533;
\item General track alumni from different cohorts: DD coef = $0.042$; OLS wage premium = 0.455;
\item Only AA alumni from the same cohort: DD coef = $0.036$; OLS wage premium = 0.293;
\item Only AA alumni from different cohorts: DD coef = $0.010$; OLS wage premium = 0.070.
\end{itemize}
Thus we can explain $(-0.098*0.533+0.042*0.455+0.036*0.293+0.010*0.070)/(-0.132) \approx 17\%$ of our main DD estimate for hourly wages.
}

Lastly, we estimate that 32 percent of the negative wage effect for top enrollees can be explained by learning mechanisms. We find that affirmative action reduced UERJ's white private high school students' proportion of correct answers on the Enade exam by 2.2 percentage points (first row in Panel B of Table \ref{tab:dd_enade}, column C).\footnote{\enskip We find a slightly larger point estimate ($-$3.0pp) in a triple-difference specification that also compares UERJ majors with more- and less-exposure to affirmative action (Appendix Table \ref{tab:dd_prog_enade}).} All else equal, white students from private high schools would have had to obtain \emph{higher} entrance exam scores to be admitted to UERJ in the cohorts with affirmative action.\footnote{\enskip However, we find limited evidence that affirmative action changed the demographics characteristics of UERJ's white private high school students who took the Enade exam (Panel A of Table \ref{tab:dd_enade}).} Thus we think that the 2.2 percentage point decline in Enade scores is, if anything, likely to underestimate the decline in scores that we would find in our top enrollee sample. We cannot estimate the relationship between hourly wages and the proportion of correct answers on the Enade exam because we do not have these two datasets linked at the individual level. As an alternative benchmark, we use \citet{reyes2022endurance}'s estimate that a one percentage point increase in the proportion of correct answers on Brazil's national college entrance exam (ENEM) is associated with a 0.0192 log point increase in early-career wages. Under the assumption that the relationship between correct answers and wages is the same for the ENEM and Enade exams, the decline Enade scores for white private high school students can explain 32 percent of our DD estimate for top enrollees' wages, i.e., $(-2.2*0.0192)/(-0.132) = 32\%$.

\end{document}